\documentclass[%
 superscriptaddress,
 reprint,
 showpacs,preprintnumbers,
 nofootinbib,
 amsmath,amssymb,
 aps,
 prd, %prd
 floatfix,
]{revtex4-1}

\usepackage[table,xcdraw]{xcolor}

\usepackage{amsmath}
\usepackage{amssymb}
\usepackage{float}

\usepackage{graphicx}% Include figure files
\usepackage{dcolumn}% Align table columns on decimal point
\usepackage{bm}% bold math
\usepackage{afterpage}
\usepackage{tikz}
\usepackage{textgreek}
\usepackage{caption}
\RequirePackage[colorlinks,citecolor=blue,urlcolor=blue,linkcolor=blue]{hyperref}
\usepackage{todonotes}

\newcommand{\pt}{p_\textrm{T}}

\newcommand{\dalphat}{\delta\alpha_\textrm{T}}
\newcommand{\dphit}{\delta\phi_\textrm{T}}
\newcommand{\dpt}{\delta \pt}

\usepackage{xspace}
\newcommand{\argenie}{\texttt{AR23\_20i\_00\_000}\xspace}

\newcommand{\affiliationCERN}{European Organization for Nuclear Research (CERN), 1211 Geneva 23, Switzerland.}

\begin{document}

\title{Benchmarking neutrino interaction models via a comparative analysis of kinematic imbalance measurements from the T2K, MicroBooNE and MINERvA experiments}% Force line breaks with \\

	\author{W.~Filali}
	\email[Contact e-mail: ]{wissal.filali@cern.ch}
	\affiliation{\affiliationCERN}

	\author{L.~Munteanu}
	\email[Contact e-mail: ]{laura.munteanu@cern.ch}
	\affiliation{\affiliationCERN}

	\author{S.~Dolan}
	\email[Contact e-mail: ]{stephen.joseph.dolan@cern.ch}
	\affiliation{\affiliationCERN}

\date{\today}% It is always \today, today,

\begin{abstract}
\noindent Recent neutrino-nucleus cross-section measurements of observables characterising kinematic imbalance from the T2K, MicroBooNE and MINERvA experiments are used to benchmark predictions from widely used neutrino interaction event generators. Given the different neutrino energy spectra and nuclear targets employed by the three experiments, comparisons of model predictions to their measurements breaks degeneracies that would be present in any single measurement. In particular, the comparison of T2K and MINERvA measurements offers a probe of energy dependence, whilst a comparison of T2K and MicroBooNE investigates scaling with nuclear target. In order to isolate the impact of individual nuclear effects, model comparisons are made following systematic alterations to: the nuclear ground state; final state interactions and multi-nucleon interaction strength. The measurements are further compared to the generators used as an input to DUNE/SBN and T2K/Hyper-K analyses. Whilst no model is able to quantitatively describe all the measurements, evidence is found for mis-modelling of A-scaling in multi-nucleon interactions and it is found that tight control over how energy is distributed among hadrons following final state interactions is likely to be crucial to achieving improved agreement. Overall, this work provides a novel characterisation of neutrino interactions whilst offering guidance for refining existing generator predictions.

\end{abstract}

\maketitle

\section{\label{sec:level1}Introduction}

The challenge of constraining systematic uncertainties in neutrino oscillation experiments operating in the GeV regime of neutrino energy is paramount~\cite{Alvarez-Ruso:2017oui, Katori:2016yel}. These uncertainties, if not properly accounted for, can significantly impact the ultimate precision of current (T2K~\cite{T2K:2011qtm} and NOvA~\cite{NOvA:2007rmc}) and future (DUNE~\cite{Abi:2020wmh,DUNE:2020ypp} and Hyper-Kamiokande~\cite{Hyper-Kamiokande:2018ofw}) long baseline neutrino oscillation experiments, as well as oscillation analyses from Fermilab's short baseline (SBN~\cite{MicroBooNE:2015bmn}) program. One of the primary sources of these uncertainties are those that cover potential mismodelling of nuclear-medium effects in neutrino-nucleus interactions within the neutrino interaction Monte Carlo (MC) event generators used as an input to neutrino oscillation analyses~\cite{T2K:2023smv,NOvA:2021nfi}. These nuclear effects include Fermi motion, which pertains to the inherent movement of nucleons within the nucleus; final state interactions (FSI), referring to the re-interaction of outgoing hadrons from an interaction with the remnant nucleus; and multi-nucleon two-particle two-hole (2p2h) interactions, in which neutrinos interact with a correlated pair of nucleons, bound via the exchange of a virtual meson. A detailed review of nuclear effects are available in Refs.~\cite{Alvarez-Ruso:2017oui,Katori:2016yel,Jachowicz:2021ieb}.

A global campaign of neutrino-nucleus cross-section measurements is underway, within which a key goal is to characterise and constrain these nuclear effects. Prior to $\sim$2016 such measurements focused mostly on the detection of final-state muons in charged-current muon neutrino interactions, but there is now also a rising interest for detecting and analysing final-state hadrons, made possible with the evolution of experimental capabilities and analysis techniques. 

In this context, variables characterising kinematic imbalances between the outgoing lepton and nucleon in pion-less neutrino-nucleus interactions have emerged as a powerful tool~\cite{Lu:2015hea, Furmanski:2016wqo, Baudis:2023tma, MINERvA:2019ope}. The power of ``transverse kinematic imbalance'' (TKI) variables comes with the examination of the transverse components of the final state particles relative to the incoming neutrino direction. Imbalances in the outgoing particles' transverse momentum vectors are thereby able to characterise nuclear effects without relying on the unknown incoming neutrino energy. Differential cross-section measurements as a function of these variables hence provide an accurate and straightforward probe of nuclear effects.

\begin{figure}
\centering
\includegraphics[width=9cm]{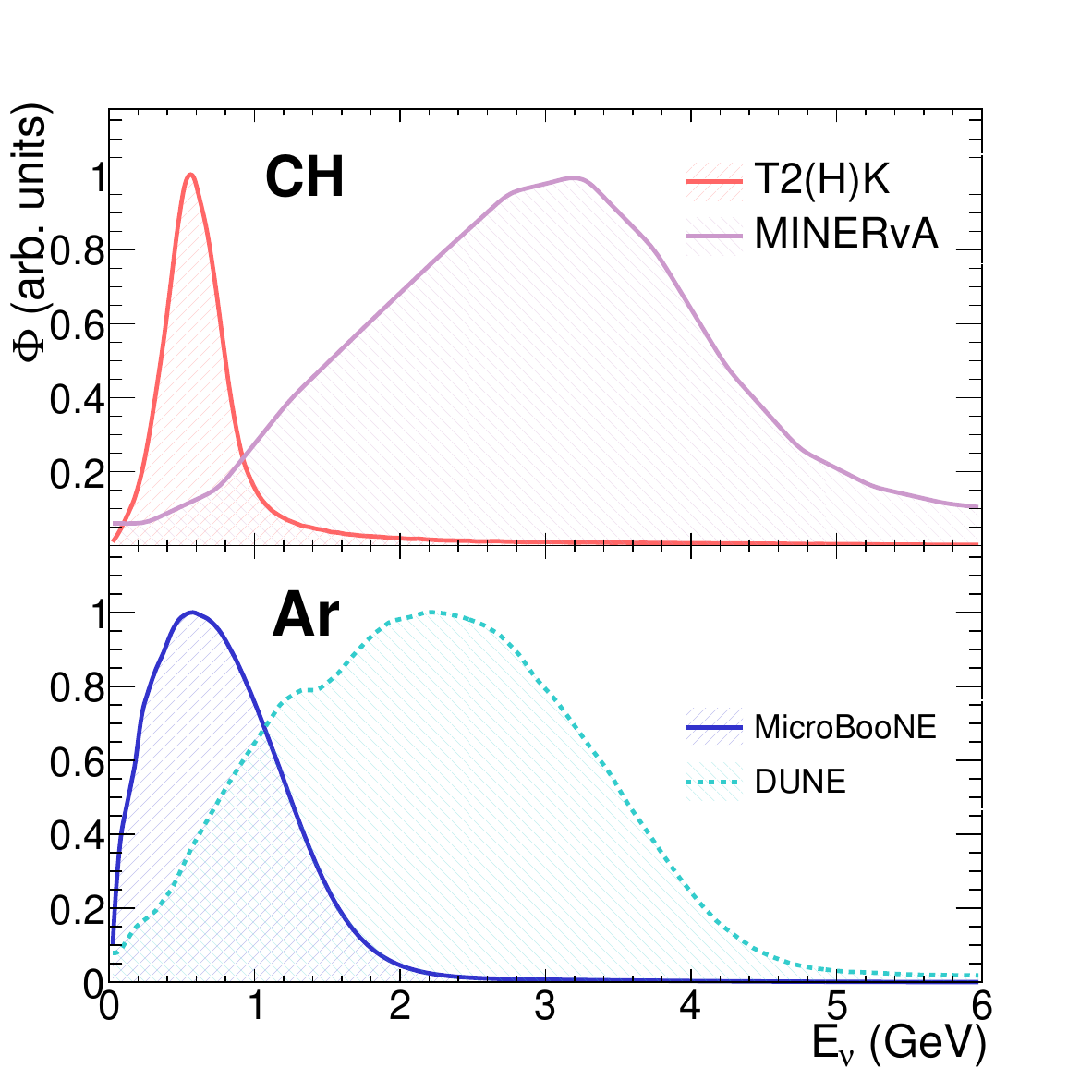}
\caption{A comparison of the shape of the incoming neutrino fluxes predictions used for the T2K~\cite{T2K:2012bge,T2K:2015sqm,t2kfluxurl}, MINERvA~\cite{MINERvA:2016iqn,minervafluxurl} and MicroBooNE analyses considered within this work, alongside the flux prediction for the future DUNE experiment~\cite{duneCDRphys,dunefluxurl}. Note that the depicted MicroBooNE flux is the same one as the one used by the MiniBooNE experiment~\cite{MiniBooNE:2008hfu}. The labels ``CH'' and ``Ar'' stand for ``hydrocarbon'' and ``argon'' and indicate the primary nuclear target used by the experiments depicted in each of the panels. 
\label{fig:fluxComp}}
\end{figure}

The T2K~\cite{T2K:2018rnz}, MicroBooNE~\cite{MicroBooNECollaboration:2023lef} and MINERvA~\cite{MINERvA:2018hba} experiments have recently produced cross-section measurements as a function of TKI variables. Whilst each result has been studied independently, and some joint analyses of the T2K and MINERvA results have been made~\cite{Dolan:2018zye,Chakrani:2023htw,GENIE:2024ufm}, there has so far not been a joint study of all three measurements. The different sensitivities of each of these experiments makes a joint study particularly interesting. T2K and MicroBooNE operate with relatively narrow-band neutrino fluxes with energies around $\sim$1~GeV, while MINERvA uses a wider-band flux extending beyond 3~GeV. A comparison of the shape of the fluxes from the three experiments can be found in \autoref{fig:fluxComp}. Additionally, the experiments differ in their target materials: whilst T2K and MINERvA use a hydrocarbon target, MicroBooNE uses argon. In this way, a combined analysis of the three measurements provides the potential to study the energy- and target-dependence of nuclear effects. In particular, due to their similar neutrino energies, a comparison of T2K and MicroBooNE measurements highlights features which probe the dependence of the aforementioned nuclear effects on the target material. Conversely, a comparison of the hydrocarbon-based measurements of T2K and MINERvA offers insight into the energy dependence of nuclear effects.

In this article, we extend the older analysis of T2K and MINERvA measurements in Ref.~\cite{Dolan:2018zye} to additionally consider the MicroBooNE measurement and to confront all three with state-of-the-art neutrino event generator predictions, including those used as an input to current oscillation analyses as well as to sensitivity studies for the next generation of experiments. We systematically alter the generator predictions by varying the modelling of one nuclear effect at a time in order to both explore each result's sensitivity and to investigate sources of model-measurement discrepancies. The variables, measurements, the models and the analysis strategy are defined in \autoref{sec:analysisstrat}; the results of the subsequent comparisons are shown and discussed in \autoref{sec:results}. The conclusions are given in \autoref{sec:conclusions}.

\section{Analysis strategy}
\label{sec:analysisstrat}

The general strategy for this analysis is to qualitatively and quantitatively compare a variety of systematically altered models to measurements of TKI variables from the T2K, MicroBooNE and MINERvA experiments. The TKI variables considered are defined in \autoref{sec:TKI}, whilst the measurements are detailed in \autoref{expData}. This includes a summary of the exact signal definition for each measurement in addition to an overview of their statistical power, correlations and the means by which models should be compared to them. In order to draw quantitative conclusions from model-measurement comparisons, a $\chi^2$ and $p$-value analysis is performed as described in \autoref{chi2analysis}. The models used to compare to the measurements are defined in \autoref{sec:models} before the systematic variations are described in \autoref{sec:systvar}.

In this analysis, we focus on measurements of charged-current neutrino interactions with nucleons where no mesons are observed in the final state, often referred to as the CC0$\pi$ topology. The dominant type of microscopic process which contributes to these topologies is the quasi-elastic (QE) process, in which the incoming muon neutrino interacts with a neutron inside the nucleus. The final state of such interactions before considering FSI would be composed of a muon and a single proton. However, the measurements considered include QE interactions on bound nucleons inside atomic nuclei, and the presence of FSI can stimulate the emission of additional nucleons. Moreover, nuclear effects permit other processes to contribute to the CC0$\pi$ topology: multi-nucleon interactions (which produce two outgoing nucleons before FSI) and resonant meson production channels (sometimes referred to as ``RES'' in this work) in which the final state meson (most often a pion) has been absorbed inside the nucleus by FSI.

\subsection{Transverse Kinematic Imbalance}
\label{sec:TKI}

\begin{figure}
\centering
\includegraphics[width=8cm]{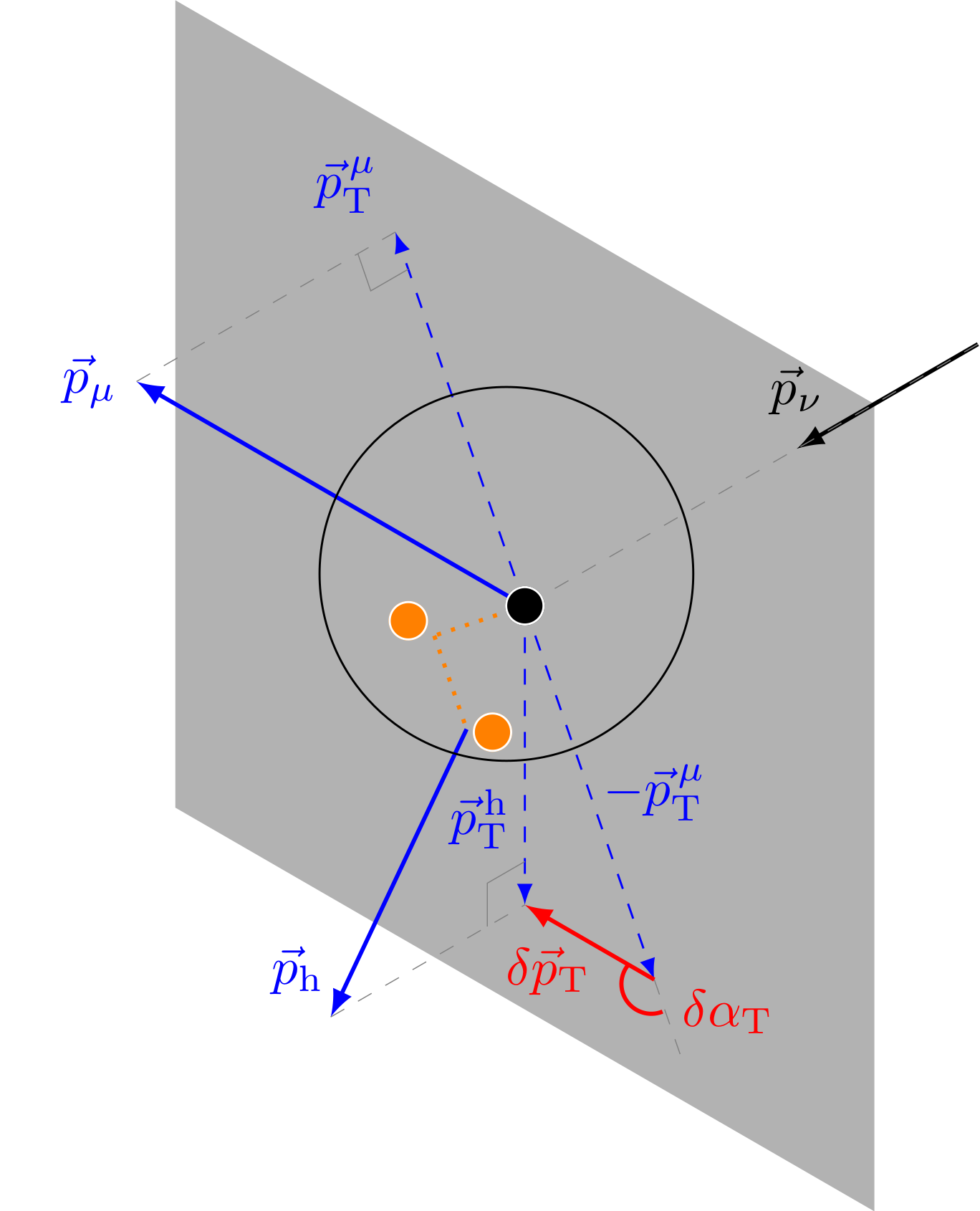}
\caption{Schematic illustration of the definition of the $\dpt$ and $\dalphat$ TKI variables for charged-current muon-neutrino interactions. The total momentum of particle $i$ is given by $\vec{p}_i$, while its transverse component with respect to the neutrino direction is represented by $\vec{p}_T^{\textrm{ }i}$. $\vec{p}_h$ stands for the momentum vector of the hadronic system, but in the CC0$\pi$ topology considered in this work this is constructed using the highest momentum outgoing proton. The black filled circle represents the initial struck nucleon; the gray plane shows the plane transverse to the incoming neutrino direction; the orange circles and dashed lines indicate possible final state interactions experienced by the outgoing hadrons. This figure is adapted from Ref.~\cite{T2K:2021naz} which was adapted from Ref.~\cite{Lu:2015hea}.} \label{fig:tkifig}
\end{figure}

In neutrino nucleus interactions, nuclear effects introduce a kinematic imbalance between the initial neutrino four-momentum and the combined four-momenta of the final-state lepton and hadronic system. This imbalance serves as a sensitive probe for nuclear effects, including Fermi motion, FSI, and 2p2h interactions. A complete four dimensional analysis of kinematic imbalance, as is used to probe nuclear ground state effects in electron scattering (e.g. in~\cite{Dutta:2003yt,CLAS:2021neh}), is challenging in neutrino experiments due to an unknown incoming neutrino energy. However, the imbalance in the plane transverse to the incoming neutrino still provides a wealth of powerful information and can be quantified by a multitude of variables that have been proposed~\cite{Lu:2015tcr,Baudis:2023tma,MINERvA:2019ope,Furmanski:2016wqo} and, in many cases, measured~\cite{T2K:2018rnz,MINERvA:2018hba,MINERvA:2019ope,MINERvA:2020anu,MicroBooNE:2021cue, MicroBooNE:2023krv}. The primary variables considered in this work are schematically represented in \autoref{fig:tkifig}. They are defined by:
\begin{equation}
\dpt = |\overrightarrow{p}_T^l + \overrightarrow{p}_T^p|, 
\label{eq:dpt}
\end{equation}

\begin{equation}
\dalphat = \arccos\frac{-\overrightarrow{p}_T^l.\delta\overrightarrow p_T}{{p}_T^l \dpt}, 
\label{eq:dat}
\end{equation}
where $\overrightarrow{p}_T^l$ and $\overrightarrow{p}_T^p$ are the transverse momenta of the outgoing muon and highest momentum (or leading) proton, respectively.

For pure charged-current quasi-elastic (QE) interactions devoid of nuclear effects (e.g. interactions on a free nucleon target), $\dpt$ would be zero. In neutrino-\textit{nucleus} interactions $\dpt$ is non-zero and its shape is sensitive to different nuclear effects~\cite{Lu:2015hea}. In the presence of Fermi motion but without FSI or 2p2h, $\dpt$ is, to a good approximation, the projection of the initial state struck nucleon momentum onto the plane transverse to the incoming neutrino, typically peaking at around 200~MeV/$c$. FSI and 2p2h tend to cause the emission of additional particles not included in \autoref{eq:dpt}, thereby giving $\dpt$ an extended tail well above the scale of Fermi motion. Very broadly, in the absence of any constraints on outgoing particle kinematics, the ``bulk'' of $\dpt$ is sensitive to Fermi motion and the ``tail'' to FSI and 2p2h.

In the presence of kinematic thresholds on the outgoing nucleon, as are included in all experimental signal definitions,  the effect of FSI is more complicated. FSI decelerates protons, which has two consequences on the $\dpt$ distribution: first, it migrates a number of protons below the detection threshold, which sculpts the visible phase space and reduces the overall within-signal-phase-space cross section (including in the bulk), and second, it increases the imbalance between the proton and the muon transverse momenta and enhances the tail of the distribution. The final distributions will be impacted by both effects simultaneously generally causing a relative increase in the size of the tail with respect to the bulk, but a lower overall cross section.

The direction of $\dpt$ with respect to the transverse projection of the outgoing lepton momentum vector is described by the angle $\dalphat$. In the absence of FSI or 2p2h, $\dalphat$ is approximately uniformly distributed due to the isotropic nature of Fermi motion. In the additional presence of FSI, it provides an interesting characterisation of the deceleration the outgoing nucleon experiences with respect to the pre-FSI kinematics. The more the outgoing nucleon is slowed down, the higher the proportion of the cross section is expected to be at high $\dalphat$ ($\dalphat>90^{\circ}$)\cite{Lu:2015hea}. 2p2h interactions also causes a shift of the $\dalphat$ distribution towards higher values, due to the highest momentum outgoing proton having only a fraction of the transferred momentum. 

The MINERvA collaboration also produced a measurement of the reconstructed nucleon momentum ($p_N$), as detailed in~\cite{Furmanski:2016wqo}. The variable $p_N$ is an estimation of the magnitude of the total momentum imbalance, which is a composite of the longitudinal and transverse momentum imbalances, $\delta p_L$ and $\dpt$ respectively. It is defined by:
\begin{equation}
p_N = \sqrt{\dpt^2 + \delta p_L^2}, 
\label{eq:pN}
\end{equation}
eere, $\delta p_L$ is the longitudinal momentum imbalance, which is expressed as:
\begin{subequations}
\begin{equation}
\delta p_L = \frac{1}{2}K - \frac{\dpt^2 + M_X^2}{2K},
\label{eq:dpL}
\end{equation}
where
\begin{equation}
K = M_A+ p_L^\mu + p_L^p - E^\mu - E^p,
\label{eq:K}
\end{equation}
and
\begin{equation}
M_X = M_A - M_n + \epsilon_n
\label{eq:Mx}
\end{equation}
\end{subequations}
where $M_A$ is the target nucleus mass, $M_n$ is the proton mass, and $\epsilon_n$ is the neutron mean excitation energy. 
The value used in this study is $\epsilon_n=27.1$ MeV, the same as in Ref.~\cite{MINERvA:2018hba}.

\subsection{Experimental measurements}
\label{expData}

The main focus of this comparative analysis is on the measurements of the missing transverse momentum $\dpt$ which has been measured by T2K, MINERvA and MicroBooNE. As explained in the previous section, this observable is uniquely suited to probe and disentangle nuclear effects due to its distinctive features (i.e. QE-dominated bulk and FSI and non-QE-dominated tail). All experiments have also measured the transverse boosting angle $\dalphat$ and the $\dphit$ angle~\cite{Lu:2015hea}. We report comparisons for the $\dalphat$ angle, which is sensitive to FSI effects, but we choose to omit comparisons to $\dphit$ as it is less overtly sensitive to nuclear effects than $\dpt$ and $\dalphat$, and additionally is more dependent on the momentum transfer to the nucleus which varies widely between experiments~\cite{Lu:2015tcr}. The MicroBooNE measurement also includes the first multi-differential measurement of $\dpt$ in different regions of $\dalphat$, which allows for a better separation of 2p2h interactions from QE interactions which have undergone FSI~\cite{T2K:2019bbb}. Finally, we also report comparisons to the $p_N$ observable measured by the MINERvA experiment.

The cross sections measured by the three experiments set signal definitions constrained to specific kinematic regions, as summarised in \autoref{tab:sigDef}. The exact way in which these ranges apply are subtly different between the three experiment's signal definitions, as is the proton momentum that is used to reconstruct the TKI variables:

\begin{itemize}
    \item For T2K, any number of protons are allowed but the proton momentum used to build the TKI variables is always that of highest momentum proton in the neutrino interaction. If the highest momentum proton or muon's momenta or angles with respect to the incoming neutrino fall outside of the kinematic ranges given in \autoref{tab:sigDef} then the interaction does not count as signal. 
    \item For MicroBooNE, only one proton is allowed inside the kinematic ranges given in \autoref{tab:sigDef} (any number of protons are allowed outside of it) and it is the momentum of this proton that goes into the calculation of the TKI variables (whether or not it is the highest momentum proton in the interaction). Note that this highest momentum proton in an interaction is not necessarily the one used to reconstruct the TKI (as it may be outside of the allowed kinematic phase space). Additionally, MicroBooNE allows interactions with final state charged pions with momentum lower than 70 MeV/$c$ to be classified as signal. 
    \item For MINERvA, any number of protons within the phase space constraints given in \autoref{tab:sigDef} are allowed and it is the highest momentum proton of these that is used to construct the TKI variables. Like in the MicroBooNE case, the highest momentum proton is allowed to fall outside of the phase space constraints.
\end{itemize}

\begin{table}[htbp]
\centering
\begin{tabular}{c|c|c|c|c}
%\hline
\small
Analysis & $p_\mu$ [GeV/$c$] & cos$\theta_\mu$ & $p_p$ [GeV/$c$]& cos$\theta_p$ \\%& Notes\\
\hline
T2K~\cite{T2K:2018rnz} & $>0.25$ & $>-0.6$ & $0.45 - 1.2$ & $>0.4$ \\
\hline
MicroBooNE~\cite{MicroBooNECollaboration:2023lef} & $0.1 - 1.2$ & - & $0.3 - 1$ & - \\
\hline
MINERvA~\cite{MINERvA:2018hba} & $1.5 - 10$ & $>0.94$ & $0.45 - 1.2$ & $>0.34$ \\

\end{tabular}
\caption{The kinematic phase space limits used in the signal definitions for T2K, MicroBooNE, and MINERvA measurements considered in this work.}
\label{tab:sigDef}
\end{table}

The comparisons of event generator model predictions to the cross-section measurements of T2K and MINERvA are relatively straightforward, since the two experiments unfold the measurement into the truth space, applying regularization techniques\footnote{It should be noted that T2K also provides an unregularised result, but that the regularisation has previously been shown to have only a very small impact on quantitative conclusions~\cite{T2K:2018rnz,Dolan:2018zye}.}. In contrast, MicroBooNE reports its measurements in a linearly transformed space with respect to the true quantity, where the cross section is smooth, using the Wiener Singular Value Decomposition Technique~\cite{Tang:2017rob}. This procedure allows MicroBooNE to produce smooth cross-section measurements with no regularisation bias at the cost of providing a measurement as a function of a variable that is smeared by a linear transformation from the true physics quantity of interest. The MicroBooNE result is thus accompanied by an additional smearing matrix, referred to as the $A_c$ matrix, encapsulating this transform. In this work, any model prediction compared to MicroBooNE's measurement has been transformed using its corresponding $A_c$ matrix~\cite{MicroBooNECollaboration:2023lef}.

\subsection{$\chi^2$ analysis}
\label{chi2analysis}
In order to quantify the agreement between the observed data and the theoretical predictions, a chi-squared ($\chi^2$) test-statistic is employed. The $\chi^2$ is defined as:

\begin{equation}
    \chi^2=\sum_{i,j}(D_i - D_i^{MC})(A^{-1}_{cov})_{ij}(D_j - D_j^{MC}),
\end{equation}

\noindent where $D_{i/j}$ and $D_{i/j}^{MC}$ represent the content of bin $i/j$ in a measurement and a generator prediction histogram, respectively, and  $A^{-1}_{cov}$ is the inverse of the covariance matrix associated with a measurement. For each of the experiments, covariance matrices are available from their respective data releases, encapsulating uncertainties and correlations in the cross-section extraction.

In addition to the $\chi^2$ values, $p$-values are also calculated to provide a more intuitive estimation of the statistical significance of the observed discrepancies between the measurements and the model predictions. Note that, like all contemporary quantitative analyses of neutrino cross-section measurements, any conclusions drawn from the $\chi^2$ and the calculation of $p$-values assume that uncertainties on the experimental measurements are well described using a multi-variate Gaussian as defined by the covariance matrix provided by each analysis. However, past and recent analyses have suggested this approximation may not always be valid, especially for results limited mostly by systematic uncertainties~\cite{Chakrani:2023htw,DAgostini:1993arp,radinu2024poster}. Since we have no way to test the validity of the assumption from the experimental data releases, we proceed assuming gaussian uncertainties but urge readers to treat quantitative conclusions cautiously (and urge experiments to detail the extent of deviations from the gaussian case).

\subsection{Models}
\label{sec:models}

To generate all the simulations used in this work, we use a variety of generators and configurations, which are described in this section. We also use the NUISANCE framework~\cite{Stowell:2016jfr} to process the simulated events from each generator using a common format. 

In order to generate interactions for comparisons to T2K and MINERvA measurements, the NEUT~\cite{Hayato:2021heg} generator was used (specifically NEUT version 5.6.2), with the official flux releases associated to the measurements~\cite{T2K:2012bge,T2K:2015sqm,t2kfluxurl,MINERvA:2016iqn,minervafluxurl}, on a hydrocarbon target. 

In CC0$\pi$ topologies in these energy ranges, the majority of the signal is populated by QE interactions. We simulate QE interactions using the Benhar SF~\cite{Benhar:1994hw} model, as is used as an input to T2K's latest neutrino oscillation measurements~\cite{T2K:2023smv}. Within NEUT, this model is used to describe the initial nuclear ground state as a function of nucleon momenta and their removal energies in a factorized approach~\cite{Hayato:2021heg}. The distribution of intial state nucleon momenta and removal energies has been derived from electron scattering data and the available phase space is broadly divided into two parts: a mean-field (MF) part, in which single nucleons are involved in the interaction, and a region corresponding to high-momentum short-range correlated (SRC) nucleon pairs, accounting for $\sim$5\% of the total cross section and in which only one nucleon participates in the interaction but another ``spectator'' nucleon is emitted alongside it. For a more detailed description of the NEUT SF model and its associated uncertainties, more details can be found in~\cite{Furmanski:2015knr,Chakrani:2023htw}.

In addition to the Benhar SF model, we also provide comparisons with two Fermi gas-based models which are also implemented in NEUT: the Local Fermi Gas (LFG) model developed by the Valencia group~\cite{Nieves:2011pp}, which includes corrections for long-range nucleon correlations based on the random phase approximation (RPA) prescription, and the global relativistic Fermi gas (RFG) following the prescription of Smith and Moniz~\cite{Smith:1972xh}. Note that in the RFG case a nucleon axial mass $M_A^{QE}$ of 1.21 GeV is used (the NEUT default) as opposed to 1.03 or 1.05 GeV for SF and LFG respectively.

A direct comparison between the MicroBooNE measurement and a NEUT SF on argon is not possible, since NEUT currently doesn't have an implementation of an argon spectral function. The NuWro event generator~\cite{Juszczak:2005zs}, on the other hand, has an argon spectral function but also contains a significantly different modelling of 2p2h and meson absorption, which would make direct comparisons with NEUT predictions difficult to interpret. To address this for the case of MicroBooNE, we simulate QE interactions on argon with the NuWro version 19.02.1 SF implementation and non-QE interactions (2p2h and RES) with NEUT to create our SF baseline prediction. This model is referred to as the ``SF*'' model throughout the remainder of this paper. The consequence of this choice is that the QE events generated with the SF* model on an argon target is also put through a different type of intra-nuclear cascade than those from NEUT generated on other targets. To assess the impact of this inconsistency, we also compare MicroBooNE predictions obtained with the LFG model in NEUT with an argon target and which undergo the NEUT FSI cascade. This is discussed in \autoref{sec:fsi}.

For 2p2h interactions, the NEUT model is based on the model by the Valencia group~\cite{Nieves:2011pp,Gran:2013kda,Schwehr2017}. Note that NEUT contains two implementations of the Valencia model, and in this work we opt to use the lookup table approach employed in recent T2K measurements \cite{T2K:2023smv}.

Resonant meson production is simulated in NEUT using the Rein-Sehgal model~\cite{Rein:1980wg}, with modifications to the axial form factor from Graczyk and Sobczyk~\cite{Graczyk:2007xk} as well as lepton mass corrections~\cite{Berger:2007rq}. RES interactions can pass selection criteria for mesonless samples in one of two ways - for the MicroBooNE samples, charged pions with momenta below 70 MeV/$c$ are allowed by the selection criteria; for all other samples, RES interactions enter CC0$\pi$ samples through meson absorption processes, a type of FSI. Since the dominant type of mesons produced in such interactions are pions, we will subsequently refer to this process as pion absorption (though it applies, in principle, to heavier mesons). 

FSI are simulated through a semi-classical intra-nuclear cascade (INC) in both NEUT and NuWro. The philosophy of the simulation is similar for both generators, but they differ in several details of the implementation (notably, in the choice of data sets used to tune the probability of each intra-nuclear process). A more detailed review of the differences between the modeling of hadron FSI in both NEUT and NuWro can be found in~\cite{Dytman:2021ohr}.

Whilst the NEUT SF model represents the baseline model used by the T2K experiment for its oscillation analysis, we also consider the \argenie~\cite{GENIE_30400tag,DUNEmodel_genworkshop} configuration from the GENIE~\cite{Andreopoulos:2009rq,Andreopoulos:2015wxa} event generator version 3.04.00~\cite{GENIE:2021npt}, which is used as the baseline input model for DUNE and SBN analyses. Like NEUT LFG, GENIE also uses the Valencia LFG model to describe QE interactions. Unlike the NEUT LFG model, the GENIE configuration uses a few different model parameters which will impact the predictions. Firstly, the $Q$-value (or the separation energy) for nuclei is chosen such that the distribution of removal energies from which the MC sampling is done covers the majority of the removal energies in the argon Spectral Function model detailed in~\cite{JeffersonLabHallA:2022cit}. Second, unlike the NEUT LFG equivalent, the GENIE \argenie model also includes high-momentum nucleons in addition to the baseline LFG prediction, whose role is to approximate the presence of SRCs~\cite{GENIE:2021npt}. The simulation parameters were also modified according to the nucleon-level tune described in~\cite{GENIE:2021zuu}. The 2p2h model used in this configuration is the SuSAv2 model~\cite{RuizSimo:2016rtu,Megias:2016fjk}, following the implementation described in~\cite{Dolan:2019bxf}. The RES model is similar to the model in NEUT, but with the aforementioned GENIE tune applied. 

An important difference between the NEUT and GENIE simulations used in this work is the modeling of FSI. The \argenie configuration employs the so-called ``\texttt{hA2018Intranuke}'' model, which is not a semi-classical cascade model. The latter applies data-driven predictions to determine the ``fates'' that hadrons undergo once produced inside the nucleus in a single step. 

The main model choices between the different generator configurations which will be described in \autoref{sec:impact_globalGen} are summarized in \autoref{tab:gencomp}. 

\begin{table*}[htbp]
\begin{tabular}{c|c|c|c}

\textbf{Configuration}   & \textbf{1p1h model}  & \textbf{2p2h model} & \textbf{FSI model} \\ \hline
NEUT SF                  & Benhar SF                      & Valencia            & NEUT cascade       \\ \hline
NEUT LFG                 & Valencia LFG                   & Valencia            & NEUT cascade       \\ \hline
GENIE \argenie & Valencia LFG + correlated tail & SuSAv2              & GENIE hA2018       \\ 
\end{tabular}
\caption{A summary of models used for each generator configuration considered in this work.}
\label{tab:gencomp}
\end{table*}

\subsection{Systematic variations}
\label{sec:systvar}

For this analysis, the reference model against which we have compared the measurements and have used as a baseline to apply variations to nuclear effects was chosen to be the NEUT Benhar Spectral Function (SF) model~\cite{Benhar:1994hw}, described in \autoref{sec:models}. 

As described in \autoref{sec:TKI}, the TKI distributions offer sensitivity to the presence and strength of different nuclear effects. In particular, the tail of $\delta p_T$ is sensitive to the presence of FSI (on the outgoing proton as well as the pions in the resonant background) and 2p2h. The $\delta \alpha_T$ distribution has unique sensitivity to FSI. We investigate the impact of these nuclear effects by varying them independently and assessing the evolution of the agreement between the data and the generator predictions. We apply these systematic variations by either reweighting the events in our simulations or regenerating events with altered parameters in the simulations.

\paragraph{Fermi motion}
The bulk of $\dpt$ is directly sensitive to the transverse component of the Fermi motion inside the nucleus. We compare the reference model (NEUT SF or SF* for argon) with the predictions from the LFG and RFG models from NEUT.

\paragraph{Total 2p2h cross section}

The total cross-section for 2p2h processes is not well-known, with theoretical models differing substantially in their prediction of it~\cite{Nieves:2011pp,Martini:2013sha,RuizSimo:2016rtu,Megias:2016fjk}. In the context of this work, we choose to first assess the impact of the total 2p2h cross section by scaling the strength of 2p2h interactions by 70\% flatly across all kinematics. This number was chosen based on the difference in the total cross-section predicted by the Valencia~\cite{Nieves:2011pp}, SuSAv2~\cite{RuizSimo:2016rtu} and Martini et al.~\cite{Martini:2013sha} 2p2h models. Of these models, the Martini et al. 2p2h model shows the largest difference in integrated cross section with respect to the Valencia model for neutrinos, and 70\% was taken as a representative size of the difference. This approach tests the impact of increasing the total cross section of 2p2h interactions, but does not build in any shape-related freedom. There is little available theoretical guidance on the plausible types of variations we can expect on the final state nucleon kinematics, and generators predict the latter based only on the available interaction phase space. Varying the shape of, for example, the outgoing proton momentum spectrum may also introduce sizable variations to the TKI distributions but, given the lack of guidance from theory on what these variations should be, we leave such studies for future work (although promising work in Refs~\cite{Sobczyk:2020dkn,Martinez-Consentino:2023hcx} may soon change this). For a discussion of the most extreme variations predicted by generators on the outgoing nucleon kinematics, see Ref.~\cite{Bathe-Peters:2022kkj}.

\paragraph{Nucleon FSI}

To gauge the impact of nucleon FSI on the features of the distributions, we perform variations where we vary the mean free path (MFP) of protons inside the nucleus by $\pm$30\%. This value was chosen based on Ref.~\cite{Niewczas:2019fro}, in which it is shown that an alteration of this size encompasses the spread of nuclear transparency measurements from various sources on a carbon target. An increase (decrease) of the MFP by 30\% corresponds to a corresponding increase (decrease) in the nucleus transparency, thereby decreasing (increasing) the probability that a proton undergoes FSI. We will refer to these alterations as ``strong/more'' and ``weak/less'' FSI for the $-30\%$ and $+30\%$ variations respectively. These alterations are applied by regenerating the simulations with altered values of the MFP, and the same approach is applied to the NEUT models (SF and LFG), as well as the QE component from NuWro in the SF* model. 

\paragraph{Pion FSI}

We take a similar approach for pion absorption events. The NEUT intra-nuclear cascade model has been tuned to world pion-nucleus scattering data~\cite{PinzonGuerra:2018rju}, and as a result the underlying parameters governing the cascade have a data-driven constraint. Since RES events can only end up in the mesonless samples used in this analysis via pion absorption processes, we vary the probability of this fate within the cascade. The variation we apply is of $\pm$43\%, on top of the tuned absorption probability used by NEUT (which is itself 40\% larger than that prescribed by Salcedo and Oset~\cite{Salcedo:1987md}), following the prescription in Ref.~\cite{PinzonGuerra:2018rju}. However, none of the T2K or MicroBooNE measurements exhibited any sizable sensitivity to this alteration (due to the lower RES rate in their energy regimes), so we will only present variations of this parameter for the MINERvA measurements.

\section{Results}
\label{sec:results}

A summary of the $p$-values obtained between each systematic variation with each measurement is available in \autoref{tab:chi2_pValues}. This high-level analysis indicates that all models, when compared to the MINERvA measurements, are statistically rejected (i.e. all $p$-values $<0.05$ with one marginal exception), but not the T2K and MicroBooNE measurements. Consequently, the focus of our quantitative analysis focuses on comparisons between T2K and MicroBooNE measurements. Despite the fact that the MINERvA measurements quantitatively exclude all models, it remains crucial to consider these results qualitatively, especially in the context of insights gained from the two other experiments.

\begin{table*}[htbp]
\centering

\begin{tabular}{l|c|c|c c|c|c c|c c}%{lrrrrrrrrr}
Measurement                           & $N_{bins}$ & SF/SF*                       & LFG                          & RFG                          & More 2p2h                     & More FSI                     & Less FSI                     & More $\pi$ abs.               & Less $\pi$ abs.               \\ 
\hline
T2K $\dalphat$                        & 8          & \cellcolor[HTML]{FFCCC9}0.01 & \cellcolor[HTML]{FFCCC9}0.00 & \cellcolor[HTML]{FFCCC9}0.00 & \cellcolor[HTML]{FFCCC9}0.00  & \cellcolor[HTML]{FFCCC9}0.00 & \cellcolor[HTML]{FFCCC9}0.02 & \cellcolor[HTML]{FFFFFF}0.06  & \cellcolor[HTML]{FFCCC9}0.02  \\
T2K $\dpt$                            & 8          & \cellcolor[HTML]{FFFFFF}0.08 & \cellcolor[HTML]{FFFFFF}0.69 & \cellcolor[HTML]{FFCCC9}0.00 & \cellcolor[HTML]{FFCCC9}0.00  & \cellcolor[HTML]{FFCCC9}0.02 & \cellcolor[HTML]{FFFFFF}0.07 & \cellcolor[HTML]{FFCCC9}0.00  & \cellcolor[HTML]{FFFFFF}0.18  \\
\hline
MINERvA $\dalphat$                    & 12         & \cellcolor[HTML]{FFCCC9}0.00 & \cellcolor[HTML]{FFCCC9}0.00 & \cellcolor[HTML]{FFCCC9}0.00 & \cellcolor[HTML]{FFCCC9}0.00  & \cellcolor[HTML]{FFCCC9}0.00 & \cellcolor[HTML]{FFFFFF}0.06 & \cellcolor[HTML]{FFCCC9}0.00  & \cellcolor[HTML]{FFCCC9}0.00  \\
MINERvA $\dpt$                        & 24         & \cellcolor[HTML]{FFCCC9}0.00 & \cellcolor[HTML]{FFCCC9}0.00 & \cellcolor[HTML]{FFCCC9}0.00 & \cellcolor[HTML]{FFCCC9}0.00  & \cellcolor[HTML]{FFCCC9}0.00 & \cellcolor[HTML]{FFCCC9}0.00 & \cellcolor[HTML]{FFCCC9}0.00  & \cellcolor[HTML]{FFCCC9}0.00  \\
MINERvA $p_N$                         & 24         & \cellcolor[HTML]{FFCCC9}0.00 & \cellcolor[HTML]{FFCCC9}0.00 & \cellcolor[HTML]{FFCCC9}0.00 & \cellcolor[HTML]{FFCCC9}0.00  & \cellcolor[HTML]{FFCCC9}0.00 & \cellcolor[HTML]{FFCCC9}0.00 & \cellcolor[HTML]{FFCCC9}0.00  & \cellcolor[HTML]{FFCCC9}0.00  \\
\hline
MicroBooNE $\dalphat$                 & 7          & \cellcolor[HTML]{FFCCC9}0.02 & \cellcolor[HTML]{FFFFFF}0.45 & \cellcolor[HTML]{FFFFFF}0.62 & \cellcolor[HTML]{FFFFFF}0.07  & \cellcolor[HTML]{FFFFFF}0.18 & \cellcolor[HTML]{FFCCC9}0.00 & \cellcolor[HTML]{FFCCC9}0.02  & \cellcolor[HTML]{FFFFFF}0.01  \\
MicroBooNE $\dpt$                     & 13         & \cellcolor[HTML]{FFFFFF}0.12 & \cellcolor[HTML]{FFFFFF}0.42 & \cellcolor[HTML]{FFCCC9}0.00 & \cellcolor[HTML]{FFFFFF}0.33  & \cellcolor[HTML]{FFFFFF}0.23 & \cellcolor[HTML]{FFCCC9}0.02 & \cellcolor[HTML]{FFFFFF}0.13  & \cellcolor[HTML]{FFFFFF}0.10  \\
MicroBooNE $\dpt$ low $\dalphat$      & 11         & \cellcolor[HTML]{FFFFFF}0.26 & \cellcolor[HTML]{FFFFFF}0.23 & \cellcolor[HTML]{FFFFFF}0.14 & \cellcolor[HTML]{FFFFFF}0.37  & \cellcolor[HTML]{FFFFFF}0.44 & \cellcolor[HTML]{FFFFFF}0.10 & \cellcolor[HTML]{FFFFFF}0.28  & \cellcolor[HTML]{FFFFFF}0.24  \\
MicroBooNE $\dpt$ mid-low $\dalphat$  & 12         & \cellcolor[HTML]{FFFFFF}0.07 & \cellcolor[HTML]{FFFFFF}0.40 & \cellcolor[HTML]{FFFFFF}0.19 & \cellcolor[HTML]{FFFFFF}0.23  & \cellcolor[HTML]{FFFFFF}0.38 & \cellcolor[HTML]{FFCCC9}0.00 & \cellcolor[HTML]{FFFFFF}0.08  & \cellcolor[HTML]{FFFFFF}0.06  \\
MicroBooNE $\dpt$ mid-high $\dalphat$ & 13         & \cellcolor[HTML]{FFCCC9}0.04 & \cellcolor[HTML]{FFFFFF}0.23 & \cellcolor[HTML]{FFCCC9}0.02 & \cellcolor[HTML]{FFFFFF}0.16  & \cellcolor[HTML]{FFFFFF}0.22 & \cellcolor[HTML]{FFCCC9}0.01 & \cellcolor[HTML]{FFCCC9}0.05  & \cellcolor[HTML]{FFCCC9}0.04  \\
MicroBooNE $\dpt$ high $\dalphat$     & 13         & \cellcolor[HTML]{FFCCC9}0.03 & \cellcolor[HTML]{FFFFFF}0.13 & \cellcolor[HTML]{FFFFFF}0.08 & \cellcolor[HTML]{FFFFFF}0.12  & \cellcolor[HTML]{FFFFFF}0.09 & \cellcolor[HTML]{FFCCC9}0.01 & \cellcolor[HTML]{FFCCC9}0.04  & \cellcolor[HTML]{FFCCC9}0.03  \\
\end{tabular}
\caption{$p$-values obtained from $\chi^2$ under a Gaussian error approximation between different models and measurements as well as to systematic variations of the SF/SF* models. $N_{bins}$ gives the number of bins for each measurement. $p$-values below 0.05, broadly indicating model rejection, are marked in red.}
\label{tab:chi2_pValues}
\end{table*}

Our detailed analysis of model-measurement comparisons begins with T2K and MicroBooNE in \autoref{sec:impact_ascaling}, exploring how mis-modelling of nuclear effects changes with nuclear target. We then compare T2K and MINERvA in \autoref{sec:impact_energydep_minerva}, allowing an exploration the energy dependence of mismodelling. We finish with a comparison of all three measurements to the models used for T2K, SBN and DUNE oscillation analyses in \autoref{sec:impact_globalGen}.

\subsection{T2K vs MicroBooNE: exploring nuclear target dependence}
\label{sec:impact_ascaling}

In this section, we focus on the comparison between T2K and MicroBooNE measurements, concentrating on the comparison of nuclear ground state models, nucleon FSI strength and 2p2h normalization. Although T2K and MicroBooNE operate at comparable neutrino beam energies, a major difference between the measurements from the two experiments lies in the nuclear target for neutrino interactions: T2K uses a hydrocarbon (CH) target, whereas MicroBooNE uses an argon (Ar) target. The comparison of these measurements therefore allows an identification potential issues with the way in which neutrino event generators predict how nuclear effects change as a function of atomic number.

\subsubsection{Breakdown by interaction channel}
\label{sec:bymode}

\begin{figure*}[htpb]
\centering
\includegraphics[width=8cm]{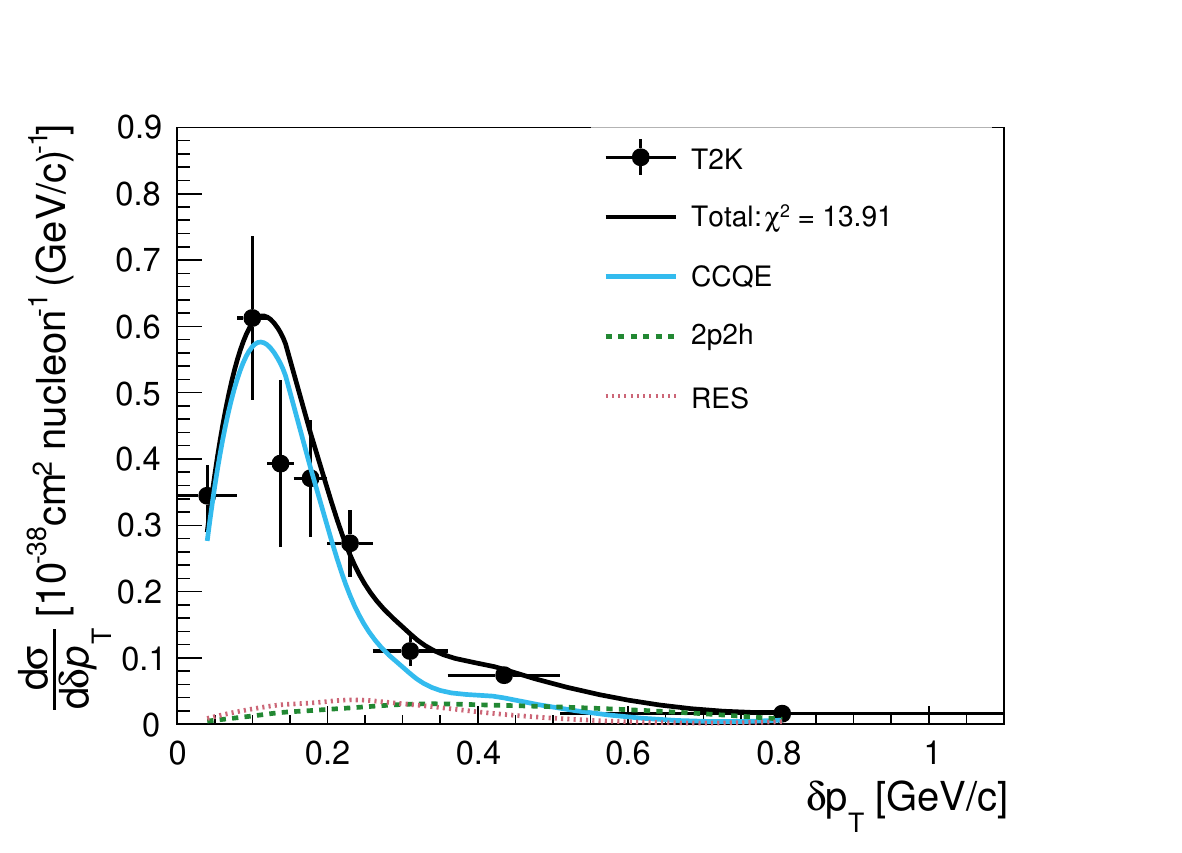}
\includegraphics[width=8cm]{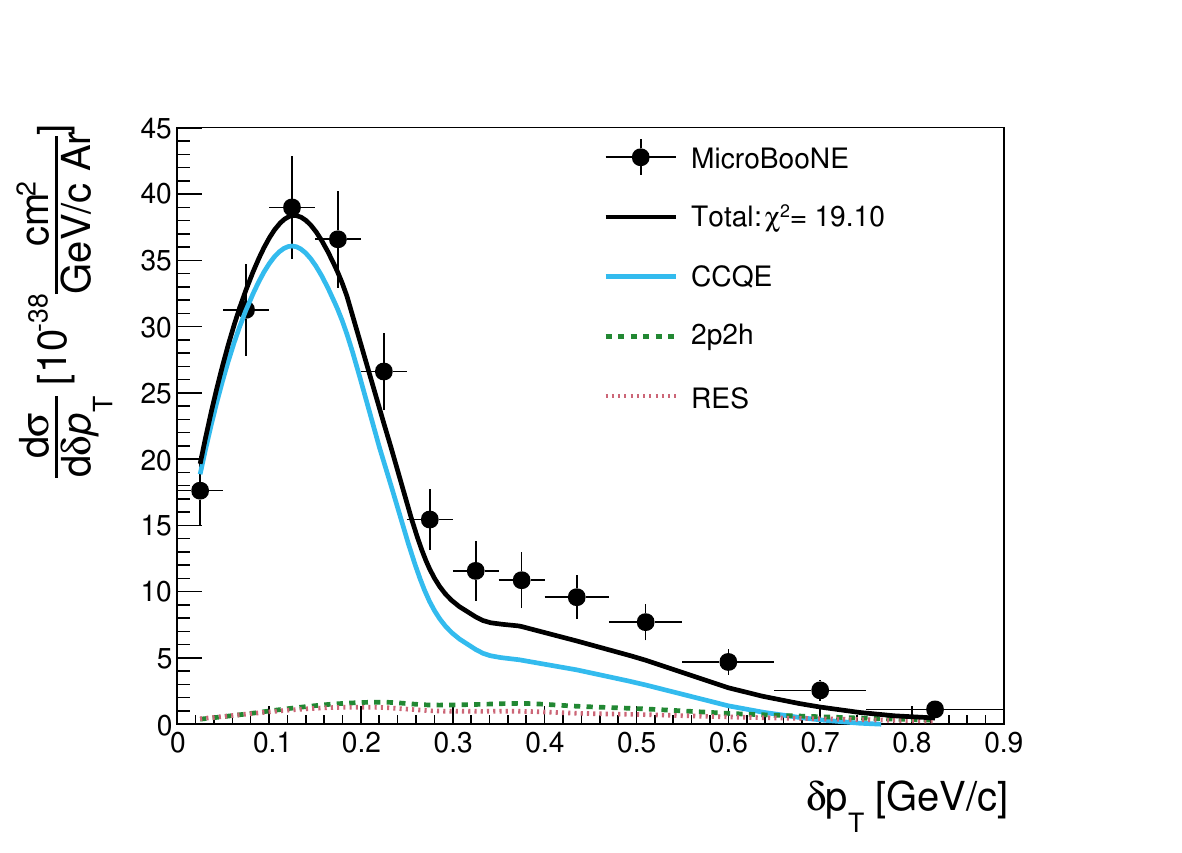}
\caption{Differential cross section as a function of $\dpt$ as predicted by the NEUT SF model for T2K (left) and the SF* model for MicroBooNE (right), compared to the respective measurements from each experiment. The QE, 2p2h and resonant (RES) contributions are highlighted.}
\label{fig:byModeT2KuBooNE}
\end{figure*}

Due to their similar neutrino beam energies, the $\dpt$ distributions for T2K and MicroBooNE have a broadly similar composition of the different interaction channels, as illustrated in ~\autoref{fig:byModeT2KuBooNE}. It is clear that the flux-integrated cross section for both measurements is dominated by the bulk, which is composed essentially of QE interactions. The relative contribution with respect to the QE-dominated bulk of 2p2h and RES processes which have undergone pion-absorption is comparable between the two experiments in the simulation, but the MicroBooNE measurement requires more strength in the tail. For both measurements, the nominal model (SF) yields a $p$-value$>$0.05, implying that neither of the measurements is capable of excluding this model. However, it is important to recall that there isn't a complete correspondence between the SF models used for these comparisons, with NuWro's FSI model being applied in the SF* case, as discussed in \autoref{sec:nucmodel}. The impact of the altered FSI treatment is further discussed in \autoref{sec:fsi}. The comparison to the MicroBooNE measurement suggests that the combination of the NuWro Ar SF model, alongside the 2p2h and RES contribution from NEUT, lacks some strength in describing the tail of the distribution. The bulk of the distribution is also slightly shifted towards lower values of $\dpt$, but this may be an artefact of working in the $A_c$ smeared space as discussed in \autoref{expData}.

\begin{figure*}[htpb]
\centering

\begin{tikzpicture}
    \draw (0, 0) node[inner sep=0] {\includegraphics[width=8cm]{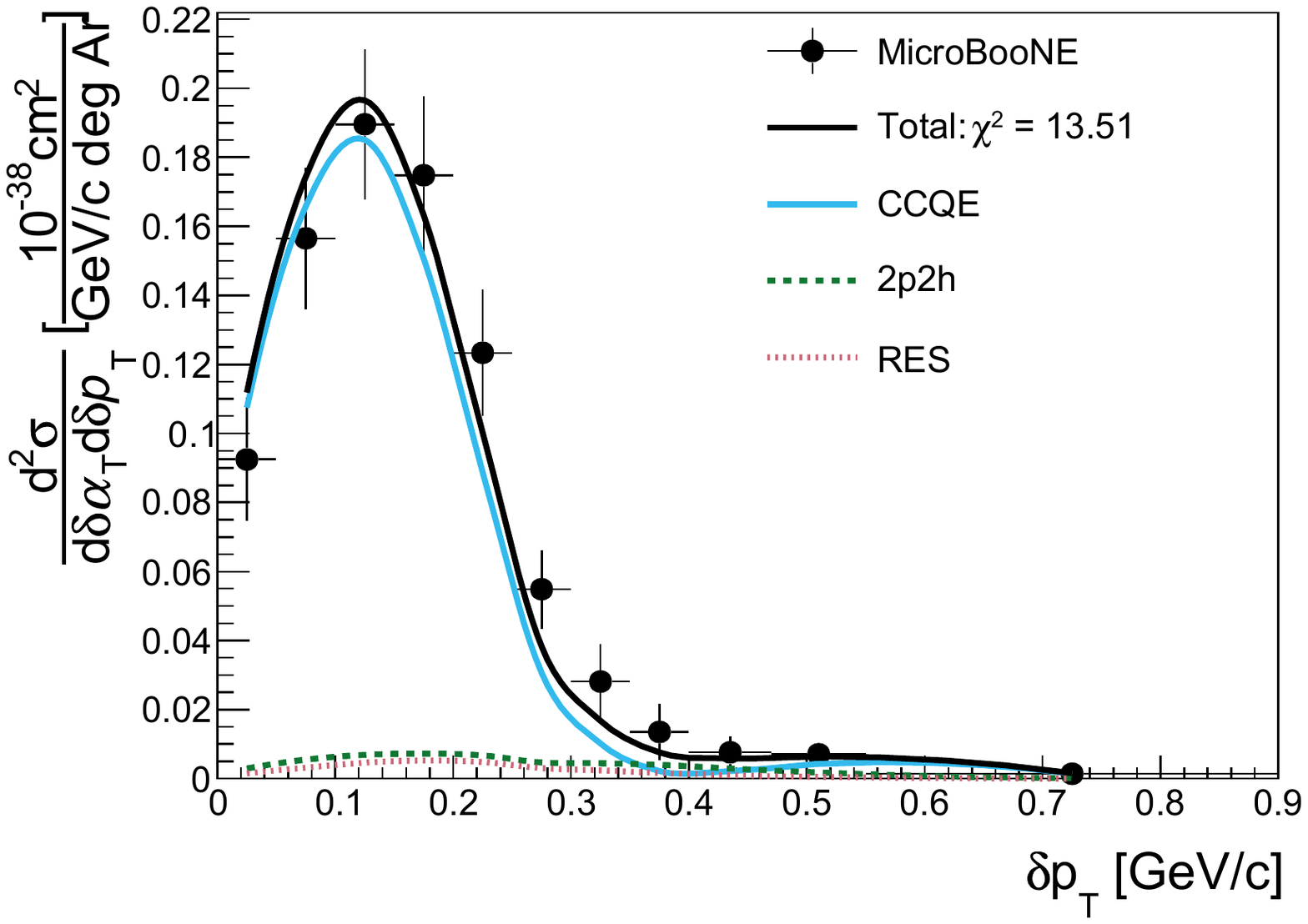}};
    \draw (2, 0) node {$0^\circ<\dalphat<45^\circ$};
\end{tikzpicture}
\begin{tikzpicture}
    \draw (0, 0) node[inner sep=0] {\includegraphics[width=8cm]{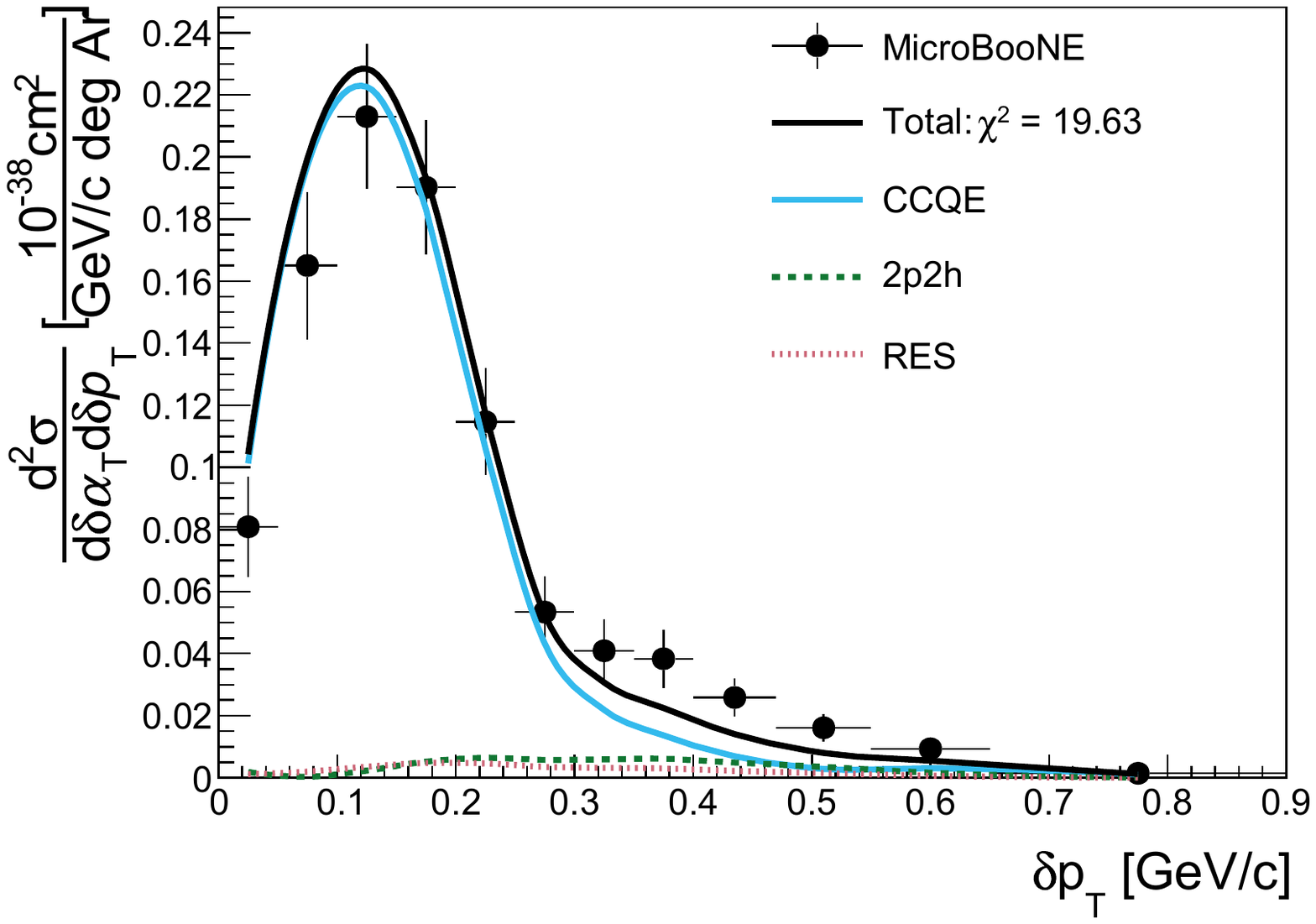}};
    \draw (2, 0) node {$45^\circ<\dalphat<90^\circ$};
\end{tikzpicture}
\begin{tikzpicture}
    \draw (0, 0) node[inner sep=0] {\includegraphics[width=8cm]{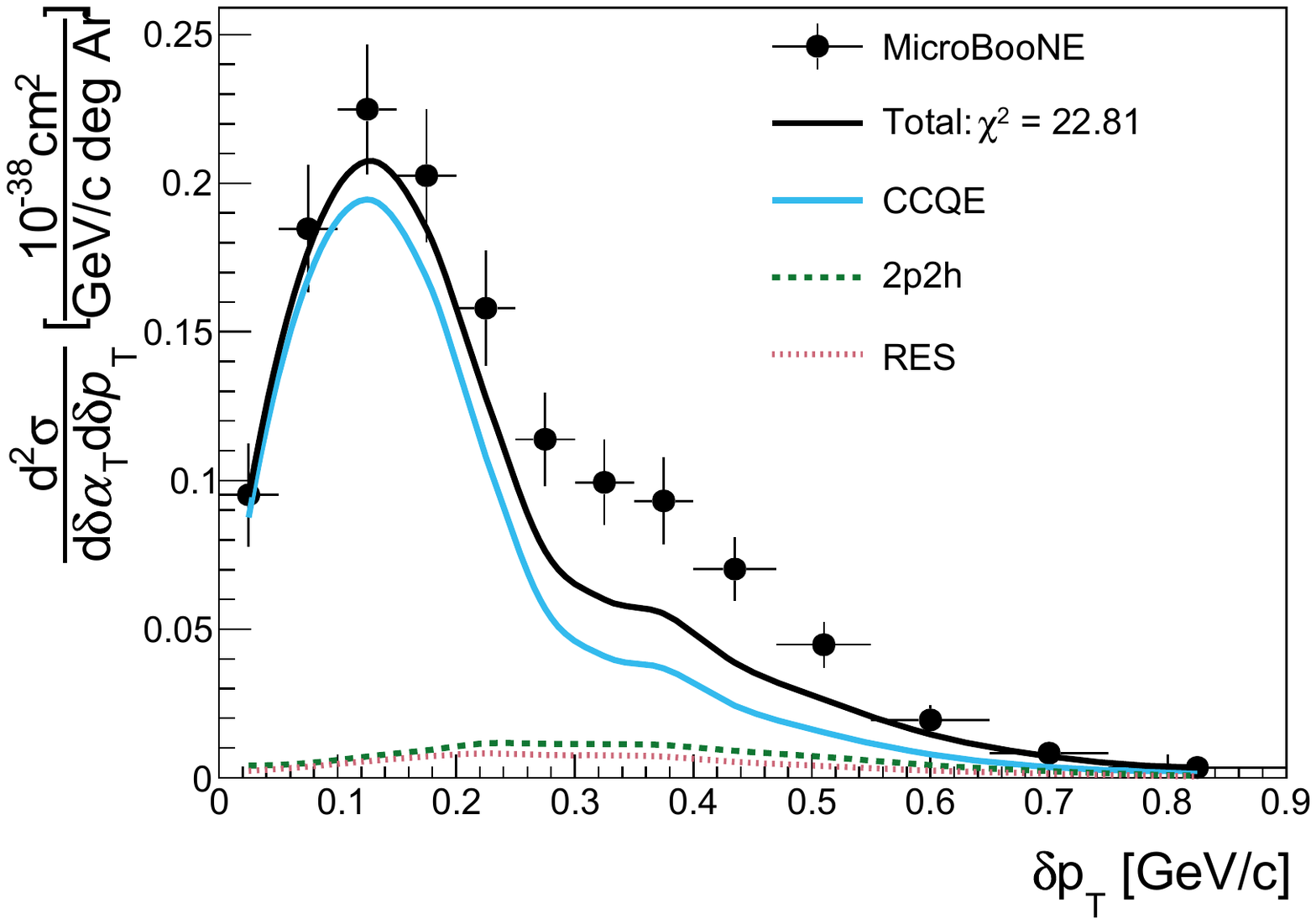}};
    \draw (2, 0) node {$90^\circ<\dalphat<135^\circ$};
\end{tikzpicture}
\begin{tikzpicture}
    \draw (0, 0) node[inner sep=0] {\includegraphics[width=8cm]{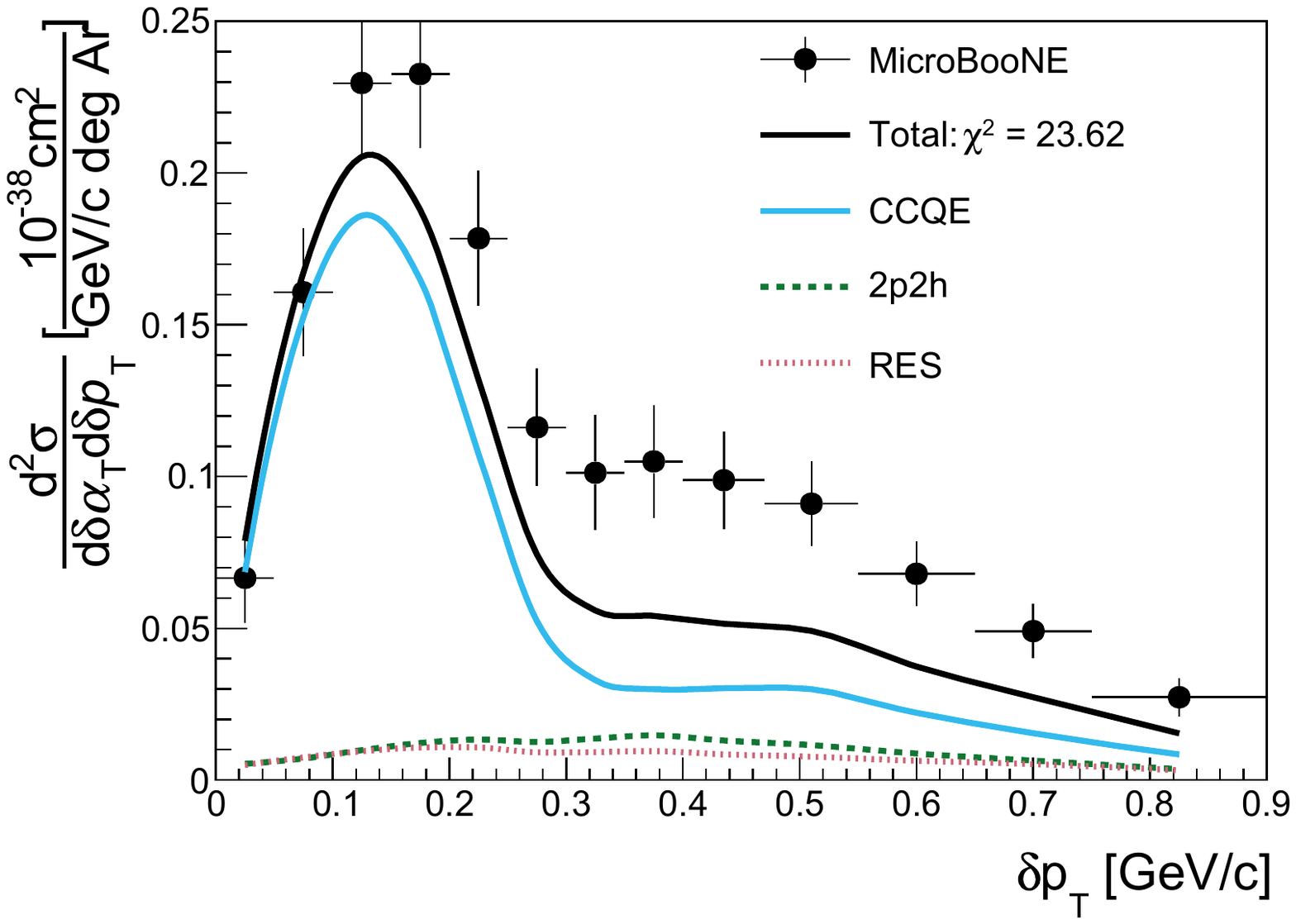}};
    \draw (1.9, 0) node {$135^\circ<\dalphat<180^\circ$};
\end{tikzpicture}

\caption{Multi-differential cross-section as a function of $\dpt$ and $\dalphat$ from MicroBooNE, compared with predictions using the combined SF* model. The figures showcase the breakdown by interaction mode into QE, 2p2h and resonant interactions (RES), spanning four regions of $\dalphat$.}
\label{fig:dptvsdatByMode}
\end{figure*}

A finer analysis of the agreement between the models and the measurement can be obtained by comparing the predictions to MicroBooNE's multi-differential measurement of $\dpt$ as a function of $\dalphat$, which is shown in \autoref{fig:dptvsdatByMode}. As described in \autoref{sec:analysisstrat}, generally large values of $\dalphat$ imply a more important role of FSI. It is therefore unsurprising to see that, in the low $\dalphat$ measurement where nuclear effects beyond Fermi motion are expected to be small, the $\dpt$ distribution is almost entirely dominated by QE interactions concentrated almost exclusively in the bulk. The small amount of remaining 2p2h and RES contributions is also shifted towards lower values of $\dpt$. In this region, the agreement between the simulation and the measurement is very good (a $p$-value of 0.26), suggesting a good description of the Fermi momentum by the Ar SF model in NuWro. At $45^\circ<\dalphat<90^\circ$, the $\dpt$ distribution exhibits a slightly more pronounced tail which is sill dominated by QE interactions, consistent with the increase in the strength of proton FSI. The 2p2h and RES contributions remain small, but are now shifted away from under the bulk of the distribution. The SF* simulation of the bulk is also slightly shifted towards slightly lower values of $\dpt$ with respect to the measurement, and it is apparent that the simulation lacks strength in the description of the tail of the measurement. As a result, the quantitative agreement between measurement and simulation is less good (giving a $p$-value of 0.07). A similar (and more pronounced) evolution can be seen at $90^\circ<\dalphat<135^\circ$ - the simulation clearly underpredicts the tail of the measurement as well as the bulk, and the quantitative agreement is poor (giving a $p$-value of 0.04). We also note the increase in the 2p2h and RES contributions across the entire distribution. Finally, in the FSI-rich region of $135^\circ<\delta\alpha_T<180^\circ$, the qualitative disagreement between simulation and the measurement is accentuated in both the tail, despite remaining quantitatively similar, and in the bulk, and we observe a slight increase in the strength of 2p2h and RES events, but neither of these are sufficient to reach the measurement. 

Overall, the relatively good agreement with MicroBooNE's low $\dalphat$ measurement, which then gets progressively worse toward larger $\dalphat$, may suggest that there is a lacking strength in either the 2p2h, RES or nucleon FSI strength for neutrino interactions on an argon target. Conversely, this does not appear to be the case for CH, as the simulation is able to accurately predict the strength of the $\dpt$ for T2K. In the following sections, we vary each of these effects individually and simultaneously for T2K and MicroBooNE in order to identify possible areas of model improvement. 

\subsubsection{Nuclear ground state}
\label{sec:nucmodel}

\begin{figure*}[htpb]
\centering
\includegraphics[width=8cm]{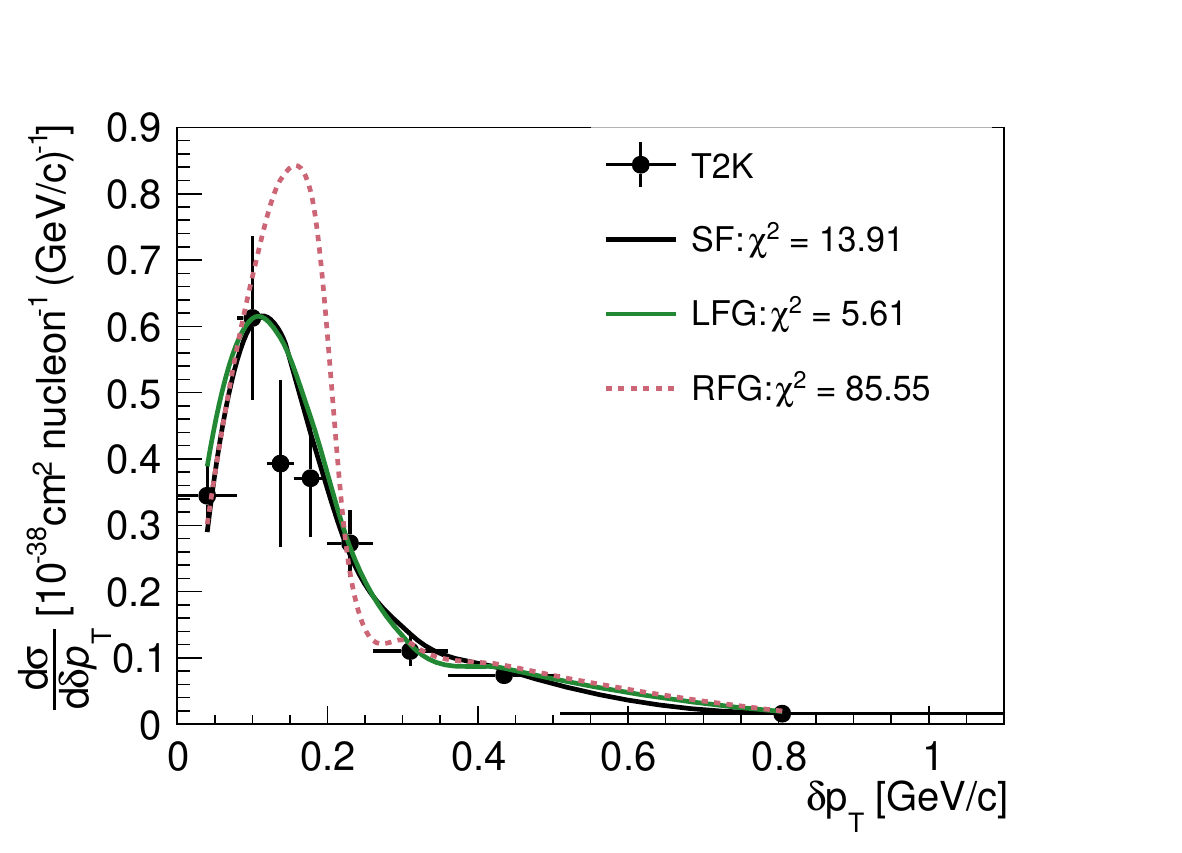}
\includegraphics[width=8cm]{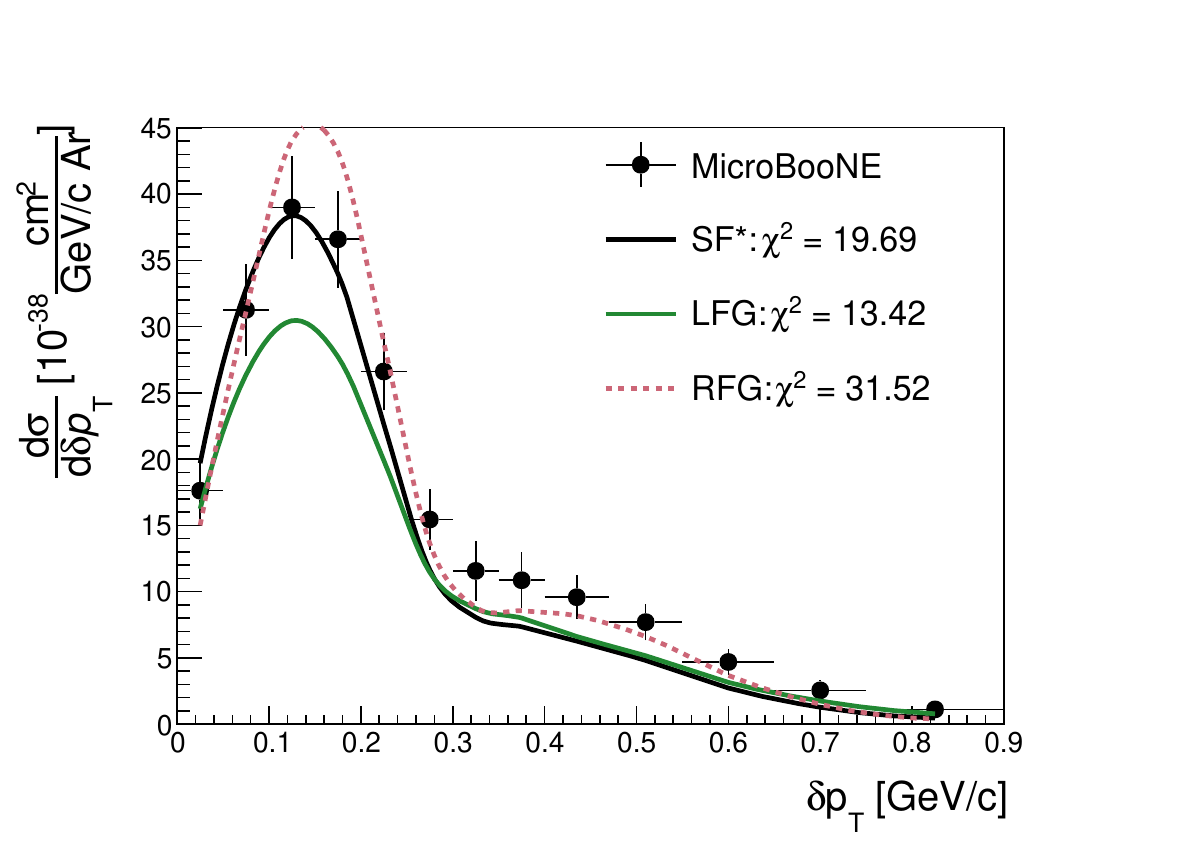} 
\caption{Differential cross section as a function of $\dpt$ using different nuclear model predictions compared with the measurements from T2K (left) and MicroBooNE (right). The different nuclear models are the NEUT SF, LFG and RFG models for T2K and the SF*, NEUT LFG and NEUT RFG models for MicroBooNE.}
\label{fig:NuclearModelT2KuBooNE}
\end{figure*}

\autoref{fig:NuclearModelT2KuBooNE} presents the comparison of nuclear ground state models for each experiment. For T2K, the SF and LFG models both align relatively well with the measurement. Conversely, in the context of MicroBooNE, the LFG model from NEUT aligns better with the measurement than the combined SF* model from NEUT and NuWro. However, this is at least partially due to the differing FSI model for QE interactions rather than the change of nuclear ground state, which is discussed further in \autoref{sec:fsi}. In both cases, the RFG model is clearly excluded by the measurement. It is notable that the RFG model predicts the expected ``Fermi cliff'' feature (the sharp drop off in strength at the Fermi momentum) for T2K but that this is washed out by the smearing applied to compare to the MicroBooNE measurement. 

\begin{figure*}[htpb]
\centering

\begin{tikzpicture}
    \draw (0, 0) node[inner sep=0] {\includegraphics[width=8cm]{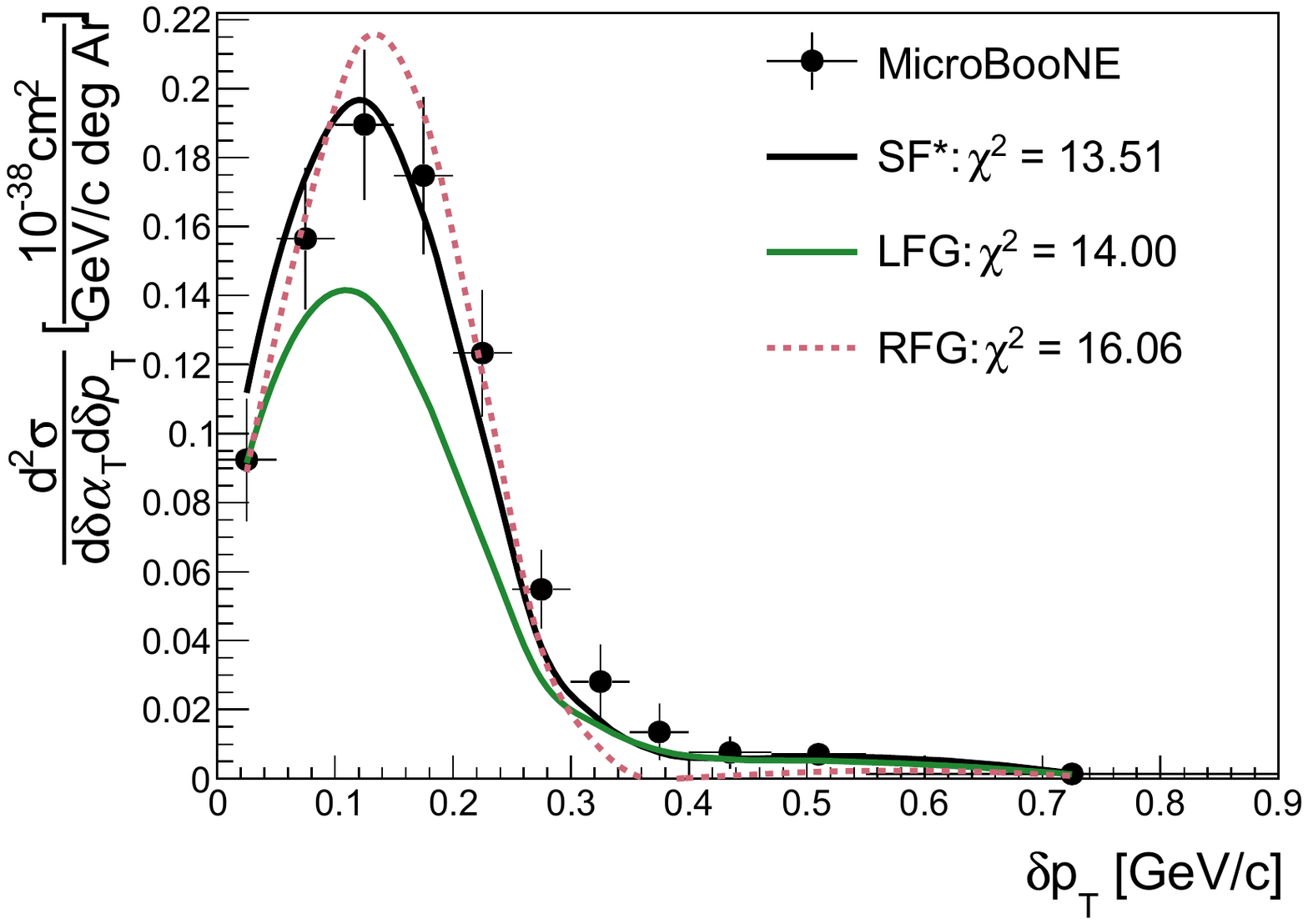}};
    \draw (2, 0) node {$0^\circ<\dalphat<45^\circ$};
\end{tikzpicture}
\begin{tikzpicture}
    \draw (0, 0) node[inner sep=0] {\includegraphics[width=8cm]{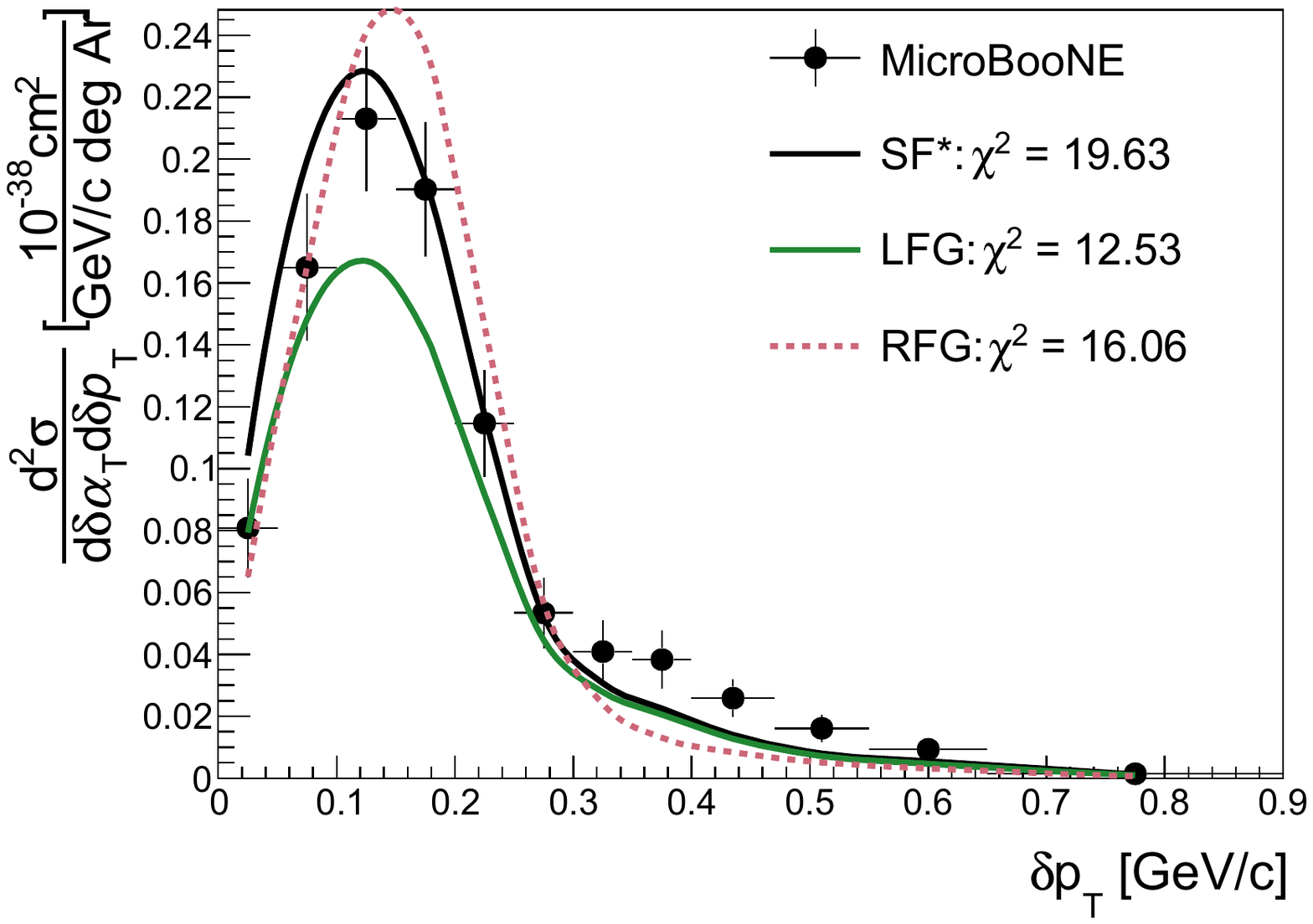}};
    \draw (2, 0) node {$45^\circ<\dalphat<90^\circ$};
\end{tikzpicture}
\begin{tikzpicture}
    \draw (0, 0) node[inner sep=0] {\includegraphics[width=8cm]{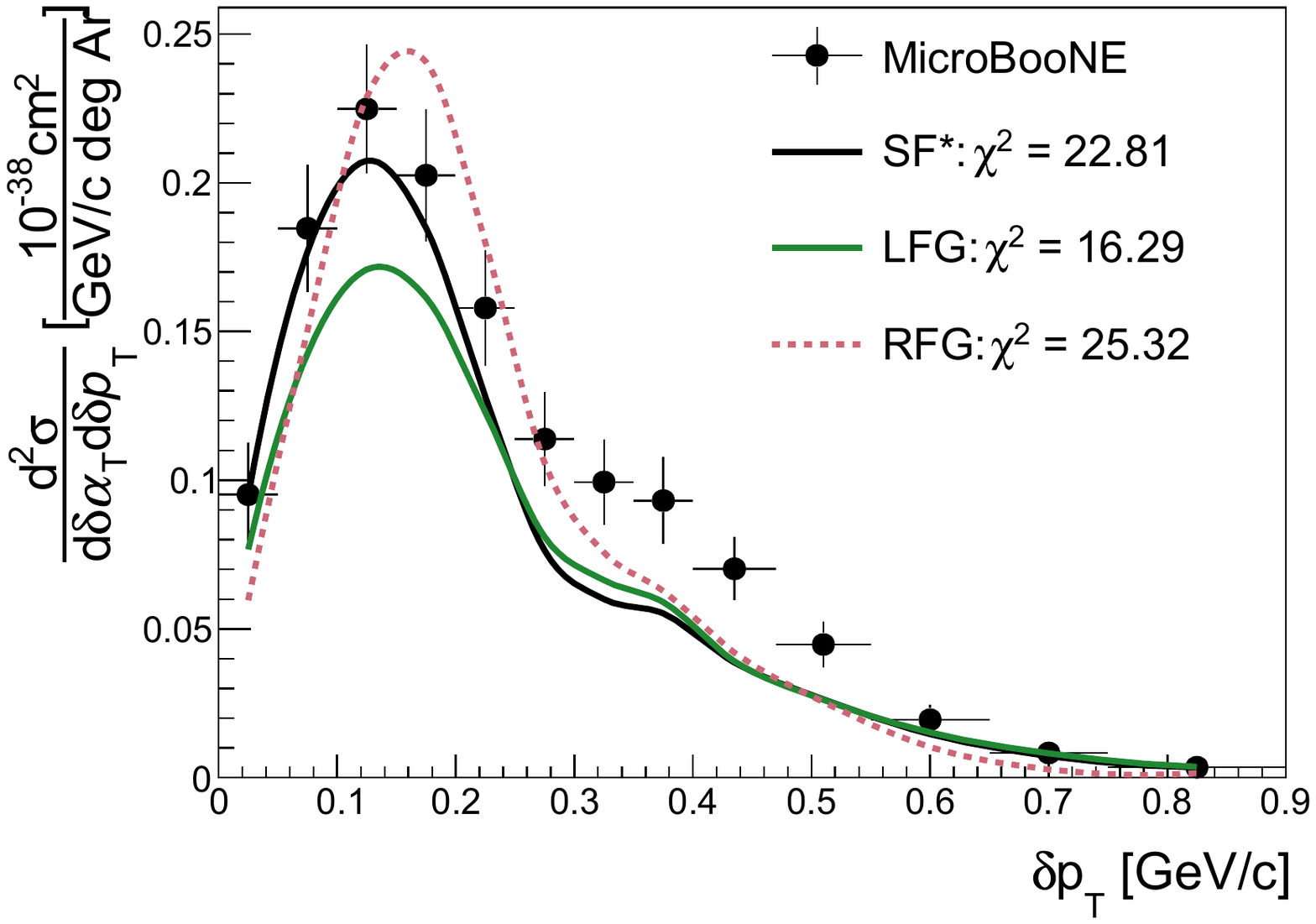}};
    \draw (2, 0) node {$90^\circ<\dalphat<135^\circ$};
\end{tikzpicture}
\begin{tikzpicture}
    \draw (0, 0) node[inner sep=0] {\includegraphics[width=8cm]{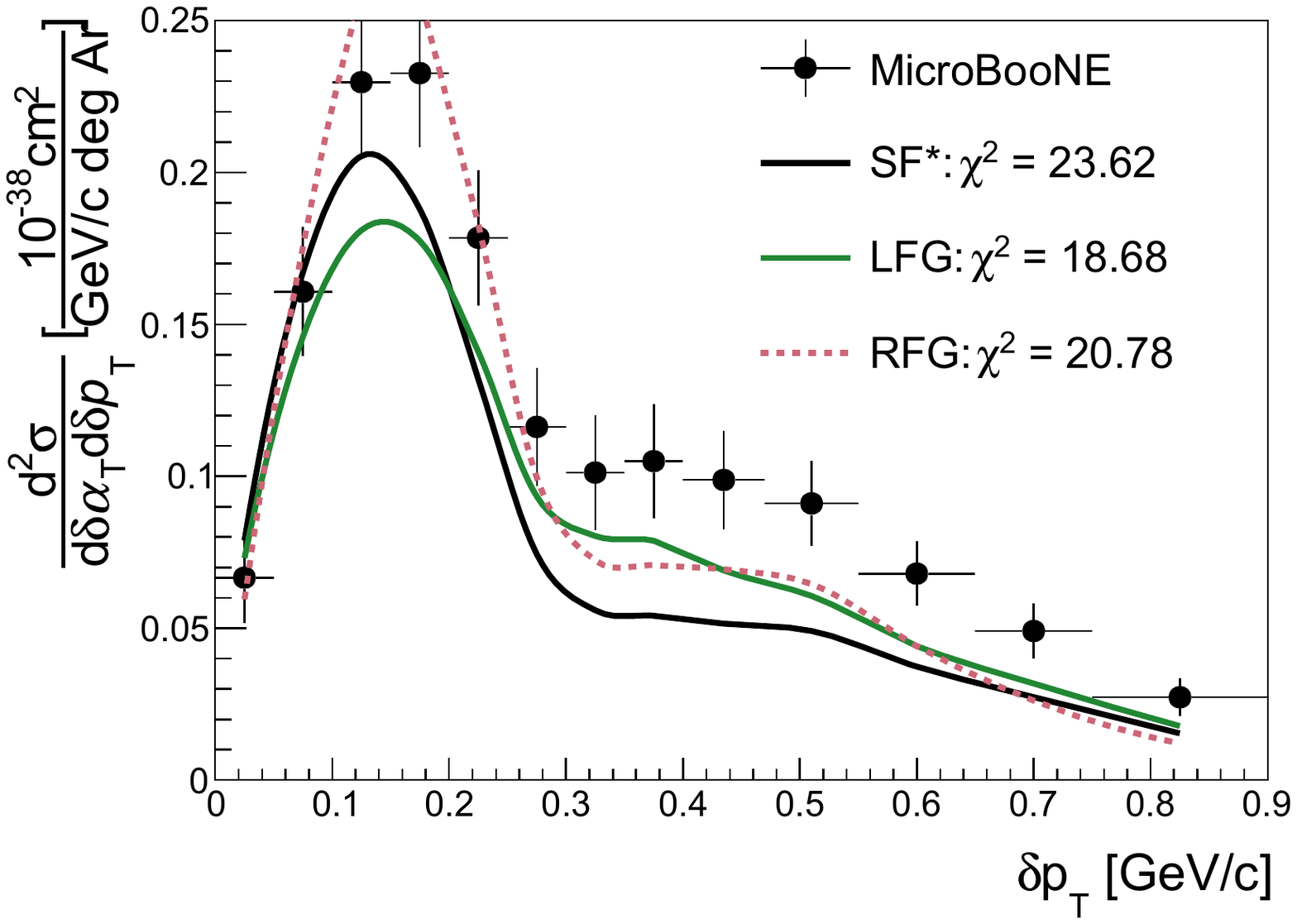}};
    \draw (1.8, 0) node {$135^\circ<\dalphat<180^\circ$};
\end{tikzpicture}

\caption{Multi-differential cross-section as a function of $\dpt$ and $\dalphat$ from MicroBooNE, compared with predictions using the SF*, NEUT LFG and NEUT RFG models, for different regions of $\dalphat$.}
\label{fig:dptvsdatNucModel}
\end{figure*}

It is interesting to note the difference in the bulk of $\dpt$ for both MicroBooNE and T2K. For T2K, the bulk predicted by the SF model broadly aligns with that predicted by the LFG model, whereas the bulk predicted by the LFG model is significantly smaller than the one for the SF* model for the MicroBooNE comparison. Whilst it is tempting to assign this to different treatments of FSI (with NEUT FSI to SF and LFG, and NuWro FSI to SF*), the two FSI models result in a very similar proportion ($\sim$75\%) of QE interactions passing MicroBooNE's signal definition and it is actually the NuWro model (affecting SF*) that migrates the larger portion of events outside of the phase space constraints (see \autoref{sec:fsi} and \autoref{tab:pre_post_fsi}). This implies an expectation for FSI not to lower the LFG prediction with respect to SF* in the $\dpt$ bulk. This is also consistent with the comparison of the models to the multi-differential MicroBooNE measurement is shown in \autoref{fig:dptvsdatNucModel}, which shows large differences between LFG and SF* in the regions where $\dalphat<90^\circ$, where FSI is expected to be less impactful. In fact, before the phase space constraints defined in \autoref{tab:sigDef}, the total cross section predicted by the SF(SF*) model is about 5\%(10\%) higher than that predicted by the LFG model for T2K(MicroBooNE). After applying the low proton momentum constraints from \autoref{tab:sigDef}, the ratio between the SF* and LFG total cross section predictions stays relatively constant for the MicroBooNE case, whereas for T2K it brings the SF prediction at about the same level in the bulk as that predicted by LFG (as is observed in \autoref{fig:NuclearModelT2KuBooNE}). This is primarily due to the more stringent T2K cuts on low momentum protons with respect to those applied by MicroBooNE. Indeed, the impact of raising the proton momentum threshold from 300 MeV/$c$ to 450 MeV/$c$ in T2K simulations lowers the total SF cross section by $\sim$35\%, and the total LFG cross section by $\sim$25\%. 
This indicates that the main driver for the suppression of the SF cross section in the case of the T2K predictions is the removal of protons whose momenta are between 300-450 MeV/$c$. 

The larger reduction of the SF cross section with respect to LFG when applying cuts that require large values of proton momentum may be expected from the fact that the low energy transfer portion of the LFG cross section (broadly corresponding to low proton momentum) is much smaller than in SF. This is because of LFG's considerations of a cross section suppression from long range nucleon correlations via the random phase approximation. At larger energy transfers (corresponding to cuts requiring higher proton momentum, as in T2K's signal definition) the LFG and SF model cross sections are more similar. We also note that applying a similar suppression to the SF* model prediction of the MicroBooNE measurement, via an optical potential correction from Ref.~\cite{Ankowski:2014yfa}, provides a cross section with a more similar normalisation to the NEUT LFG model (both before and after applying the MicroBooNE kinematic phase space constraints). This potentially suggests that the larger prediction for the MicroBooNE $\dpt$ measurement bulk from SF* is related to the use of a calculation based almost solely on the plane wave impulse approximation (PWIA) and in particular without any consideration of the nuclear potential that the outgoing nucleon experiences or of long range nucleon correlations. The MicroBooNE measurement is more sensitive to physics beyond PWIA than T2K or MINERvA, thanks to its lower proton tracking threshold giving access to interactions with lower energy transfers. 

\autoref{fig:dptvsdatNucModel} additionally shows that, at low $\dalphat$ values, the performance of the LFG and SF* models is roughly equivalent ($p$-values of 0.23 and 0.26 respectively), and better than that of the RFG model ($p$-value of 0.14). With increasing $\dalphat$, it appears that the best agreement overall is achieved by the LFG model, which applies the NEUT intra-nuclear cascade simulation, unlike the SF* model which uses the NuWro FSI model. At $135^\circ<\dalphat<190^\circ$, we can see that both the LFG and RFG models show higher predictions in the tail of the distribution and show better agreement with the measurement compared to the SF* model. In the low $\dalphat$ region (top row of \autoref{fig:dptvsdatNucModel}), the visual disagreement between the bulks predicted by the SF* and LFG models appears significant, whereas in the bottom row the models show a more consistent description of the bulk. However, the difference should not only be interpreted visually due to the large correlations in particular in the tails of the distributions and to the fact that the measurement is reported in the smeared space. 

\begin{figure}[htpb]
\centering
\includegraphics[width=10cm]{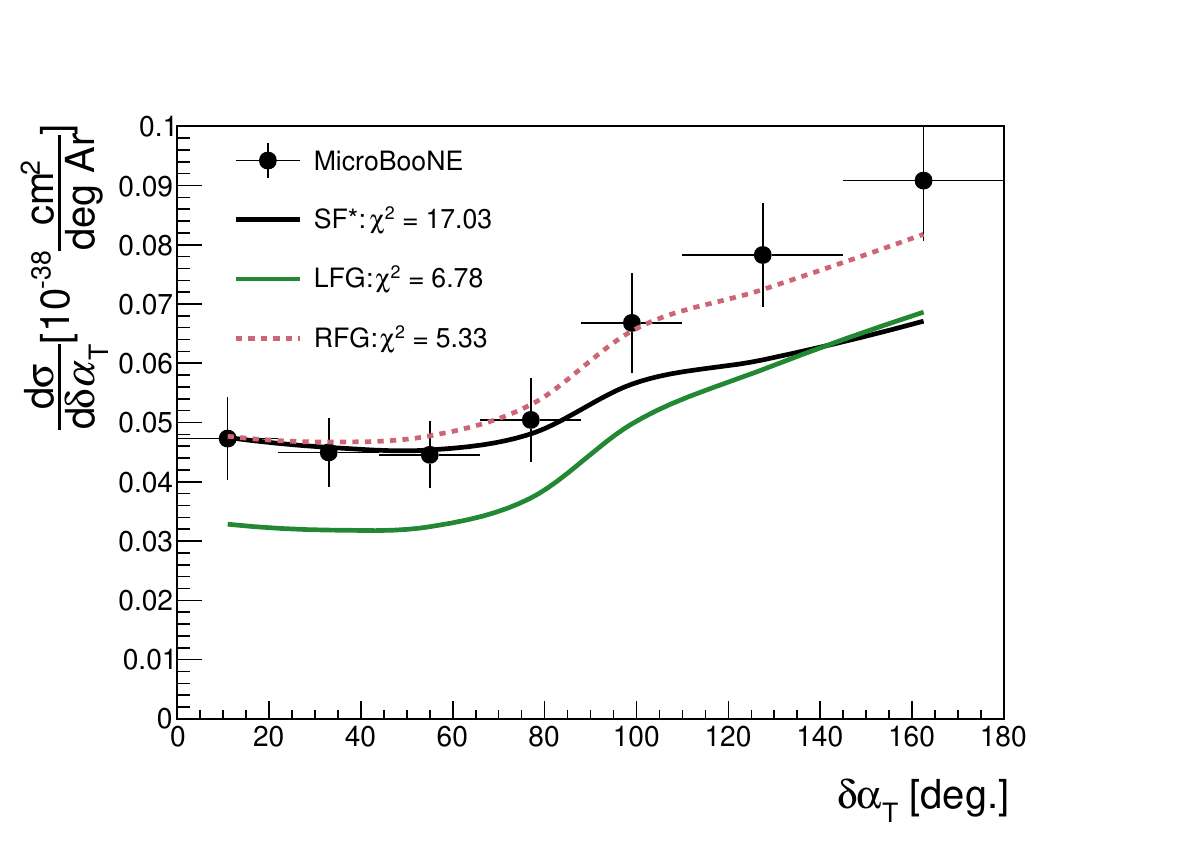}
\caption{$\dalphat$ measurement from MicroBooNE, compared with the cross section predictions from the combined SF*, NEUT LFG and NEUT RFG simulations.}
\label{fig:dalphat_uboone_nucmodels}
\end{figure}

It is additionally useful to examine the agreement between the different models with MicroBooNE's $\dalphat$ measurement, shown in \autoref{fig:dalphat_uboone_nucmodels}. We can note that, although the normalization of the LFG and RFG distributions is different, their shape as a function of $\dalphat$ is similar, and both are quite different from that predicted by the SF* model. This is not surprising, as the FSI model applied to protons using the LFG and RFG models is identical (NEUT intra-nuclear cascade), and different from the NuWro FSI model applied to the SF* model. We also note the better quantitative agreement shown in \autoref{tab:chi2_pValues} between the purely NEUT-based simulations compared to the NuWro FSI model and the MicroBooNE measurement. This is further discussed in \autoref{sec:fsi}.

Overall, it appears that LFG provides the best description of the T2K and MicroBooNE measurements. Although both SF/SF* and LFG provide acceptable $p$-values in the measurements of $\dpt$, LFG is better suited to describing MicroBooNE's measurement of $\dalphat$ and, in the high $\dalphat$ bins, its multi-differential measurement of the two. However, we repeat that this preference at higher $\dalphat$ may be driven by the differing FSI models. The RFG model is excluded by both the T2K and MicroBooNE $\dpt$ measurements.

\subsubsection{2p2h}
\label{sec:2p2h}

As discussed in \autoref{sec:analysisstrat} and \autoref{sec:bymode}, the effect of 2p2h interactions shows up predominantly in the tails of the $\dpt$ distribution, and generally more at high $\dalphat$. In this section, we increase the total cross section of 2p2h interactions by 70\% uniformly across neutrino energy, which brackets the variations predicted by available models in neutrino generators as discussed in \autoref{sec:TKI}. 

\begin{figure*}[p]
\centering
\includegraphics[width=8cm]{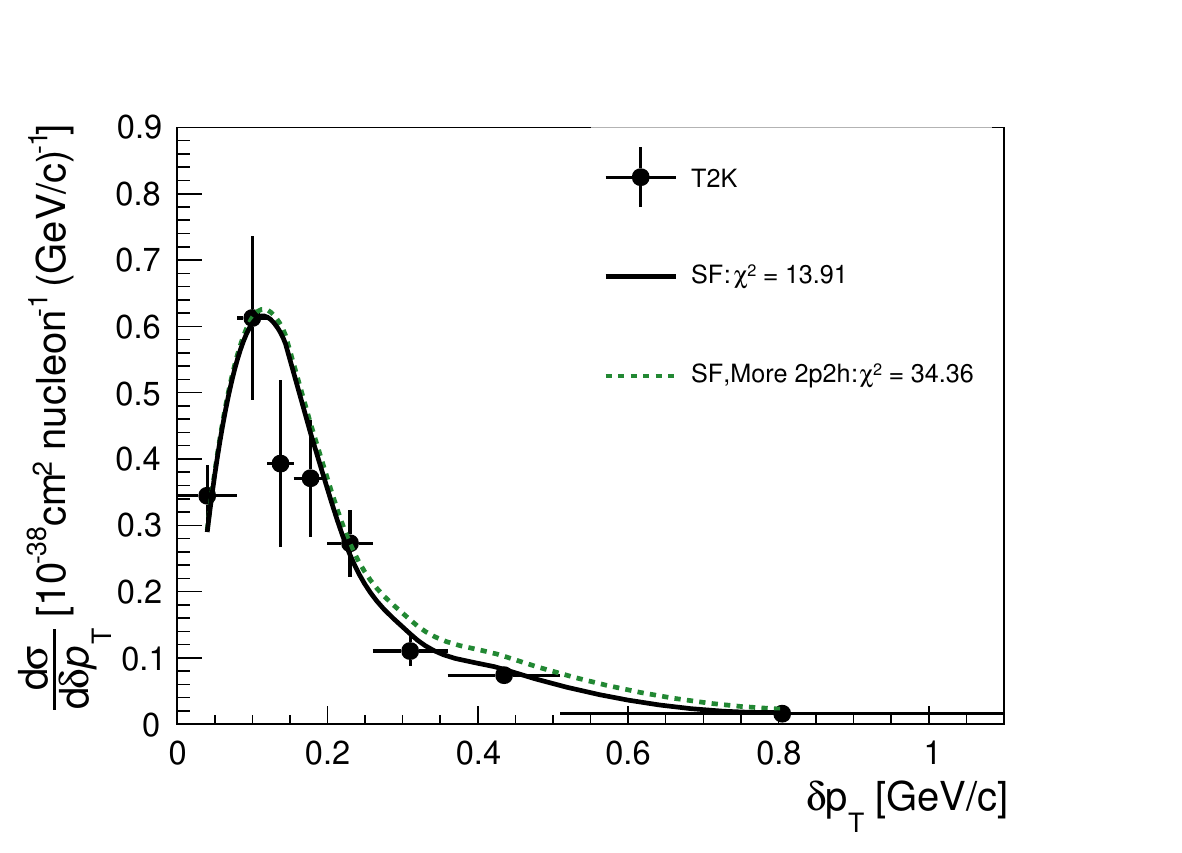}
\includegraphics[width=8cm]{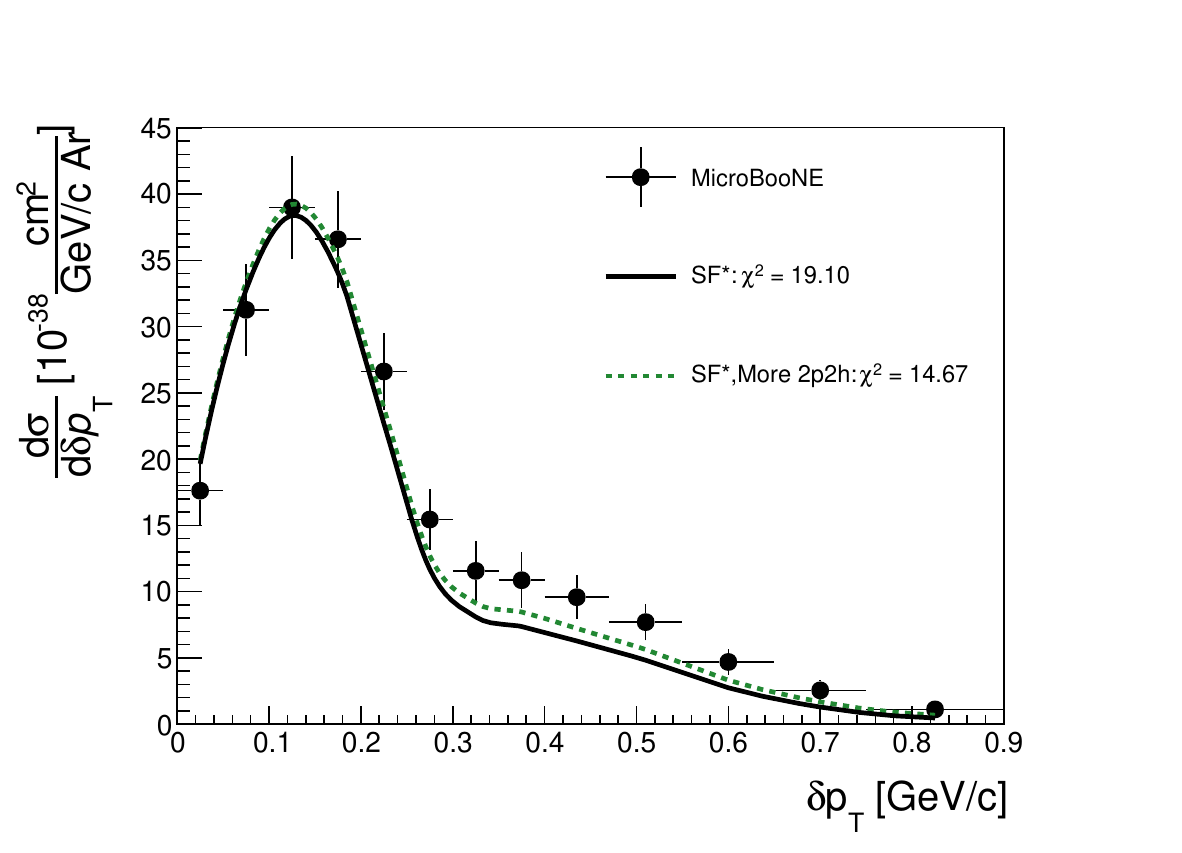}
\caption{Differential cross section measurement as a function of $\dpt$ from T2K (left) and MicroBooNE (right), compared with predictions using the NEUT SF and the combined SF* model, respectively. The measurements are compared with the same simulations where the total cross section of 2p2h processes has been increased by 70\% uniformly across neutrino energies (labeled as ``More 2p2h'' in the legends).}
\label{fig:t2k_uboone_2p2h}
\end{figure*}

\autoref{fig:t2k_uboone_2p2h} shows the agreement between the nominal models and the modified simulations with the measurements from T2K and MicroBooNE. The T2K measurement clearly disfavors increasing the strength of 2p2h interaction, whereas the MicroBooNE measurement shows the opposite preference. It is apparent that the main effect of increasing the strength of 2p2h interactions is to increase the strength of the tail of the $\dpt$ distribution, but both samples are highly dominated by QE interactions, as shown in \autoref{fig:byModeT2KuBooNE}, and thus the effect is limited. 

\begin{figure*}[p]
\centering
\begin{tikzpicture}
    \draw (0, 0) node[inner sep=0] {\includegraphics[width=8cm]{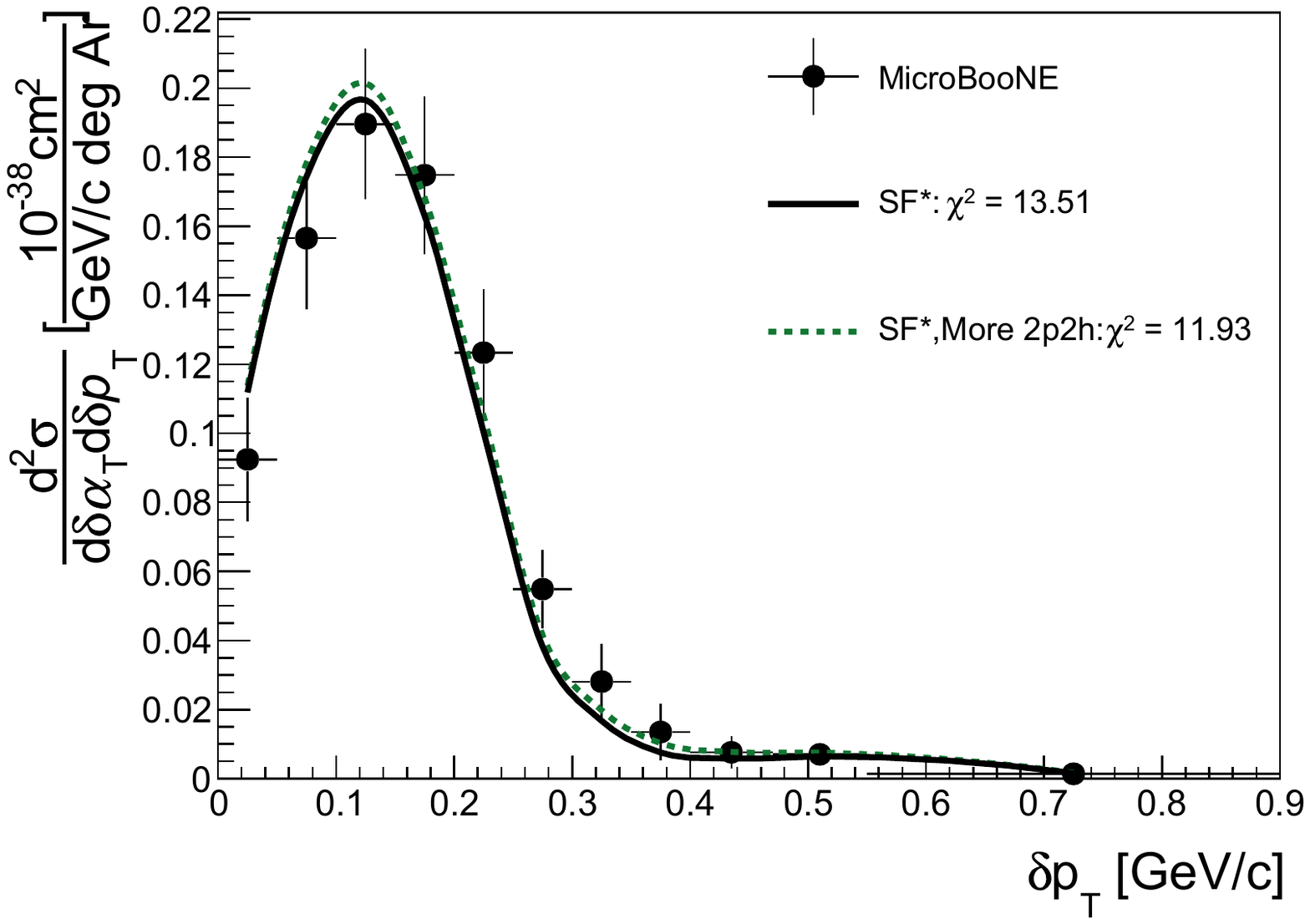}};
    \draw (2, 0) node {$0^\circ<\dalphat<45^\circ$};
\end{tikzpicture}
\begin{tikzpicture}
    \draw (0, 0) node[inner sep=0] {\includegraphics[width=8cm]{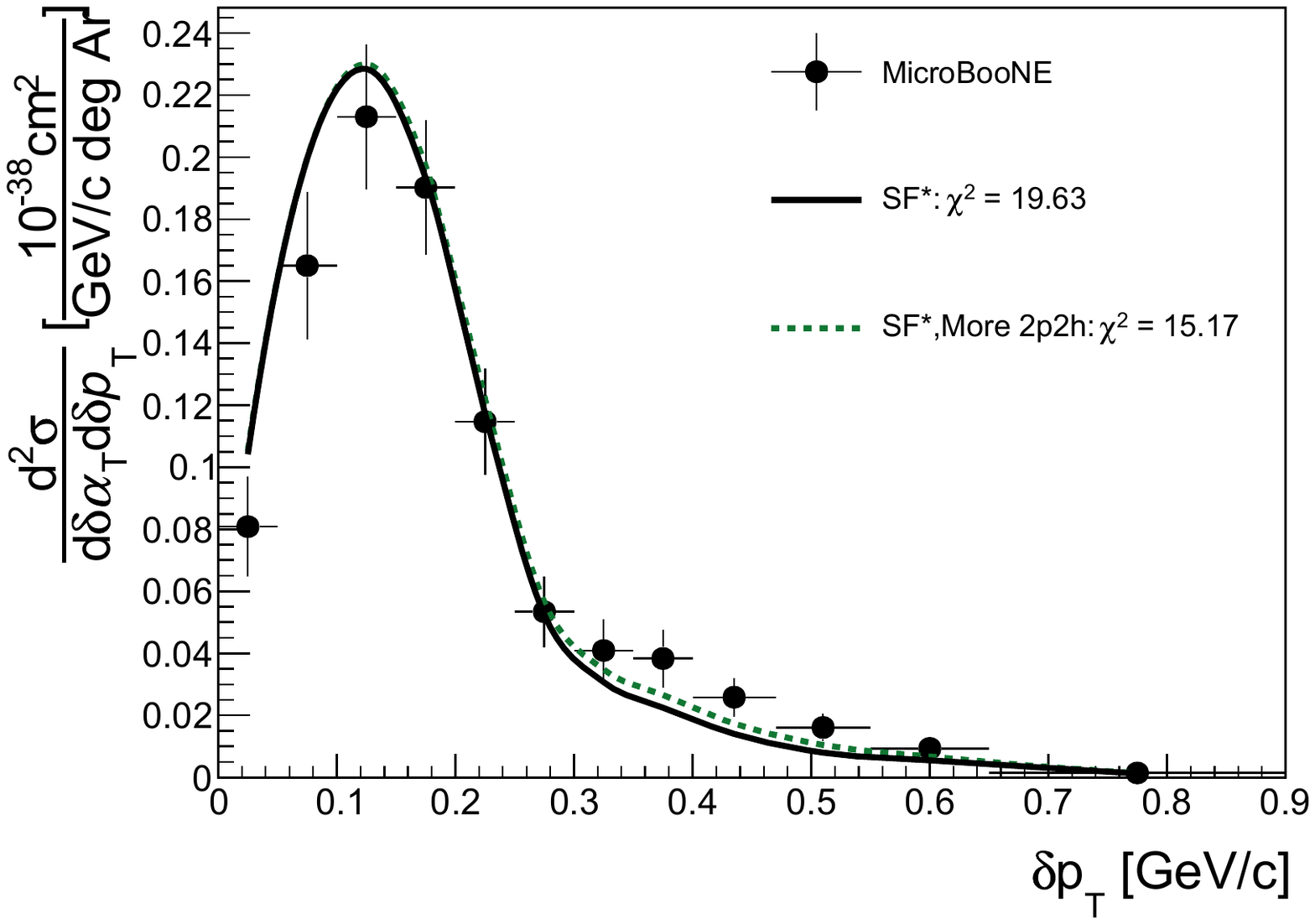}};
    \draw (2, 0) node {$45^\circ<\dalphat<90^\circ$};
\end{tikzpicture}
\begin{tikzpicture}
    \draw (0, 0) node[inner sep=0] {\includegraphics[width=8cm]{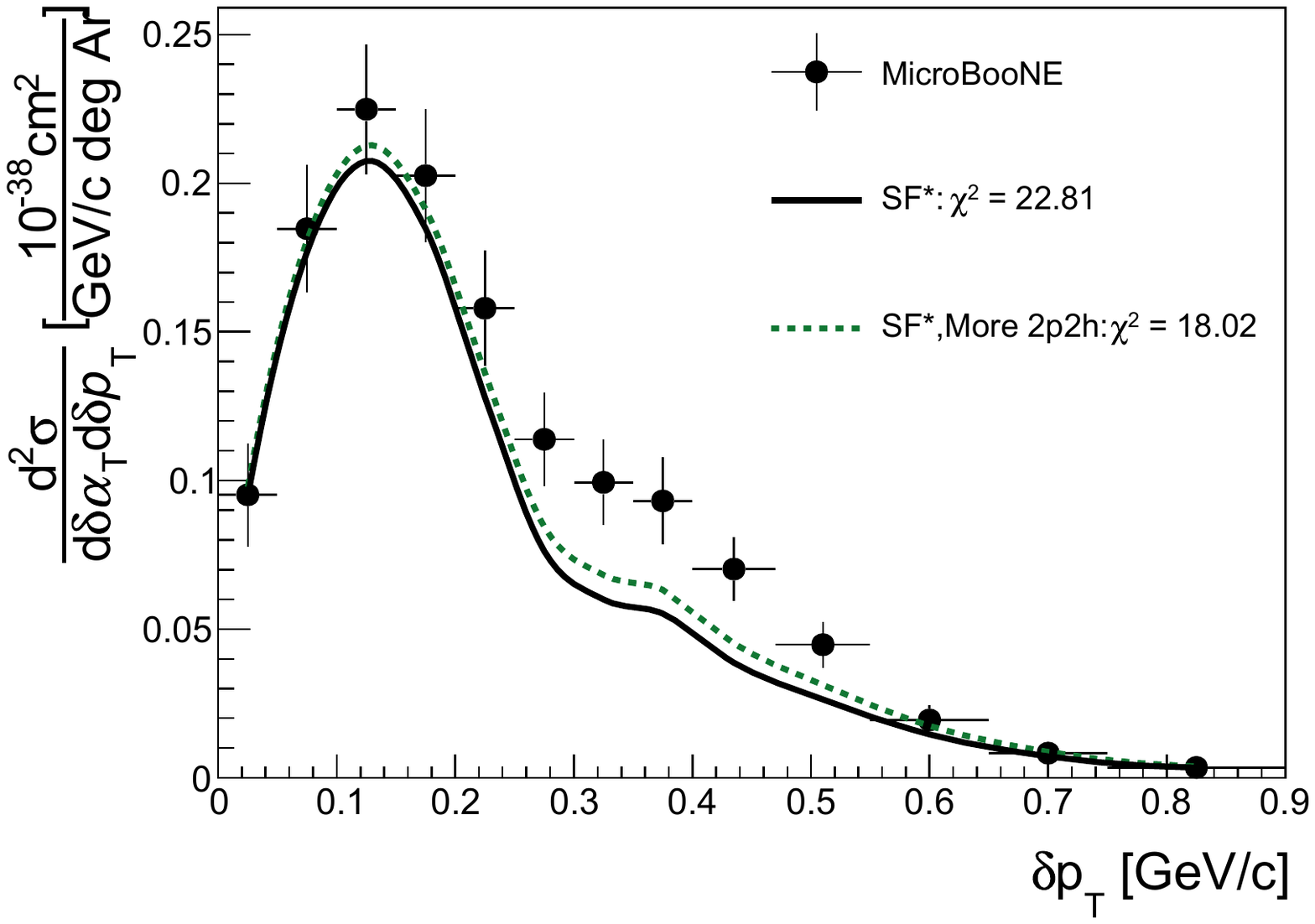}};
    \draw (2, 0) node {$90^\circ<\dalphat<135^\circ$};
\end{tikzpicture}
\begin{tikzpicture}
    \draw (0, 0) node[inner sep=0] {\includegraphics[width=8cm]{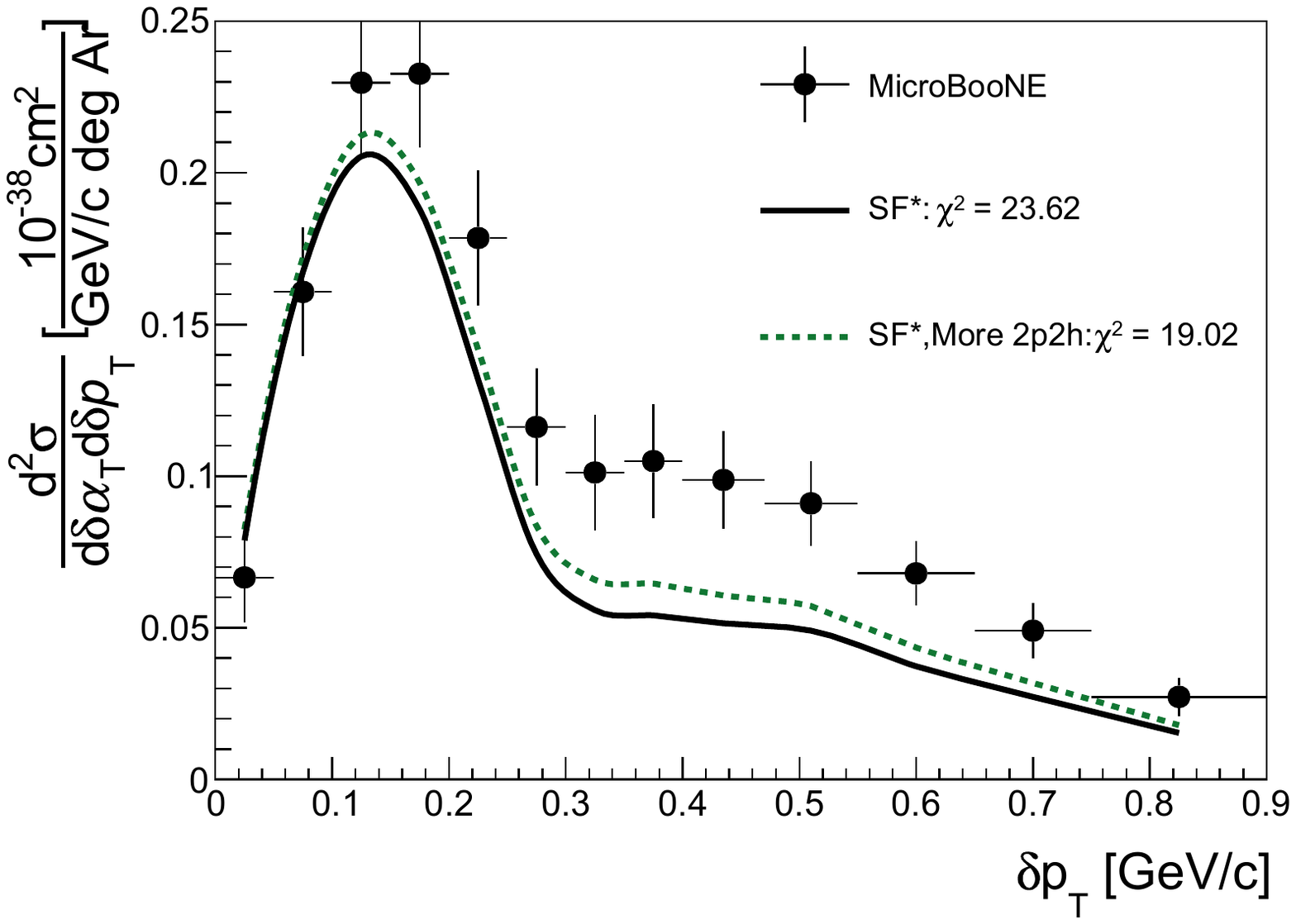}};
    \draw (1.8, 0) node {$135^\circ<\dalphat<180^\circ$};
\end{tikzpicture}

\caption{Multi-differential cross-section as a function of $\dpt$ and $\dalphat$ from MicroBooNE, compared with predictions using the SF* model, and the same model in which the 2p2h interaction cross section is increased by 70\% uniformly across neutrino energies (labelled as ``More 2p2h'' in the legends).}
\label{fig:dptvsdat2p2h}
\end{figure*}

Since the tail of the $\dpt$ distribution has contribution from both 2p2h processes as well as QE interactions where the protons have undergone FSI, it is useful to attempt to separate these effects with the multi-differential MicroBooNE measurement. \autoref{fig:dptvsdat2p2h} shows the evolution of the agreement between the two simulations as a function of the measured value of $\dalphat$. For all measurements, it is apparent that increasing the strength of 2p2h interactions improves the agreement with the measurement, although this trend is less pronounced at low $\dalphat$, which is expected since there is less 2p2h in general (see \autoref{fig:dptvsdatByMode}). However, even the large increase of 2p2h considered here is far from sufficient to describe the tails of the MicroBooNE measurement at large $\dalphat$.

While it is likely that the amount of 2p2h contribution is not uniquely responsible for the disagreement between the MicroBooNE measurement and simulations, the comparisons do seem to suggest that there may be an issue in the modeling of the scaling of the 2p2h cross section with atomic number when considering the nominal simulation's reasonable agreement with the T2K measurement. In addition to having a much larger atomic number, argon is, unlike carbon, a non-isoscalar nucleus. The NEUT implementation of the Valencia 2p2h model uses precomputed  tensors tables for a number of isoscalar nuclei to calculate the total 2p2h cross section. In the absence of such a table for a specific nucleus (which is the case for argon), NEUT uses the table for the available nucleus with the closest atomic number and scales the cross section to the atomic number of argon. Isoscalarity may have a direct impact on the rate of 2p2h interactions (as it modifies the fraction of $np$ and $nn$ initial state nucleon pairs), and other models, such as GiBUU~\cite{Buss:2011mx}, predict that the scaling of the cross section depends on the difference between the numbers of protons and neutrons inside the nucleus~\cite{Dolan:2018sbb}. 

\subsubsection{FSI}
\label{sec:fsi}

In this section, we report studies on the effect of nucleon FSI variations. As stated in \autoref{sec:TKI}, FSI affects the TKI distributions, and in particular those of $\dpt$, in a complex way, modifying both the tail and the bulk. Moreover, whilst some aspects of generator FSI modeling can be benchmarked by sophisticated theory calculations~\cite{Franco-Patino:2022tvv,Franco-Patino:2023msk,Nikolakopoulos:2024mjj,Nikolakopoulos:2022qkq}, no microscopic model can predict the kinematics and multiplicities of all outgoing hadrons. This makes generators' FSI models, with their limited predictive power and theoretical grounding, the only means of calculating the fully exclusive CC0$\pi$ cross section. 

We begin by assessing the impact of varying the nucleon MFP which, as discussed in \autoref{sec:systvar}, is changed by 30\% as motivated by nucleon transparency measurements. Whilst this changes generator predictions, it does not cover the full plausible variation from FSI on the TKI distributions of interest. Consequently, we then study how different FSI models can alter outgoing hadron kinematics differently (even if their transparency predictions remain similar) and assess the potential impact on T2K and MicroBooNE measurements. 

\begin{center}
\textit{Varying the nucleon FSI strength}
\end{center}

\begin{figure*}[htbp]
\centering
\includegraphics[width=8cm]{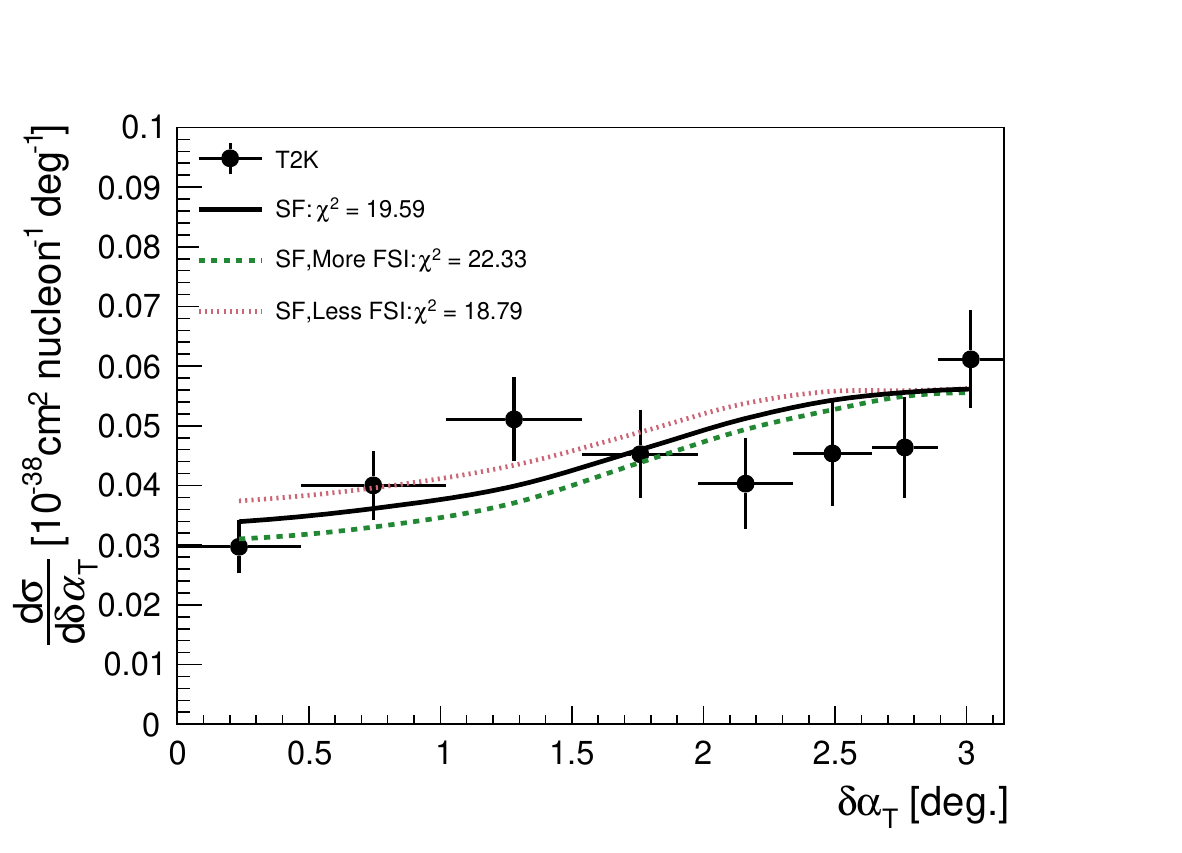}
\includegraphics[width=8cm]{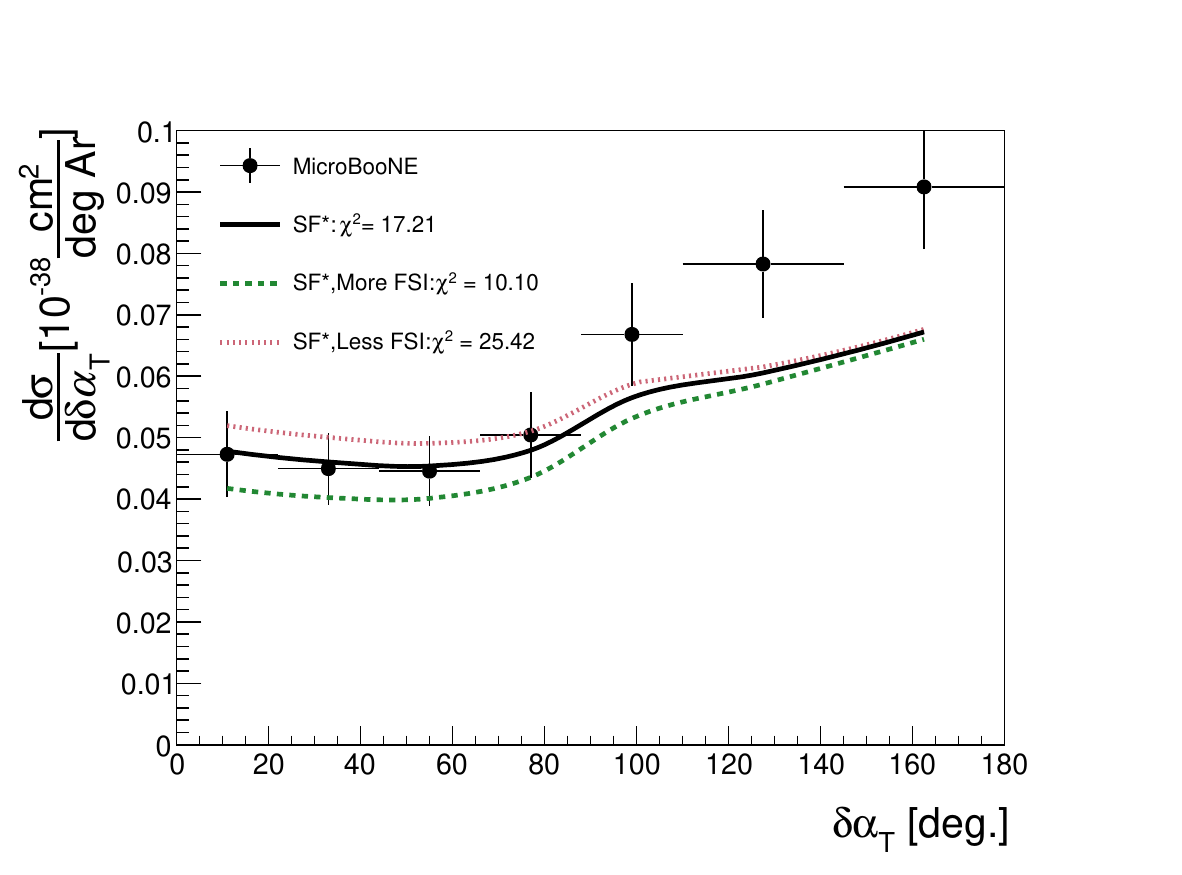}
\caption{Differential cross section measurements as a function of $\dalphat$ from T2K(left) and MicroBooNE(right), compared with predictions using the NEUT SF model for T2K, and the combined SF* model for MicroBooNE. The effects of adjusting the nucleon mean free path by -(+)30\% are displayed and labeled as ``More(Less) FSI''.}
\label{fig:t2k_uBoone_FSI_dalphat}
\end{figure*}

\begin{figure*}[htbp]
\centering
\includegraphics[width=8cm]{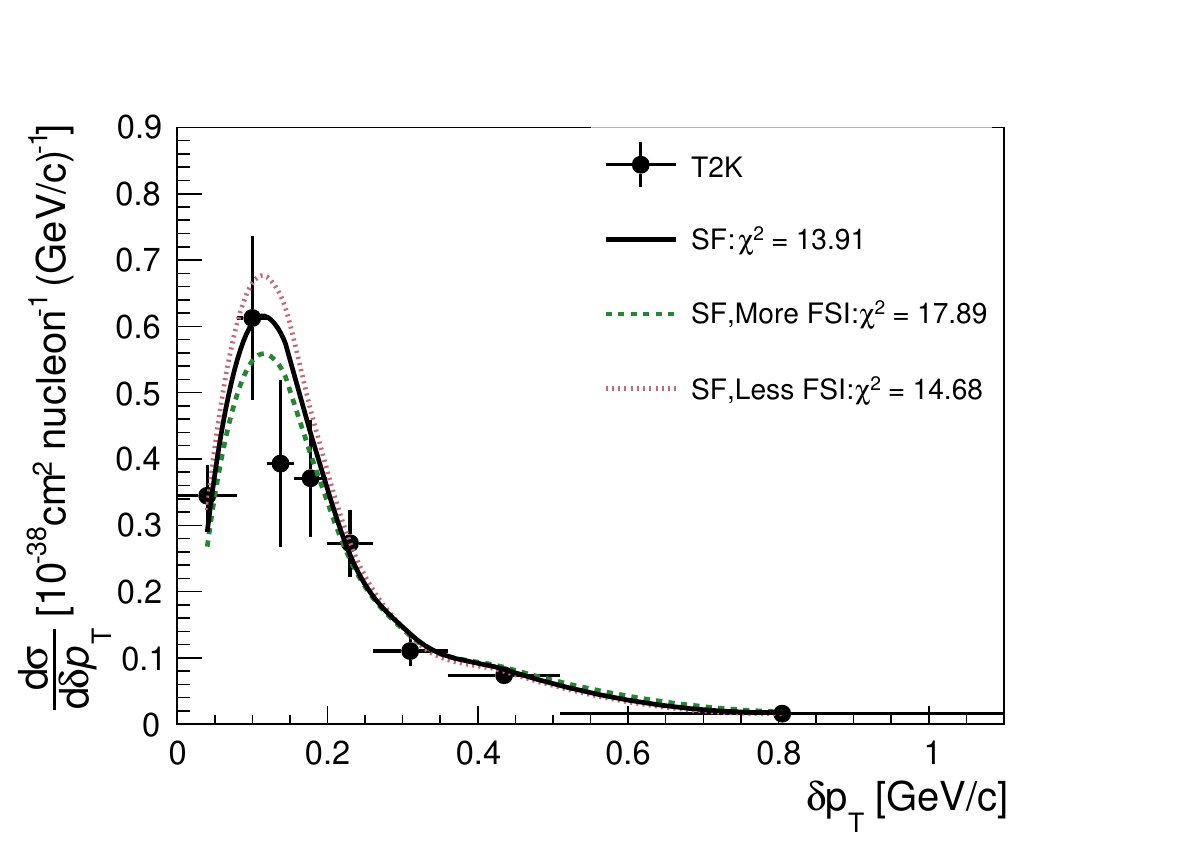}
\includegraphics[width=8cm]{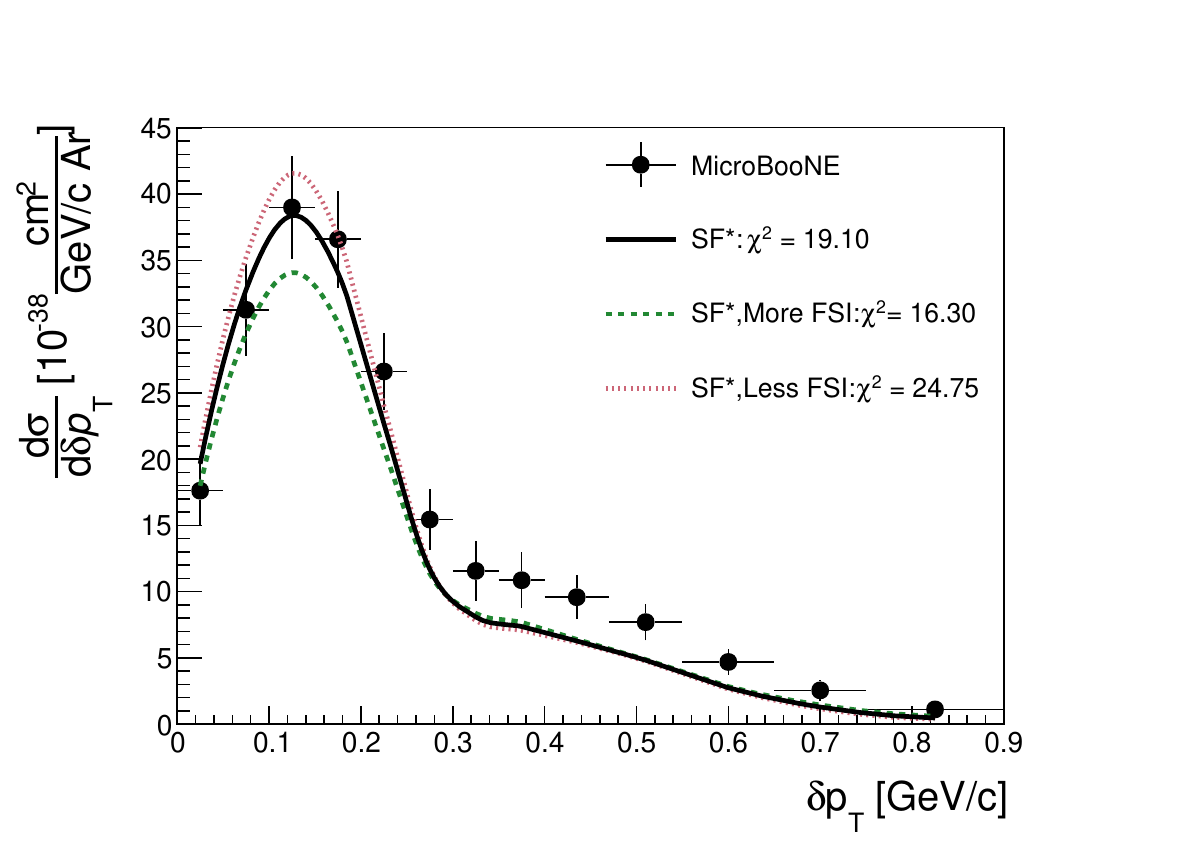}
\caption{Differential cross section measurement as a function of $\dpt$ from T2K(left) and MicroBooNE(right), compared with predictions using the NEUT SF model for T2K, and the combined SF* model for MicroBooNE. The effects of adjusting the nucleon mean free path by -(+)30\% are displayed and labeled as ``More(Less) FSI''.}
\label{fig:t2k_uBoone_FSI}
\end{figure*}

\begin{figure*}[htpb]
\centering

\begin{tikzpicture}
    \draw (0, 0) node[inner sep=0] {\includegraphics[width=8cm]{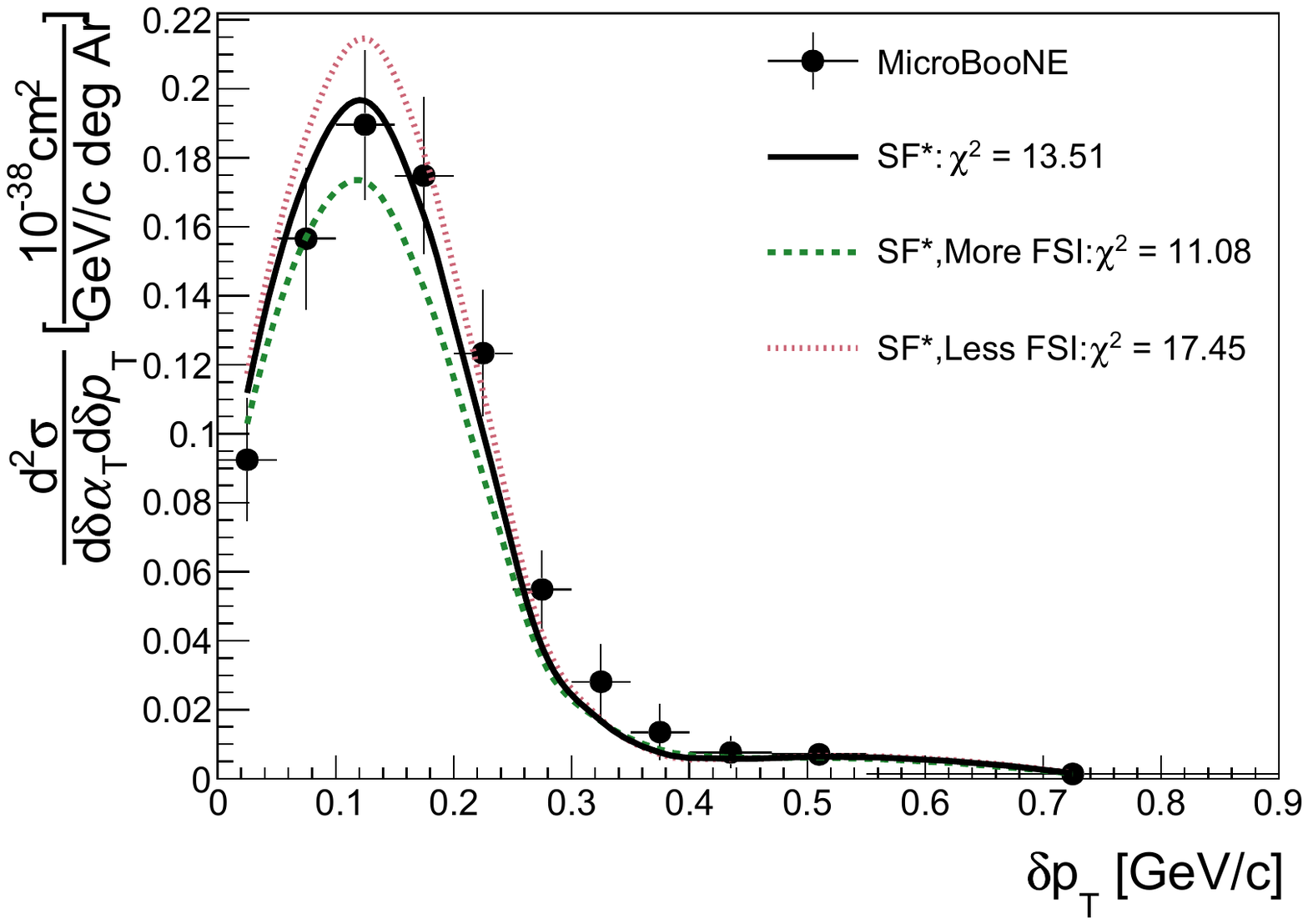}};
    \draw (2, 0) node {$0^\circ<\dalphat<45^\circ$};
\end{tikzpicture}
\begin{tikzpicture}
    \draw (0, 0) node[inner sep=0] {\includegraphics[width=8cm]{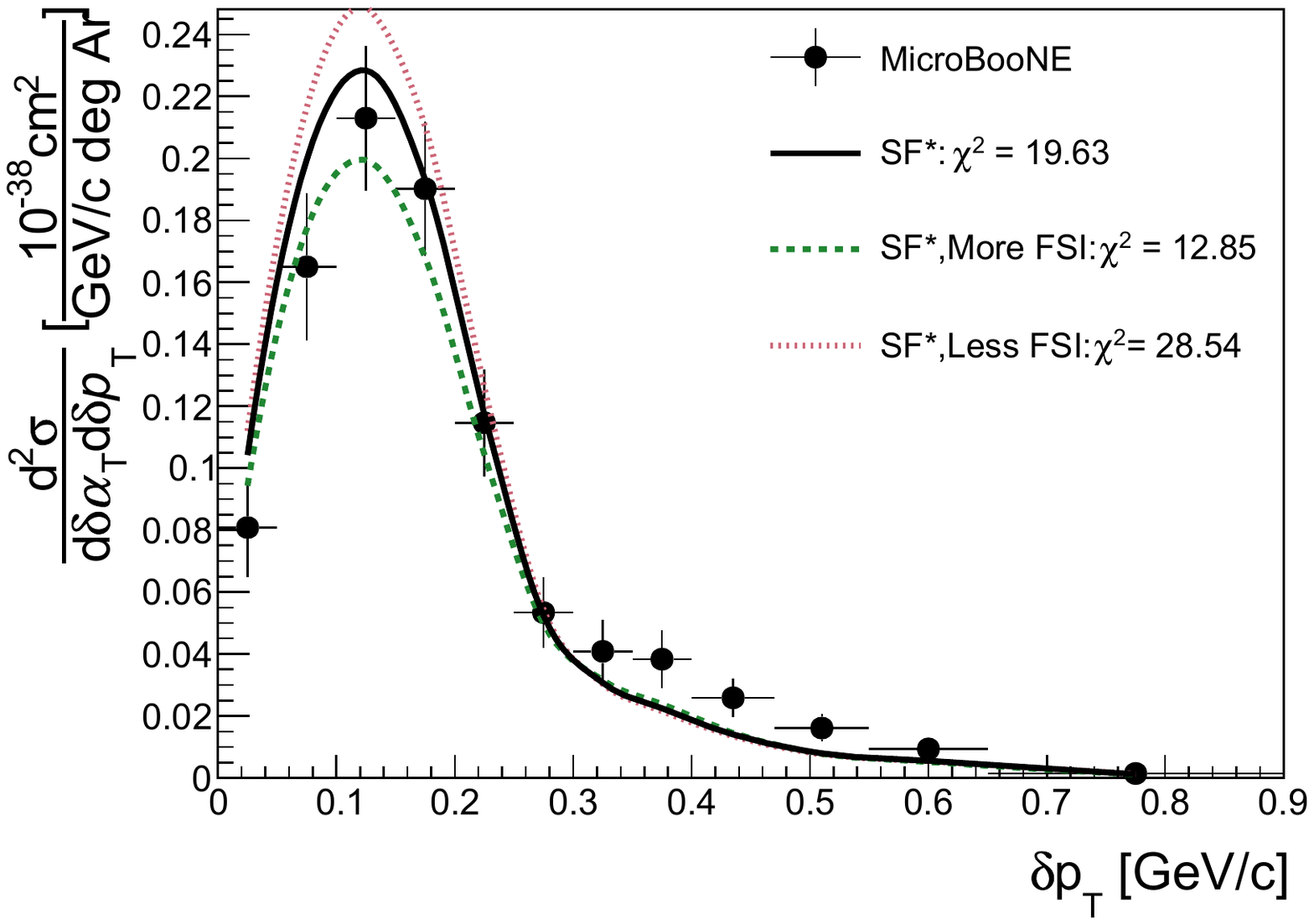}};
    \draw (2, 0) node {$45^\circ<\dalphat<90^\circ$};
\end{tikzpicture}
\begin{tikzpicture}
    \draw (0, 0) node[inner sep=0] {\includegraphics[width=8cm]{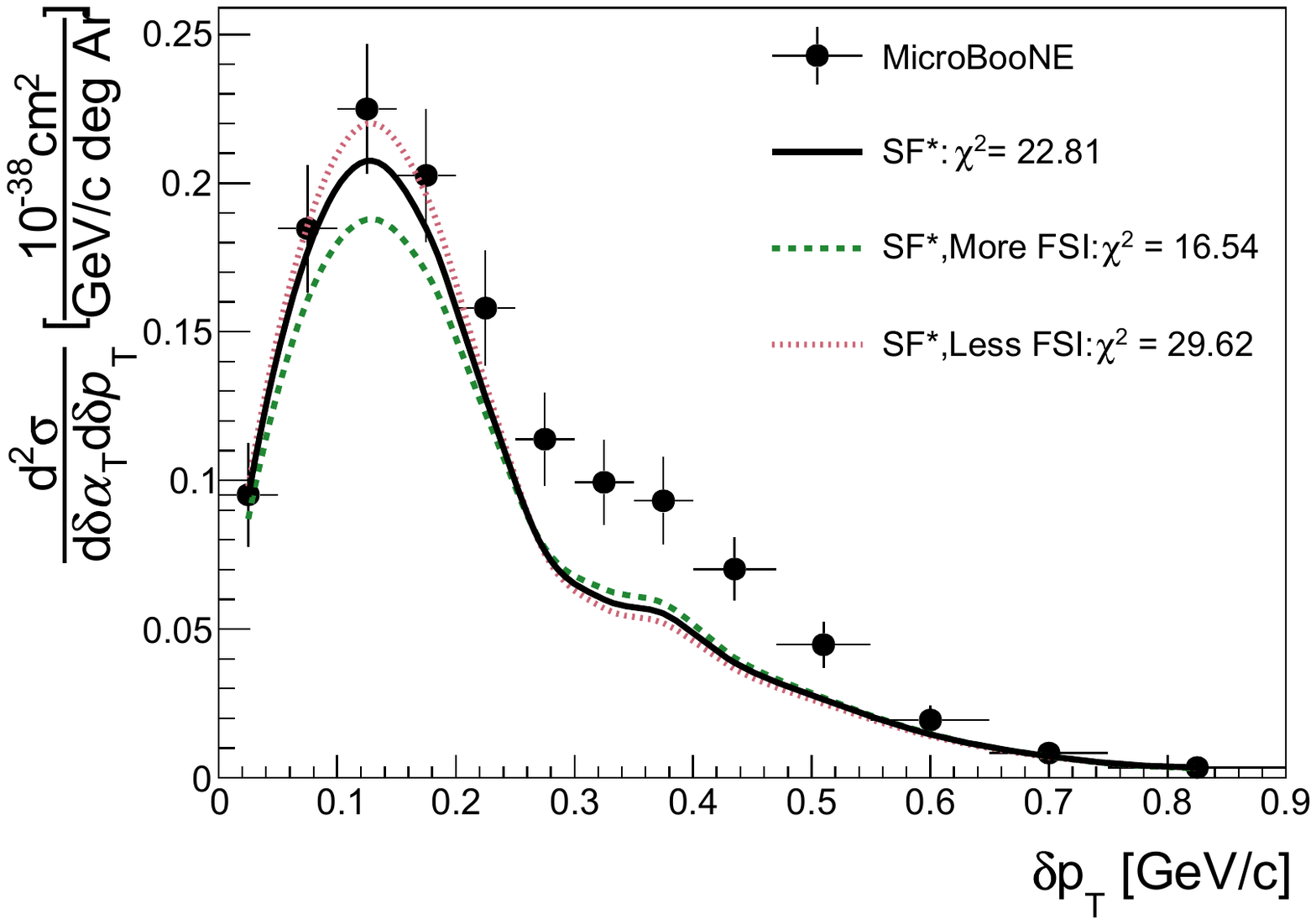}};
    \draw (1.8, 0) node {$90^\circ<\dalphat<135^\circ$};
\end{tikzpicture}
\begin{tikzpicture}
    \draw (0, 0) node[inner sep=0] {\includegraphics[width=8cm]{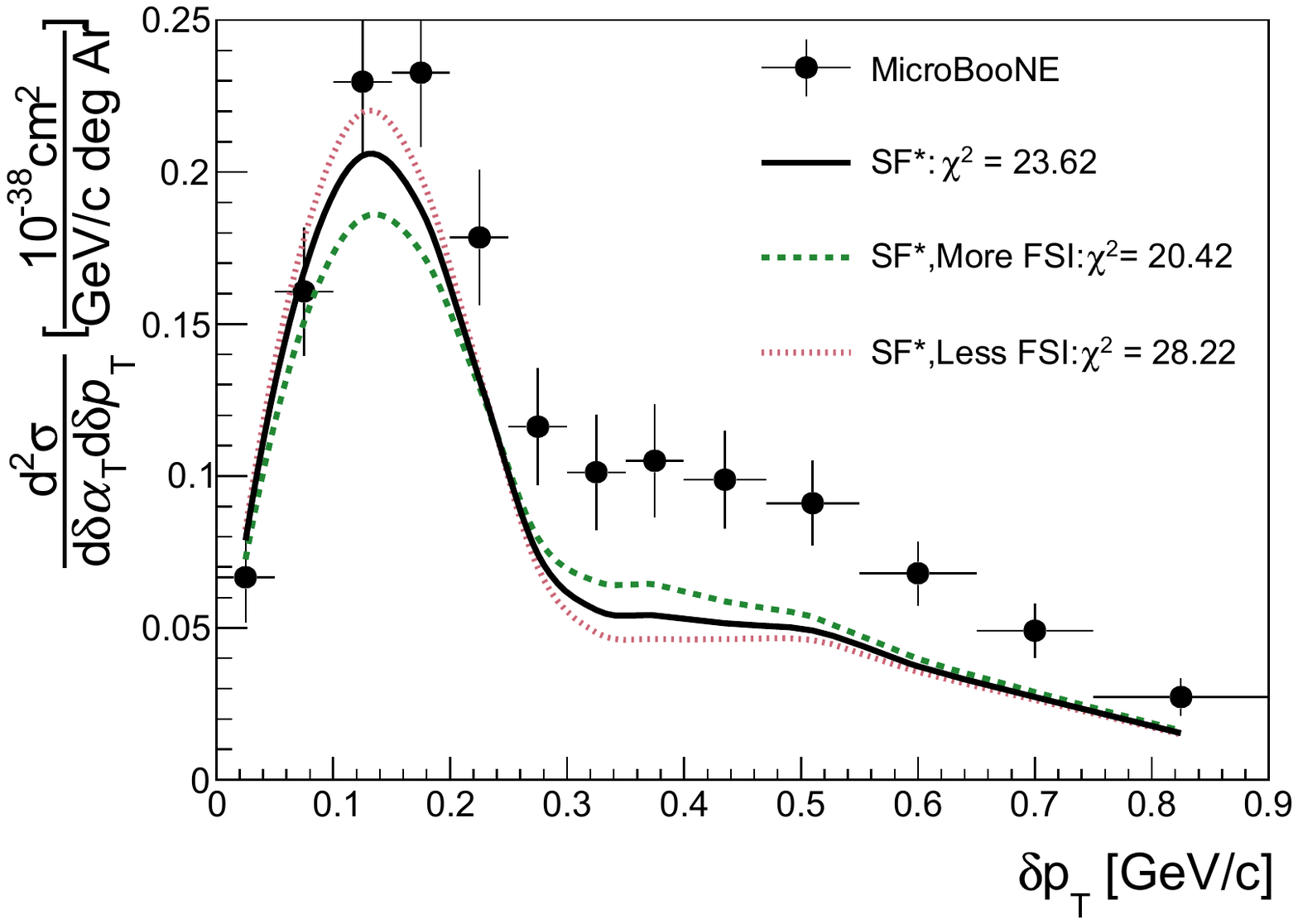}};
    \draw (1.8, 0) node {$135^\circ<\dalphat<180^\circ$};
\end{tikzpicture}
\caption{Multi-differential measurement as a function of $\dpt$ and $\dalphat$ from MicroBooNE compared with predictions using the combined SF* model. The effects of adjusting the nucleon mean free path by -(+)30\% are displayed and labeled as ``More(Less) FSI''.}
\label{fig:uboone_dptvsdatFSI}
\end{figure*}

\noindent We examine the impact of nucleon FSI by varying the intrinsic MFP inside the intra-nuclear cascades applied in both NEUT and NuWro. We begin by assessing the impact of altering FSI on the $\dalphat$ measurements from T2K and MicroBooNE, shown in \autoref{fig:t2k_uBoone_FSI_dalphat} alongside measurements of $\dpt$ and the multi-differential measurement in \autoref{fig:t2k_uBoone_FSI} and \autoref{fig:uboone_dptvsdatFSI} respectively. Note that the MFP is changed inside the generators while keeping the baseline models (NEUT SF for carbon and the SF* model for argon) fixed. As a result, the baseline simulations are intrinsically different, with SF* using NuWro for CCQE interactions. We explore this difference in more detail later in the section. 

It can be immediately seen from \autoref{tab:chi2_pValues}, \autoref{fig:t2k_uBoone_FSI_dalphat}, \autoref{fig:t2k_uBoone_FSI} and \autoref{fig:uboone_dptvsdatFSI} that the MicroBooNE measurement displays a considerable preference for a larger MFP (more FSI), conversely to the T2K measurement. \autoref{fig:t2k_uBoone_FSI_dalphat} additionally highlights that the alteration of the FSI strength changes the total predicted cross section within the allowed phase space of the defined signal, with a more pronounced change at lower values of $\dalphat$. 

\autoref{fig:t2k_uBoone_FSI} demonstrates that modifying FSI primarily impacts the bulk of $\dpt$ for both T2K and MicroBooNE. The cross section in the tail stays broadly constant in an absolute sense but, as expected, the relative contribution of the tail compared to the bulk increases with more FSI. As discussed in \autoref{sec:TKI}, the relatively fixed cross section in the tail from changing FSI strength comes from the combination of the two effects of FSI shifting proton momentum to values of higher $\dpt$ and FSI migrating protons under detection threshold (changing the normalisation of the cross section within the experimental signal definitions). In the MicroBooNE case, these two effects are further compounded by the smearing between the bulk and tail.

Comparing FSI variations to MicroBooNE's multi-differential measurement of $\dpt$ as a function of $\dalphat$, presented in \autoref{fig:uboone_dptvsdatFSI}, is a particularly useful tool to isolate the impact of nucleon FSI, as the separate evolution of FSI in $\dalphat$ and $\dpt$ allows some disambiguation from other nuclear effects, as was initially proposed in Ref.~\cite{T2K:2019bbb}. In all bins of $\dalphat$, the measurements seem to suggest that an enhancement of FSI strength is preferred. At low and lower-intermediate values of $\dalphat$ (regions with lower impact of FSI), the most visible effect of the proton FSI alterations is on the bulk of the distributions, from the aforementioned movement of events inside (outside) of the signal definition with decreasing (increasing) FSI strength. In this region, the evolution of the $p$-values with FSI strength disfavor a weakening of the FSI strength. At higher intermediate and high values of $\dalphat$, the impact of FSI becomes more visible on the tails of the $\dpt$ distributions. As the value of $\dalphat$ increases, we note that the cross section in the tail rises in both the simulations and the measurements. However, this increase is more drastic in the measurement compared to the simulation -- indeed, the rate at which the relative tail contribution rises in the simulation is insufficient to reproduce the rate at which it increases in the measurement. The $p$-values in \autoref{tab:chi2_pValues} indicate a consistent preference for increasing the strength of nucleon FSI on argon. 

It is interesting to note that, while all $\dalphat$ bins show the same overall preference for an enhancement of FSI, the balance between the two effects previously discussed (reduction of the bulk due to protons being out of phase space, on one hand, and enhancement of the tail, on the other hand) is different between the different $\dalphat$ regions. The multi-differential MicroBooNE measurement of $\dpt$ in bins of $\dalphat$ offers a particularly powerful combination of kinematic imbalance variables which allows us to lift the degeneracies between nuclear effects, and similar measurements in the future from other experiments will prove to be invaluable to further disentangle these effects.

\begin{center}
\textit{Impact of FSI model on predicted kinematics}
\end{center}

\begin{figure*}[htpb]
\centering

\begin{tikzpicture}
    \draw (0, 0) node[inner sep=0] {\includegraphics[width=8cm]{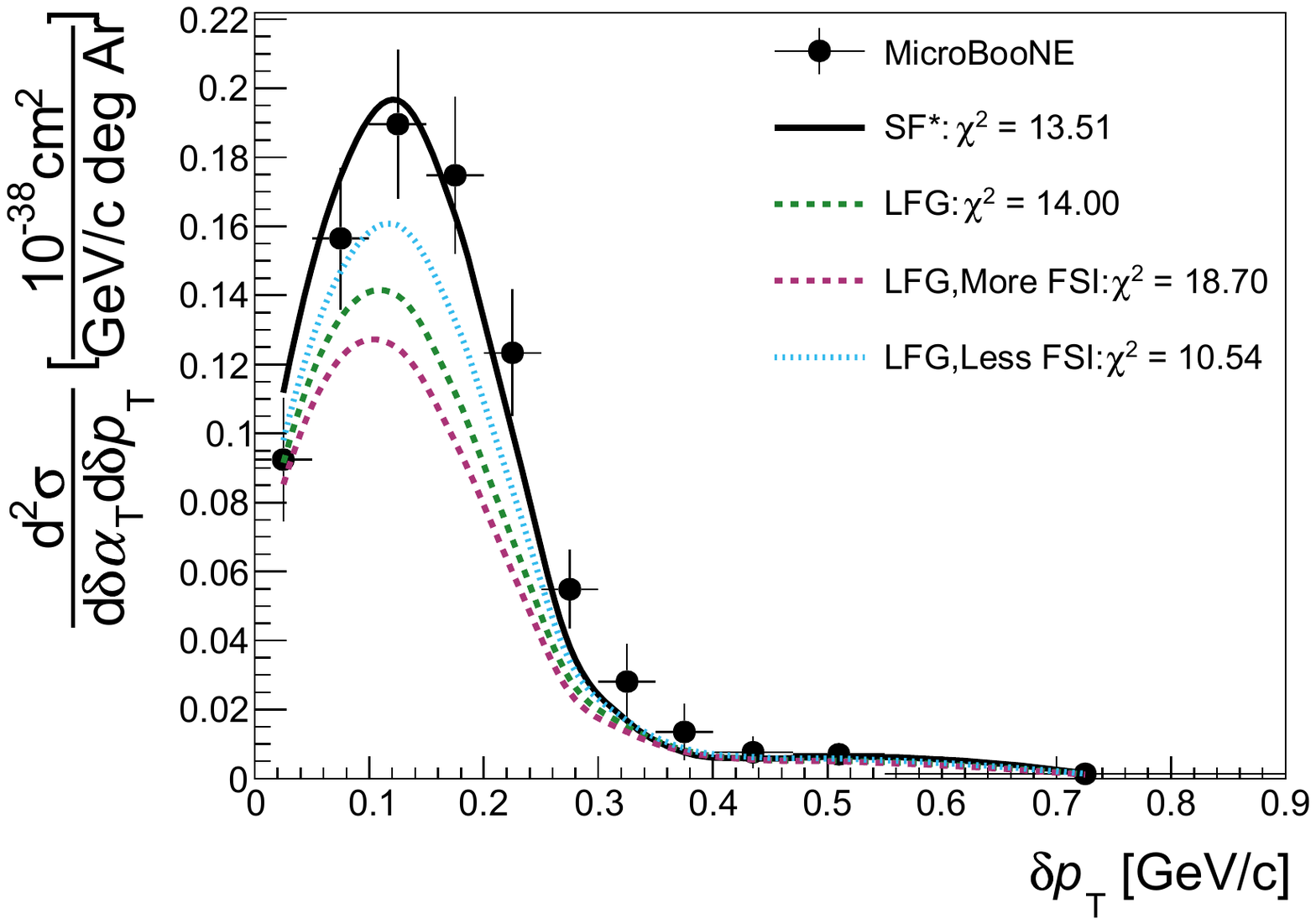}};
    \draw (2, 0) node {$0^\circ<\dalphat<45^\circ$};
\end{tikzpicture}
\begin{tikzpicture}
    \draw (0, 0) node[inner sep=0] {\includegraphics[width=8cm]{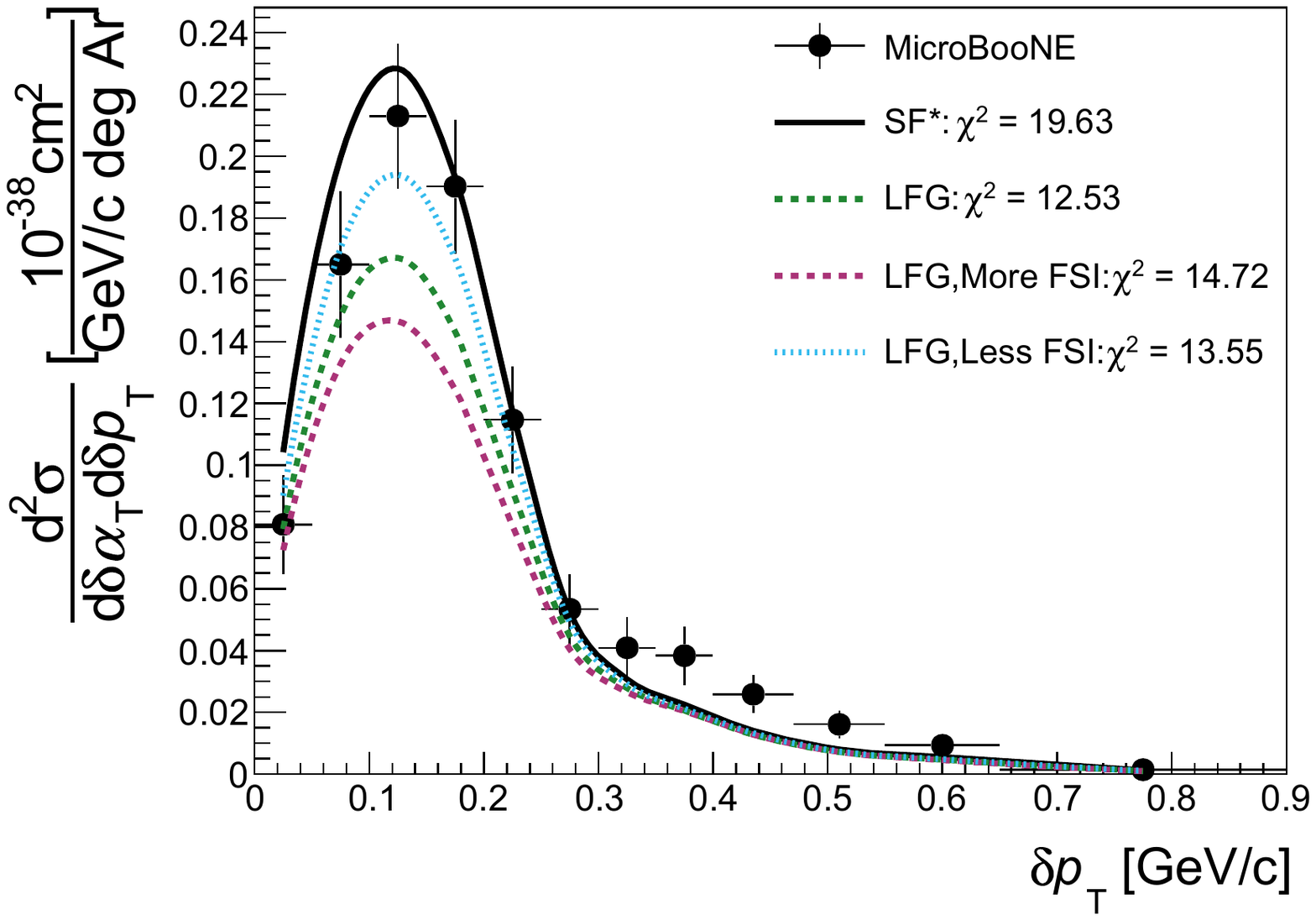}};
    \draw (2, 0) node {$45^\circ<\dalphat<90^\circ$};
\end{tikzpicture}
\begin{tikzpicture}
    \draw (0, 0) node[inner sep=0] {\includegraphics[width=8cm]{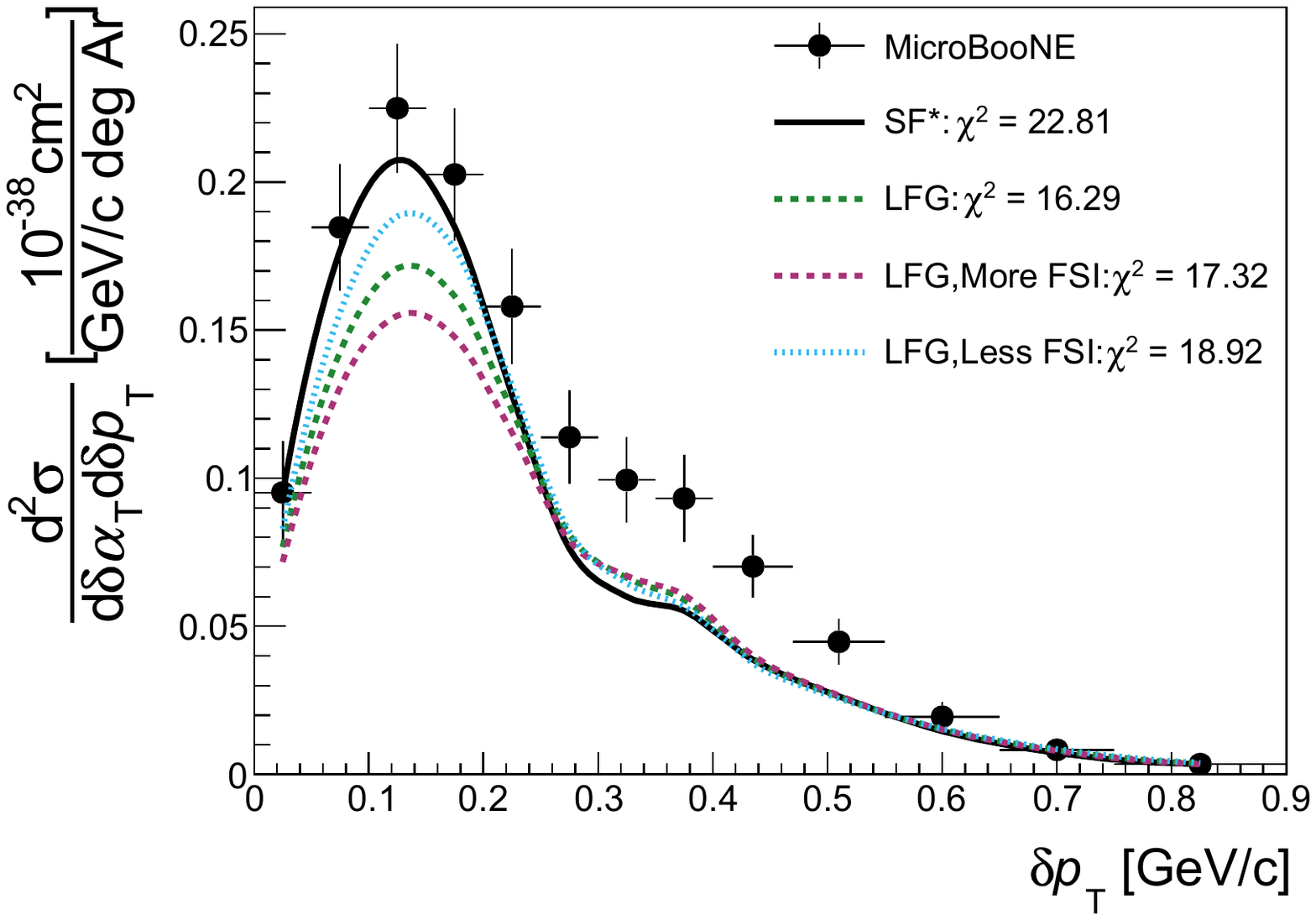}};
    \draw (2, 0) node {$90^\circ<\dalphat<135^\circ$};
\end{tikzpicture}
\begin{tikzpicture}
    \draw (0, 0) node[inner sep=0] {\includegraphics[width=8cm]{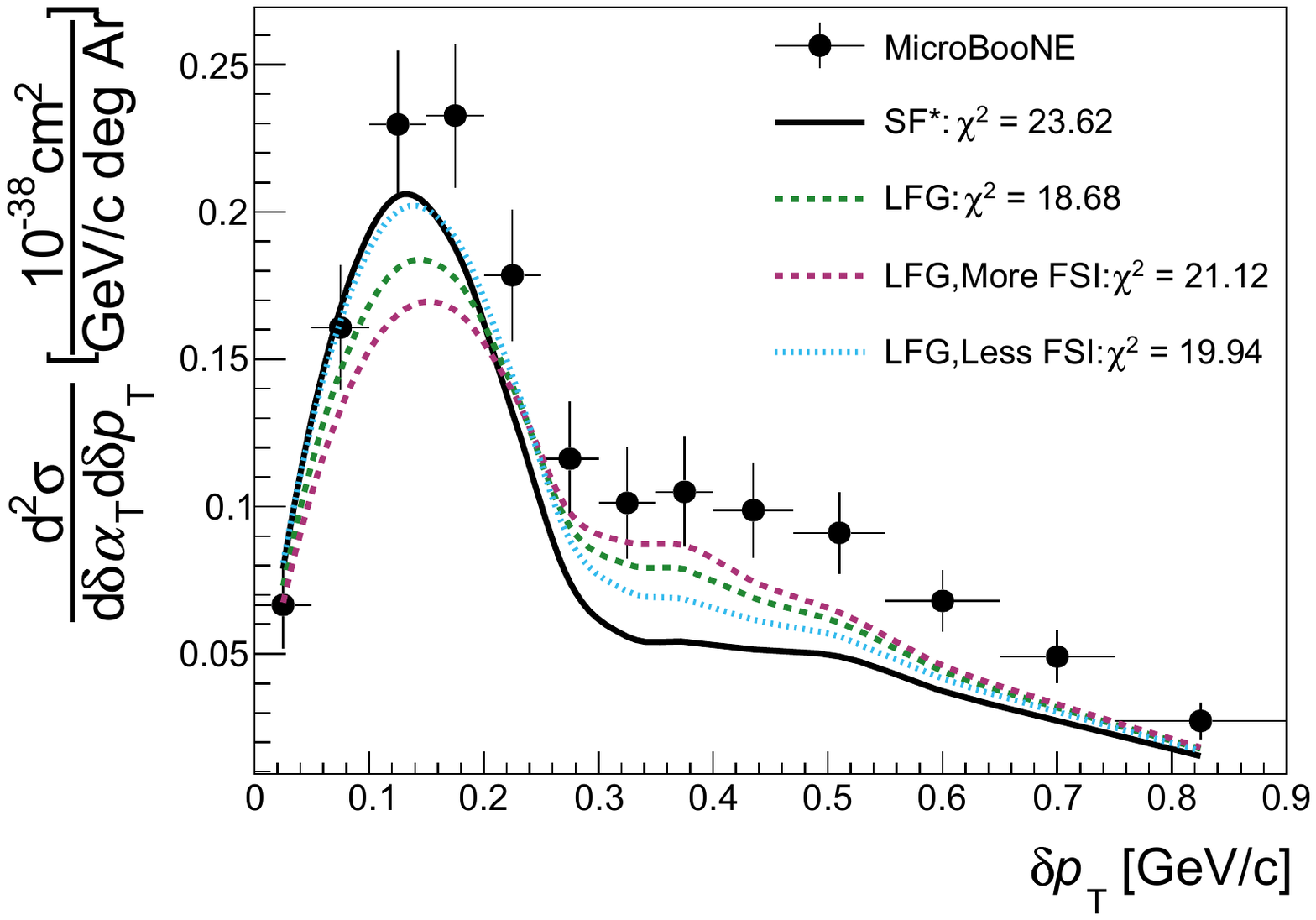}};
    \draw (1.8, 0) node {$135^\circ<\dalphat<180^\circ$};
\end{tikzpicture}

\caption{Multi-differential measurement as a function of $\dpt$ and $\dalphat$ from MicroBooNE compared with predictions using the NEUT LFG model. The effects of adjusting the nucleon mean free path by -(+)30\% are displayed and labeled as ``More(Less) FSI''.}
\label{fig:uboone_dpt_lfgfsi}
\end{figure*}

\noindent As discussed in the previous sections, it is often not possible to draw a direct comparison between the NEUT SF and SF* models due to the fact that the latter uses QE events generated with the NuWro event generator on argon and thus applies a different set of intra-nuclear cascade processes than those in NEUT. Furthermore, \autoref{fig:dalphat_uboone_nucmodels} highlights that the FSI predictions from NuWro are disfavored (i.e. $p$-value$<$0.05) by the MicroBooNE $\dalphat$ measurements and also alter the shape of the final state nucleon kinematics in a different way than those in NEUT. In order to lift the ambiguity caused by this inconsistency, we confront the MicroBooNE multi-differential measurement with a prediction from NEUT using the LFG model on argon in \autoref{fig:uboone_dpt_lfgfsi}, to which we apply the same 30\% variation in MFP at generation time. 

Through comparing \autoref{fig:uboone_dptvsdatFSI} and \autoref{fig:uboone_dpt_lfgfsi} it is clear that the use of NEUT's FSI model model helps to recover some of the missing strength in the $\dpt$ tail at large $\dalphat$, improving quantitative agreement with the measurement, but that strength remains missing at intermediate $\dalphat$. Similarly to the findings from \autoref{fig:dalphat_uboone_nucmodels}, the alteration of the shape of the $\dpt$ distribution is the main driver for improved agreement. In contrast to the FSI variations on the SF* model, which generally showed preference for stronger FSI, variations of FSI on the LFG model prefer to leave FSI unchanged (although variations in both directions are allowed). 

\begin{figure}[htpb]
\centering
\includegraphics[width=10cm]{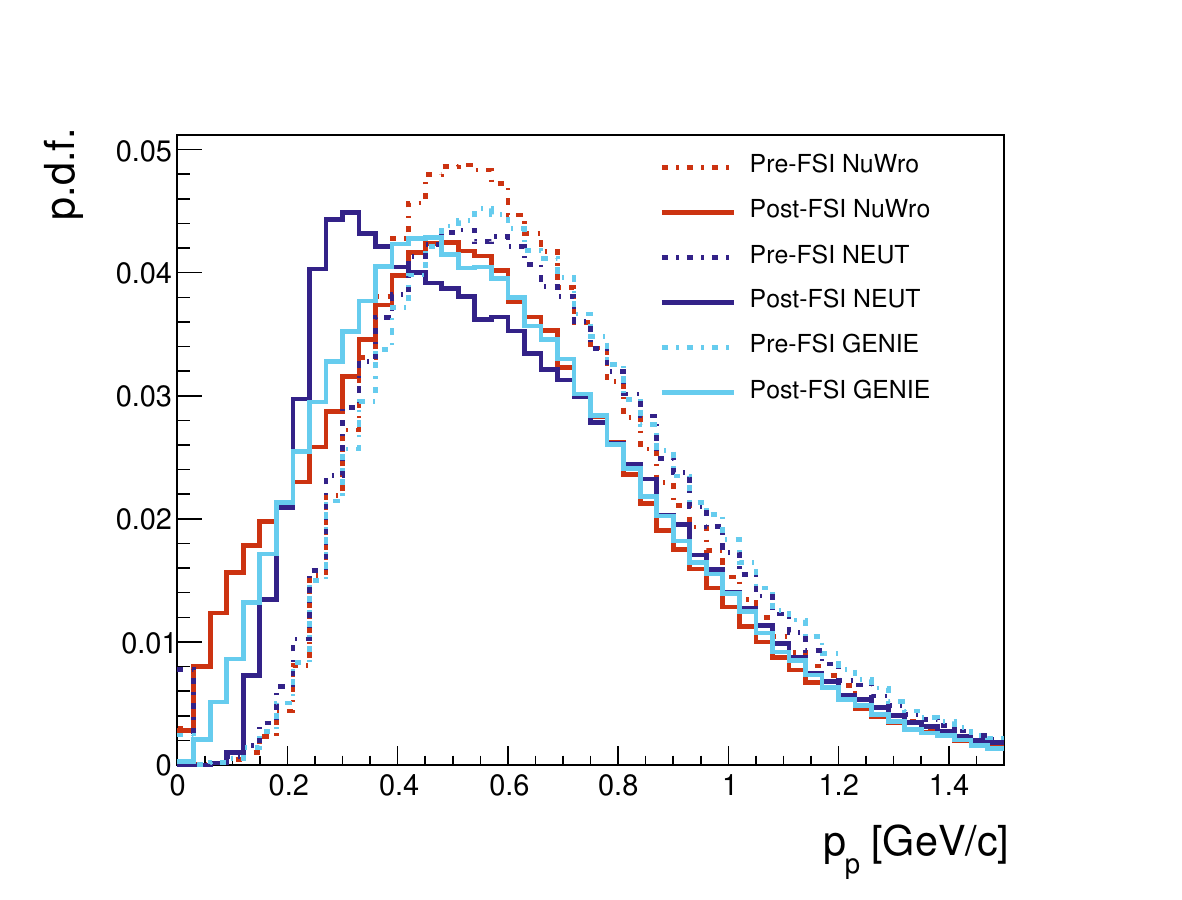}
\caption{Distribution of predicted MicroBooNE leading proton momentum for simulations of QE events with (solid lines) and without (dashed lines) FSI from different generators. The two scenarios are labelled as ``post-FSI'' and ``pre-FSI'' respectively.}
\label{fig:uboone_proton_momdist}
\end{figure}

A comparison of \autoref{fig:uboone_dptvsdatFSI} and \autoref{fig:uboone_dpt_lfgfsi} additionally serves to highlight the differences between the NEUT and NuWro FSI models -- a variation of $\pm$30\% of the MFP in NEUT is not sufficient to cover the nominal prediction from NuWro, despite the fact that the transparencies encoded in both NEUT and NuWro are within 30\% of one another~\cite{Dytman:2021ohr}. The difference between the NuWro and NEUT simulations is therefore still related to FSI, but goes beyond the impact of nuclear transparency. The modeling of FSI processes introduces alterations to the predicted particle kinematics beyond what variations of the MFP can cover. This is demonstrated in \autoref{fig:uboone_proton_momdist}, which shows the impact of different FSI models on the outgoing proton kinematics from QE interactions generated on an argon target using the MicroBooNE flux. Whilst in all cases the effect of FSI is to shift the leading momentum proton distribution to lower values, the size of the shift differs substantially between models.

The impact of the FSI model on the proportion of QE interactions that fall into the MicroBooNE kinematic constraints on the outgoing proton momentum (300-1000 MeV/$c$, see \autoref{tab:sigDef}) is quantified in \autoref{tab:pre_post_fsi}. The number of interactions which have migrated outside of the signal region as a fraction of the events inside the signal region before FSI is given by the quantity $\delta_{FSI}$:

\begin{equation}
    \delta_{FSI} = (N^{\text{signal}}_{\text{pre-FSI}} - N^{\text{signal}}_{\text{post-FSI}} )/ N^{\text{signal}}_{\text{pre-FSI}},
\end{equation}
where $N^{\text{signal}}_{\text{post-FSI}}$ and $N^{\text{signal}}_{\text{pre-FSI}}$ are the number of events contained within proton momentum range of the MicroBooNE signal definition, with and without applying FSI respectively, and $N^{\text{signal}}_{\text{pre-FSI}}$ is the total number of events in this signal region before FSI. We additionally define the quantities $\rho^{\text{signal}}_{\text{post-FSI}}$ and $\rho^{\text{signal}}_{\text{pre-FSI}}$ as the fraction of events inside the signal region for simulations with and without FSI respectively as follows:

\begin{align}
    \rho^{\text{signal}}_{\text{post-FSI}} &= N^{\text{signal}}_{\text{post-FSI}}/N^{\text{total}},\\
    \rho^{\text{signal}}_{\text{pre-FSI}} &= N^{\text{signal}}_{\text{pre-FSI}}/N^{\text{total}},
\end{align}

where $N^{\text{total}}$ is the total numbers of simulated CCQE events.

\begin{table}[htpb]
\centering
\begin{tabular}{c|c|c|c}

\textbf{Model} & \textbf{$\rho^{\text{signal}}_{\text{pre-FSI}}$} & \textbf{$\rho^{\text{signal}}_{\text{post-FSI}}$} & \textbf{$\delta_{FSI}$} \\
\hline
NEUT (LFG) & 81.3\% & 74.3\% & 8.5\% \\
\hline
GENIE (\argenie) & 81.2\% & 75.2\% & 7.3\% \\
\hline
NuWro (SF*) & 83.8\% & 73.6\% & 12.1\% \\

\end{tabular}
\caption{The fraction of CCQE events inside the signal region for simulations with and without FSI using the MicroBooNE flux on an Argon target, alongside the number of events which migrated outside of the MicroBooNE signal region as a fraction of the total number of pre-FSI events in the signal region. The quantities are defined in detail in the text.}
\label{tab:pre_post_fsi}
\end{table}

From \autoref{tab:pre_post_fsi}, we can see that the vast majority of QE events fall within the signal region, but, as expected, the effect of FSI is to cause a decrease in this fraction. Crucially, as showcased in \autoref{fig:uboone_proton_momdist}, this migration does not happen in the same way for all generators, and it is apparent that the migration in the case of the SF* model is the largest. This further supports what was highlighted in \autoref{fig:uboone_dpt_lfgfsi}, i.e. that there are effects beyond those covered by variations of the MFP which drive the discrepancy between the two generators in the case of the MicroBooNE measurement. 

\subsection{Impact of neutrino energy dependence: comparisons to MINERvA measurements}
\label{sec:impact_energydep_minerva}

As discussed in \autoref{sec:impact_ascaling} and shown in \autoref{tab:chi2_pValues}, the $\chi^2$ values obtained by comparing to the MINERvA measurements are all much higher than the number of analysis bins, yielding $p$-values which exclude all considered models. Nonetheless, it is still valuable to compare the simulations to the MINERvA measurements in order to extract qualitative trends which might indicate potential issues in the modeling of nuclear effects. The MINERvA measurements considered in this work are all performed on a plastic scintillator target, like those of T2K. However, as shown in \autoref{fig:fluxComp}, the MINERvA flux is at significantly higher neutrino energy than that of T2K, so a comparison between the T2K and the MINERvA experiments' measurements allows a study of potential mismodeling of nuclear effects which vary with neutrino energy.

\begin{figure}[htpb]
\centering
\includegraphics[width=10cm]{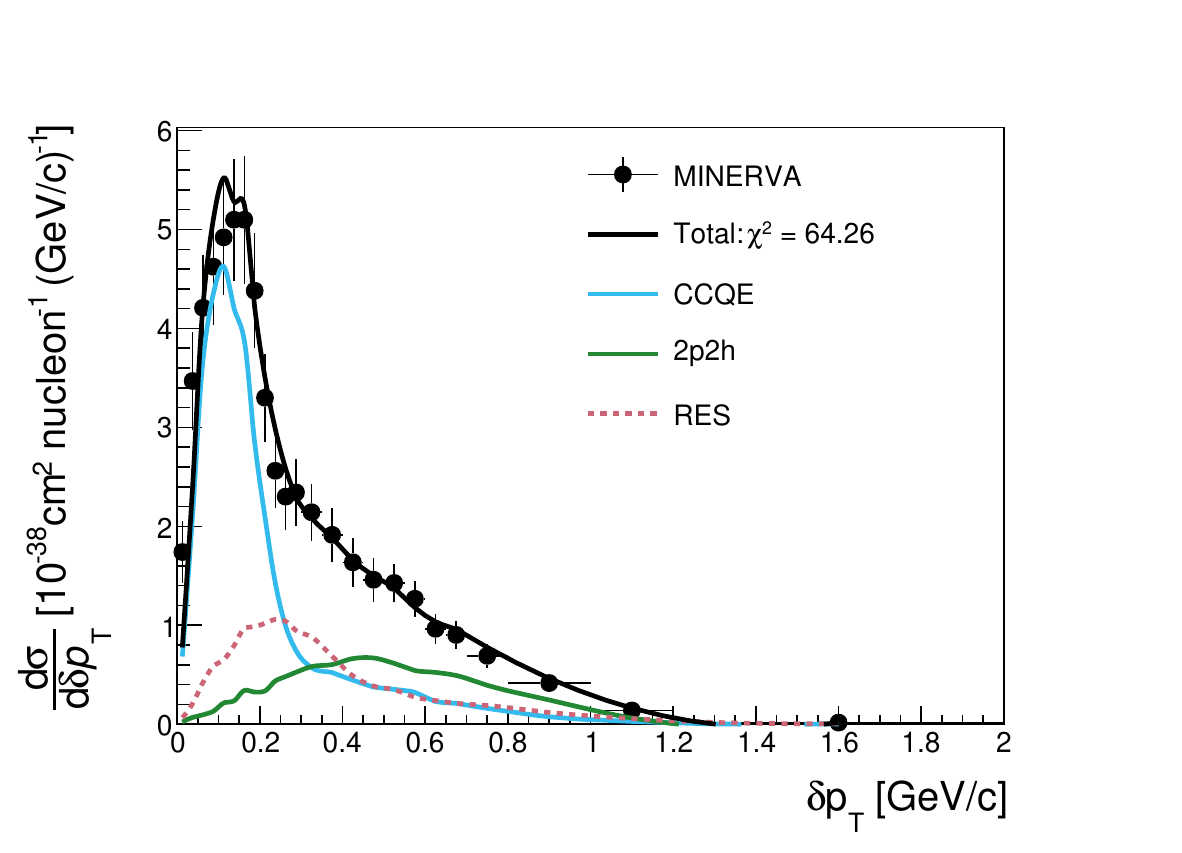}\caption{Differential cross section measurement as a function of $\dpt$ from MINERvA, compared to the NEUT SF model predictions from NEUT on a carbon target. The different channel contributions to the total cross sections are highlighted in the legend.}
\label{fig:minerva_bymode}
\end{figure}

We begin by considering the breakdown by interaction channel. \autoref{fig:minerva_bymode} shows the contributions of the different interaction channels to the total cross section as a function of $\dpt$. It is clear that the MINERvA prediction has a significantly enhanced tail compared to the T2K prediction shown in \autoref{fig:byModeT2KuBooNE}. Unlike T2K, MINERvA sees a significant proportion of RES events which have undergone pion absorption, as well as 2p2h interactions. The RES contribution is also significantly more shifted under the bulk, indicating that a lower proportion of the momentum of the hadronic system is in the absorbed pions, or that the accompanying protons are less likely to under go FSI. 

\begin{figure*}[htpb]
\centering
\includegraphics[width=8cm]{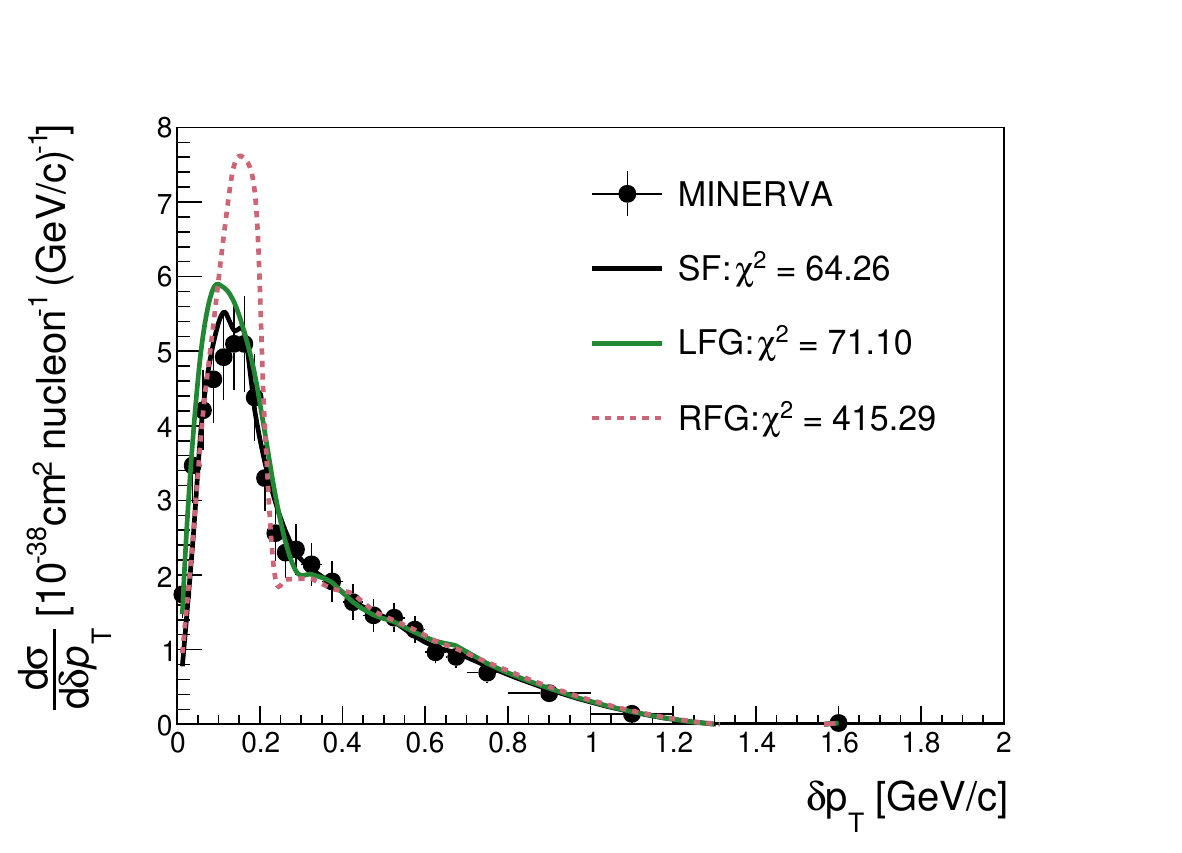}
\includegraphics[width=8.5cm]{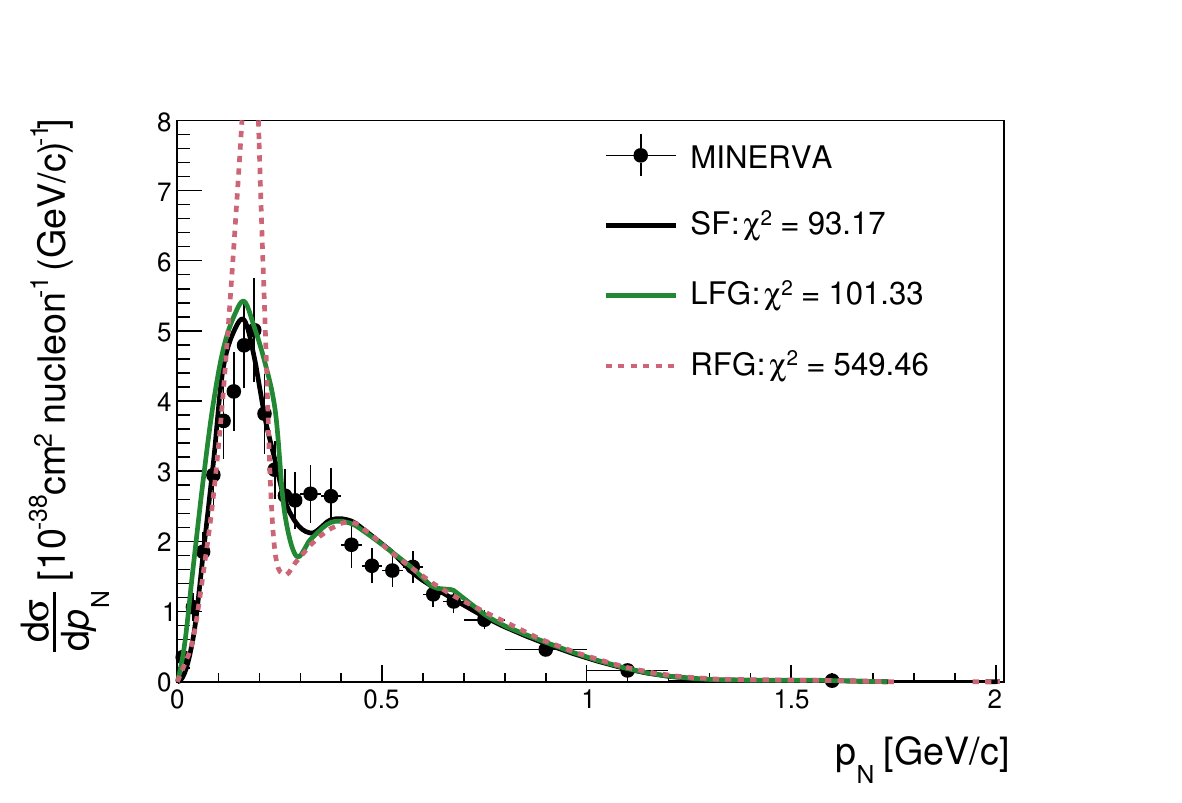}
\caption{Differential cross section measurement as a function of $\dpt$ (left) and $p_N$ (right) from MINERvA, compared to different ground state predictions using different nuclear models in NEUT: SF, LFG, and RFG.}
\label{fig:minerva_pn_dpt}
\end{figure*}

NEUT predictions for different nuclear ground state models are compared to the MINERvA $\dpt$ and $p_N$ measurements in \autoref{fig:minerva_pn_dpt}. Qualitatively, the SF model has the best agreement with both measurements, in particular due to the better normalisation of the $\dpt$ bulk and of the $p_N$ transition region between bulk and tail. For all models, the tails of the $\dpt$ distributions look similar, as most of the contribution in this region is given by non-QE events, whose modeling doesn't change. The largest apparent changes are in the bulk, where all models overestimate (RFG most significantly) the predictions but with different shapes. 

\begin{figure}[htpb]
\centering
\includegraphics[width=10cm]{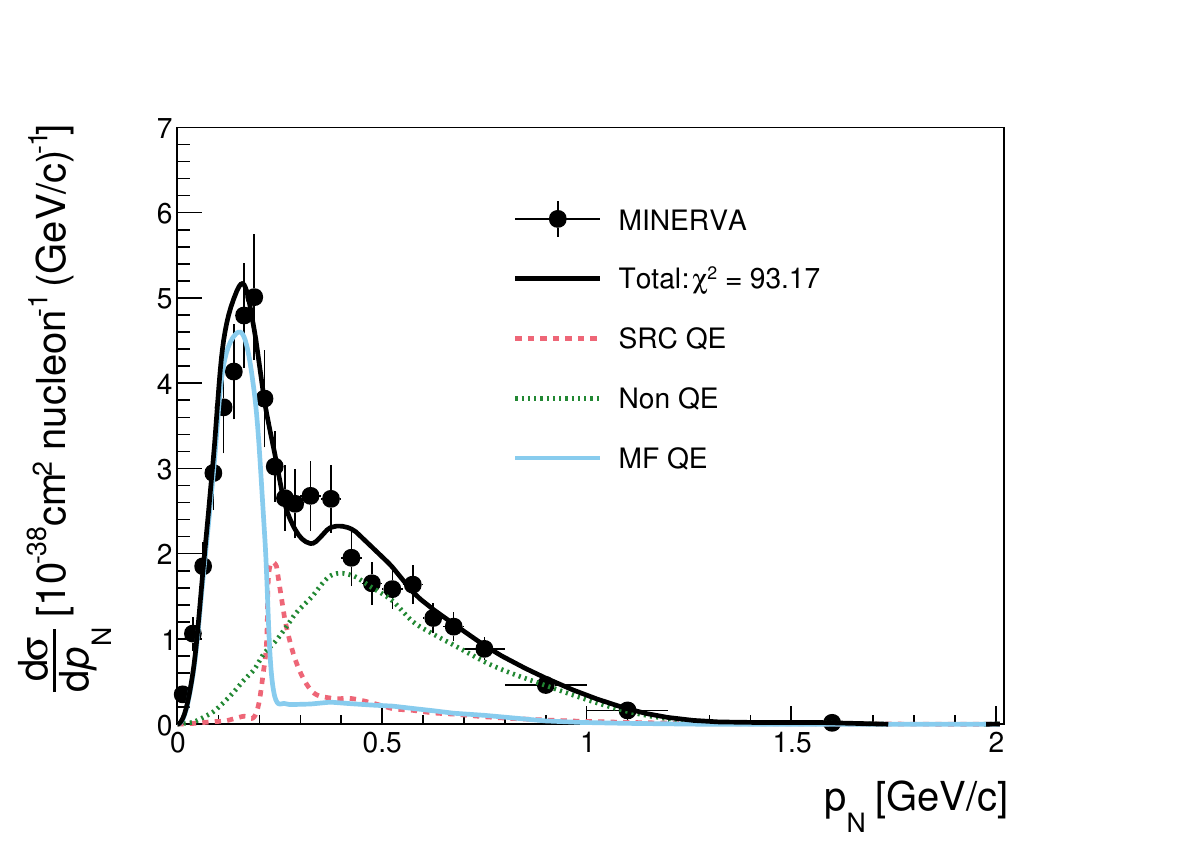}
\caption{Differential cross section measurement as a function of $p_N$ from MINERvA, compared to the NEUT SF prediction. The latter is divided into three components: a QE mean-field contribution (``MF QE''), a QE contribution from short range correlated pairs (``SRC QE'') and all other non-QE components (``Non QE'').}
\label{fig:minerva_srcs}
\end{figure}

The CCQE predictions from NEUT's LFG and RFG models contain a single outgoing proton before FSI but, as noted in \autoref{sec:models}, the SF model produces two proton states due to SRCs. The $\dpt$ measurements from T2K and MicroBooNE discussed so far offer little sensitivity to distinguish SRCs from the dominant mean field interactions, but it is informative to examine the measurement of the inferred initial state nucleon momentum by the MINERvA experiment in this context. \autoref{fig:minerva_srcs} shows the QE and non-QE contributions to the NEUT SF simulation, where the QE contribution is further subdivided into a mean-field  part and an SRC part, according to the categorization used in the NEUT implementation and described in \autoref{sec:models} (for further details on the way this categorization is done in NEUT, see~\cite{Chakrani:2023htw, Furmanski:2015knr}). This demonstrates that the reasonable description of the $p_N$ transition region between bulk and tail comes from the SRC nucleons in the SF model. Although the overall SF prediction does not agree quantitatively with the measurement, this comparison serves the purpose of highlighting the potential need for an adequate model of SRCs inside the nuclear medium to describe MINERvA's measurement.
%\lm{Stephen, check the paragraph above}
%\lm{note to stephen to add some citations}

At MINERvA energies, the 2p2h cross section predicted by the Valencia model is saturated, so their production is much more prevalent than at T2K or MicroBooNE energies. Similarly, the RES contribution is also larger due to MINERvA's higher energies but, as seen in \autoref{fig:minerva_bymode}, is also shifted further from the tail and under the bulk of the $\dpt$ distribution making the impact of varying 2p2h more apparent in the tail. The fact that the non-QE distributions have different shapes in $\dpt$ for the T2K and MINERvA measurements makes the analysis of the two useful to disambiguate nuclear effects. However, it should be noted that the details of different signatures of the non-QE contribution heavily relies on the modeling of poorly understood hadron kinematics.

\begin{figure}[htpb]
\centering
\includegraphics[width=10cm]{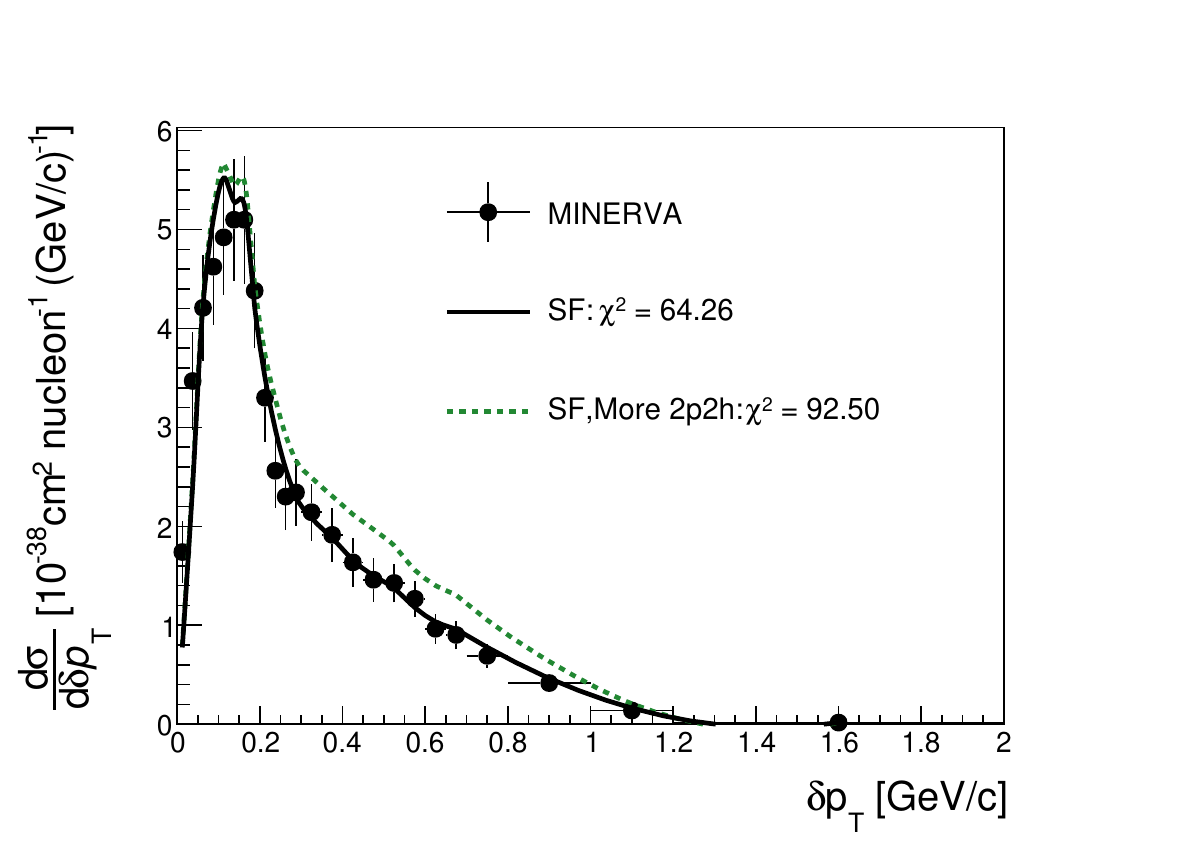} 
\caption{Differential cross section measurement as a function of $\dpt$ from MINERvA compared to the nominal NEUT SF prediction and an increase of the 2p2h cross section by 70\% uniformly across neutrino energies.}
\label{fig:minerva_dpt_2p2h}
\end{figure}

The impact of varying the strength of 2p2h interactions is shown in \autoref{fig:minerva_dpt_2p2h}. The MINERvA $\dpt$ measurement disfavors a large increase in the normalization of 2p2h interactions when other effects stay fixed. As can be seen in \autoref{tab:chi2_pValues}, this is consistent with the trend seen in the T2K measurement in \autoref{fig:t2k_uboone_2p2h}, and opposite to the MicroBooNE measurements' preferences shown in \autoref{fig:t2k_uboone_2p2h} and \autoref{fig:dptvsdat2p2h}.

\begin{figure}[htpb]
\centering
\includegraphics[width=10cm]{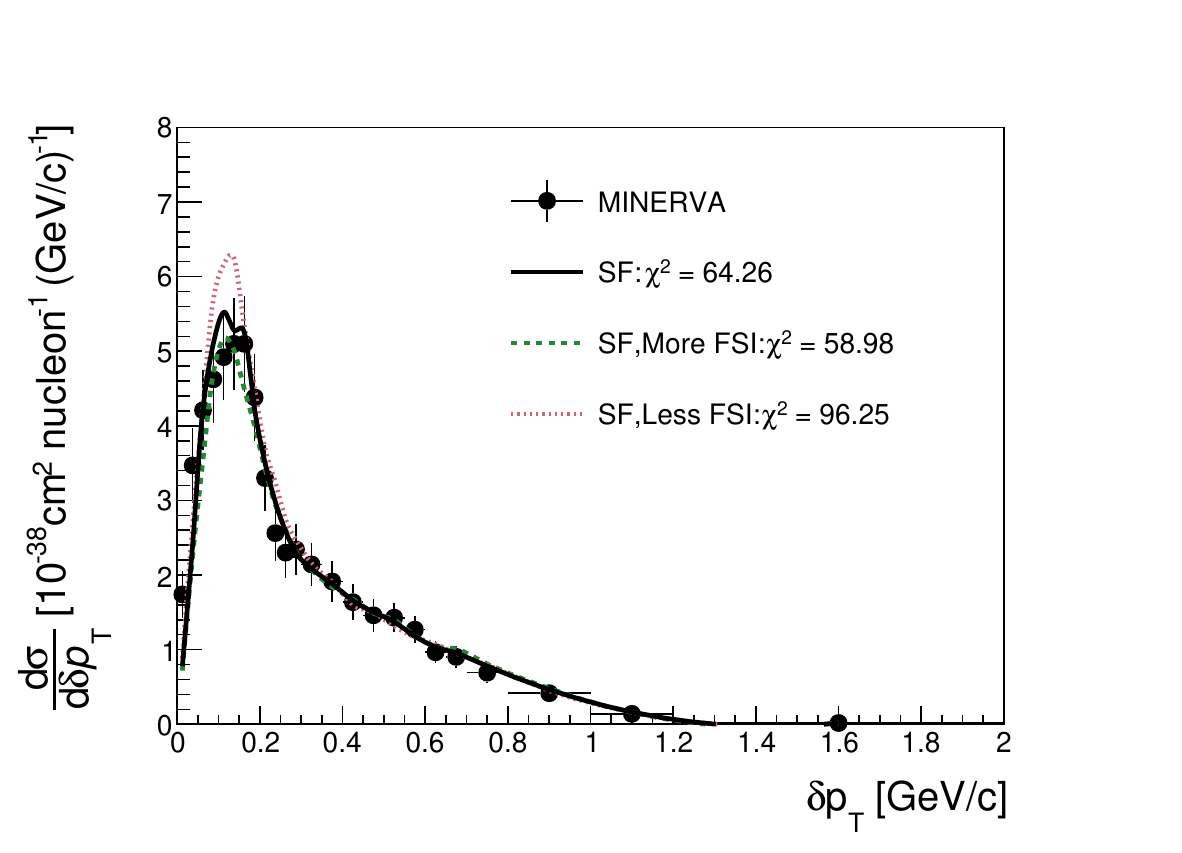}
\caption{Differential cross section measurement as a function of $\dpt$ from MINERvA compared to the NEUT SF prediction and variations of the nucleon mean free path. The effects of adjusting the nucleon mean free path by -(+)30\% are displayed and labeled as ``More(Less) FSI''.}
\label{fig:minerva_dpt_nucfsi}
\end{figure}

A similar observation can be made about the impact of nucleon FSI, depicted in \autoref{fig:minerva_dpt_nucfsi}. As in the case of the T2K measurement and opposite to the trend displayed by the MicroBooNE measurements, both shown in \autoref{sec:fsi}, the MINERvA measurement seems to disfavor an enhancement of the proton MFP. 

\begin{figure}[htpb]
\centering
\includegraphics[width=10cm]{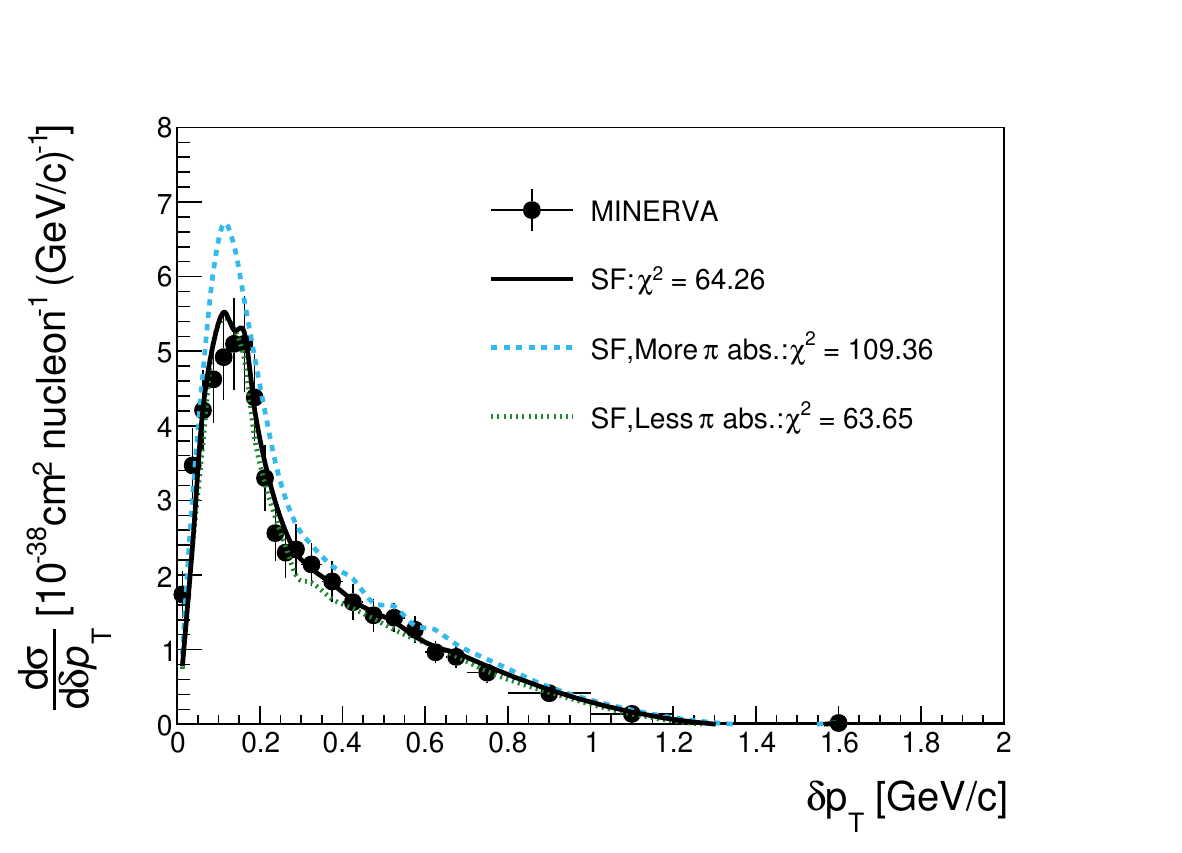}
\caption{Differential cross section measurement as a function of $\dpt$ from MINERvA compared to the NEUT SF prediction and variations of the pion absorption probability. The effects of adjusting the pion absorption probability as described in \autoref{sec:systvar} are displayed and labeled as ``More(Less) $\pi$ abs''.}
\label{fig:minerva_dpt_piabs}
\end{figure}

Finally, it is interesting to consider the impact of pion absorption processes in the context of the MINERvA measurement. As previously showcased in \autoref{fig:minerva_bymode}, MINERvA sees a significantly enhanced contribution from resonant interactions in which the pion has been absorbed, which is small for the lower-energy fluxes of T2K and MicroBooNE (as shown in \autoref{fig:byModeT2KuBooNE}). We compare the MINERvA measurement with the NEUT prediction in which we vary the cross section of pion absorption processes as described in \autoref{sec:systvar}, and show the results in \autoref{fig:minerva_dpt_piabs}. The same variation for the both T2K and MicroBooNE measurements shows only a small impact on the $\chi^2$ values, whereas in the case of MINERvA, an increase in the pion absorption probability is disfavored. 

\subsection{Global generator comparisons}
\label{sec:impact_globalGen}

Throughout this work, we have varied one nuclear effect at a time while keeping the baseline model (NEUT SF or SF* for argon) fixed. In this section, we compare predictions where we change all elements of the underlying simulations from the GENIE and NEUT generators. The NEUT event generator is currently used by the T2K collaboration for oscillation analyses (e.g.~\cite{T2K:2023smv}) and will likely be used by the future Hyper-K experiment, whereas DUNE plans to conduct its sensitivity studies with the GENIE generator's \argenie model~\cite{DUNEmodel_genworkshop}. The latter is also used by the SBN experiments.

\begin{figure}
\centering
\includegraphics[width=9cm]{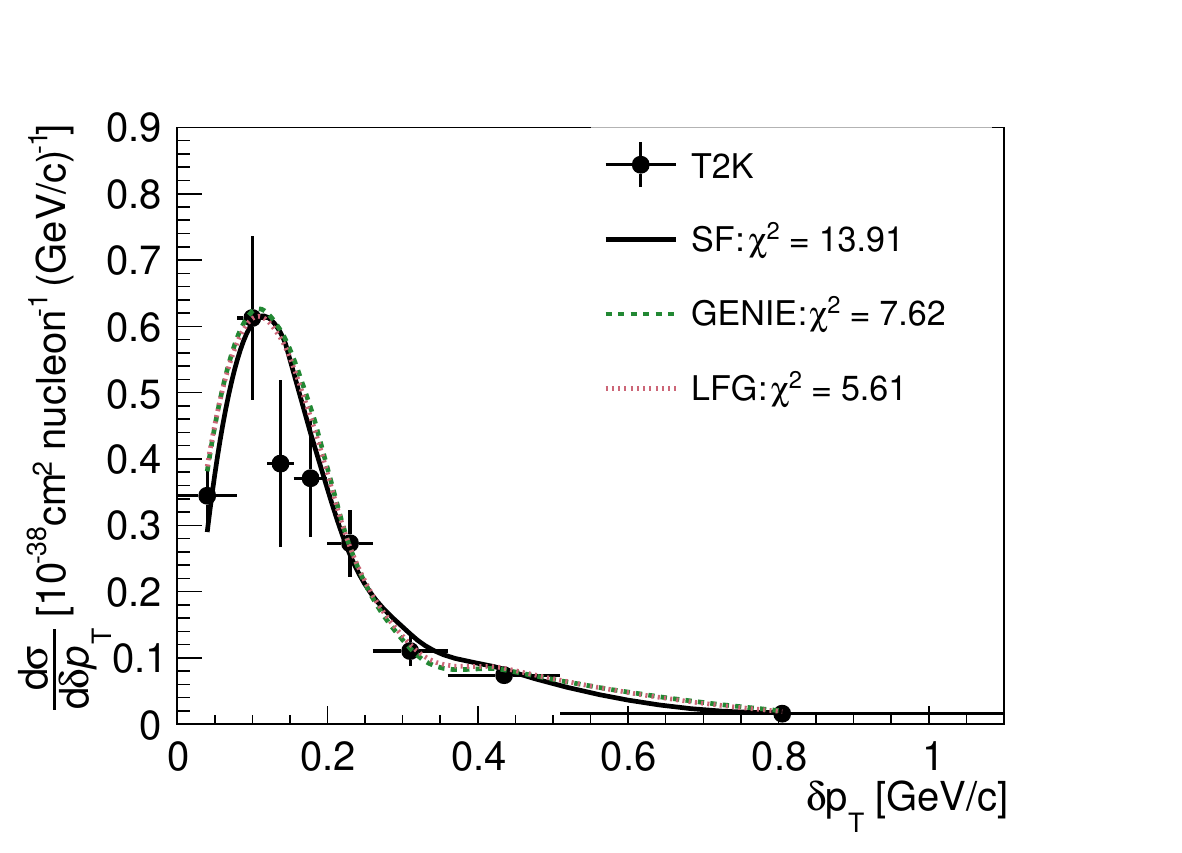}\\
\includegraphics[width=9cm]{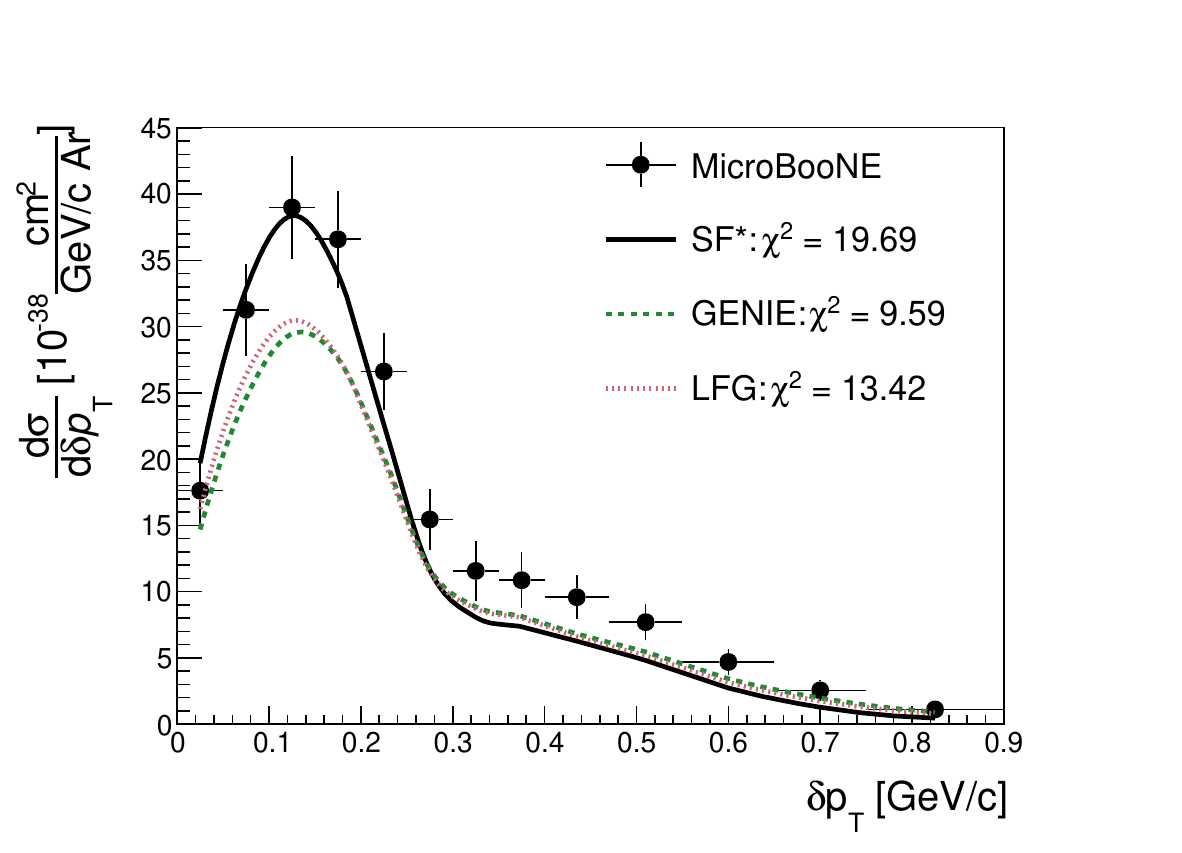}\\
\includegraphics[width=9cm]{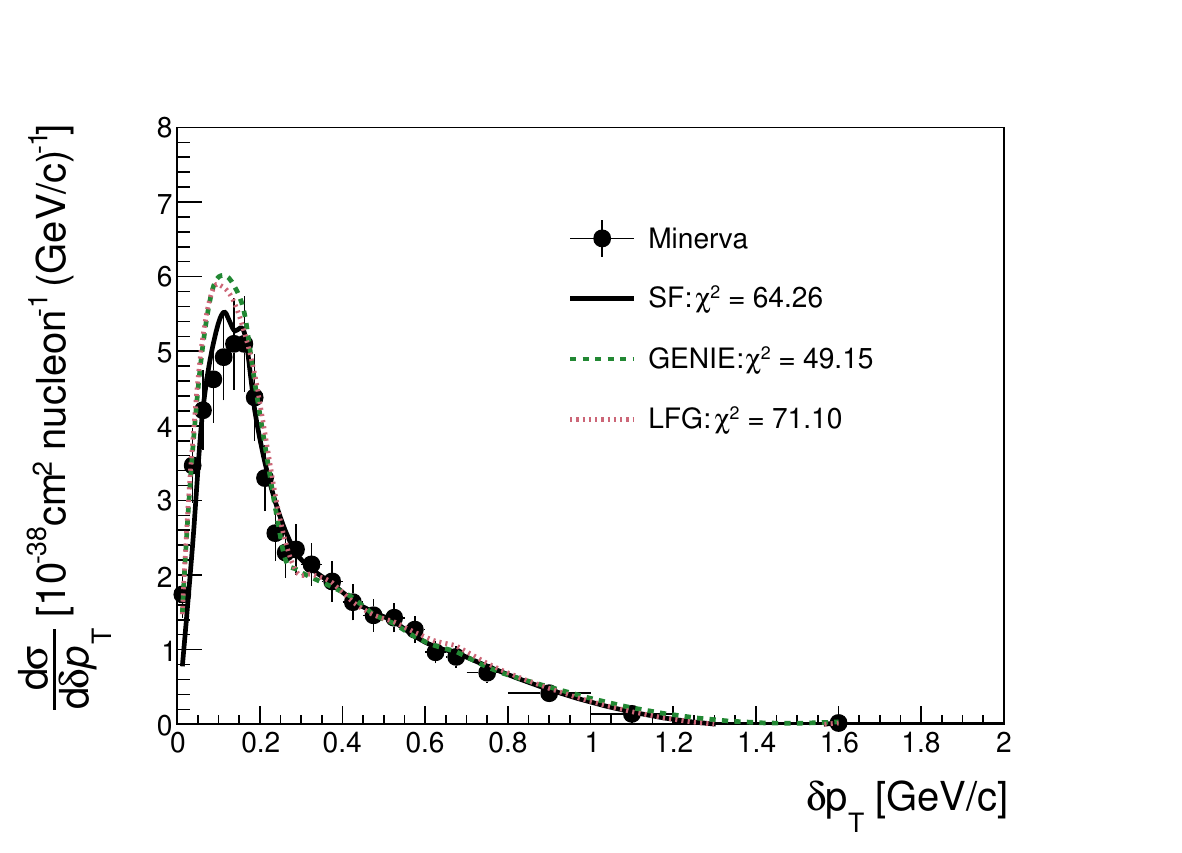}\\
\caption{Differential cross section measurement as a function of $\dpt$ from T2K (top), MicroBooNE (middle) and MINERvA (bottom), compared with the SF (SF* for MicroBooNE), NEUT LFG and the GENIE \argenie predictions.}
\label{fig:gencomp_all}
\end{figure}

\begin{figure*}[htpb]
\centering

\begin{tikzpicture}
    \draw (0, 0) node[inner sep=0] {\includegraphics[width=8cm]{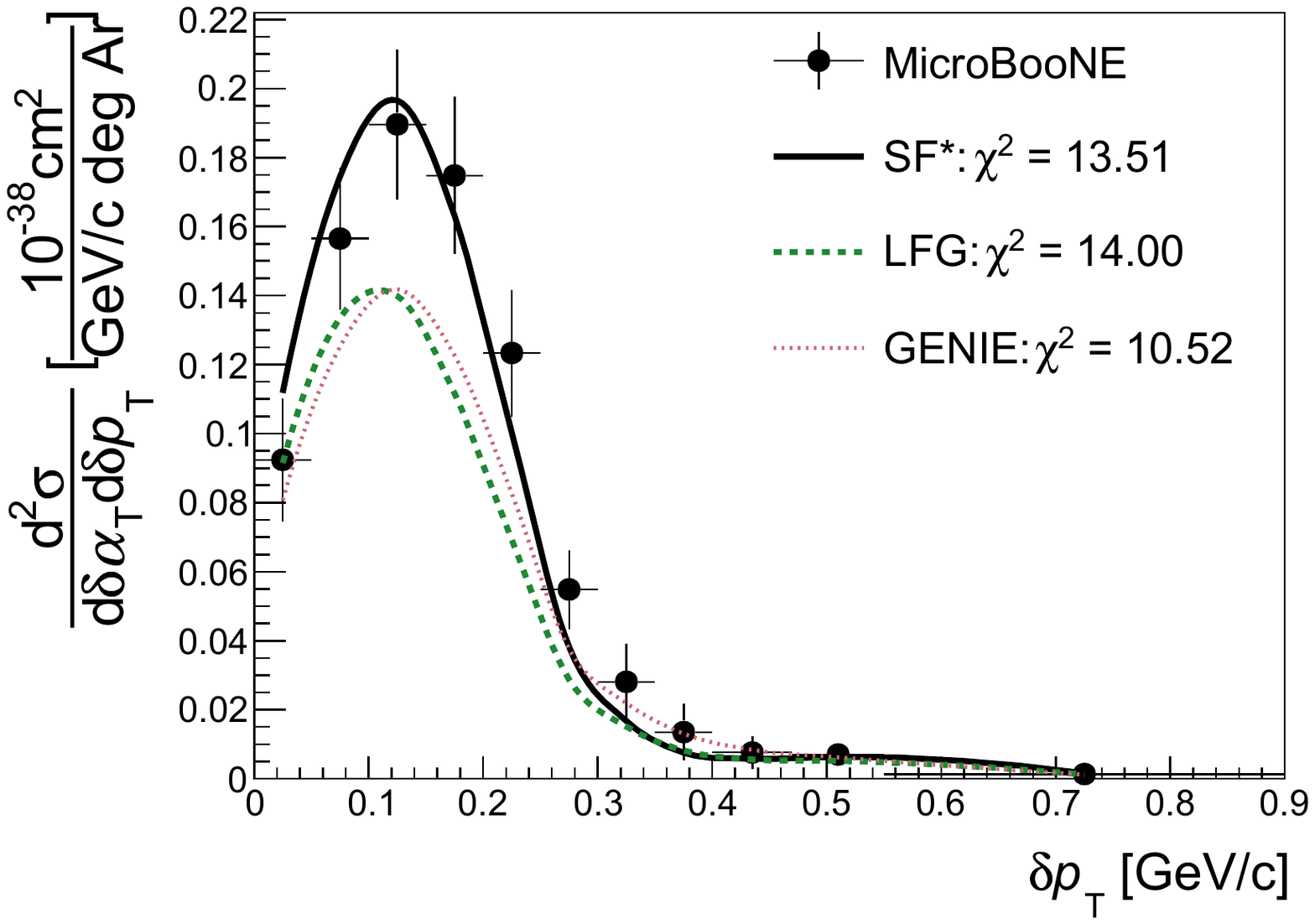}};
    \draw (2, 0) node {$0^\circ<\dalphat<45^\circ$};
\end{tikzpicture}
\begin{tikzpicture}
    \draw (0, 0) node[inner sep=0] {\includegraphics[width=8cm]{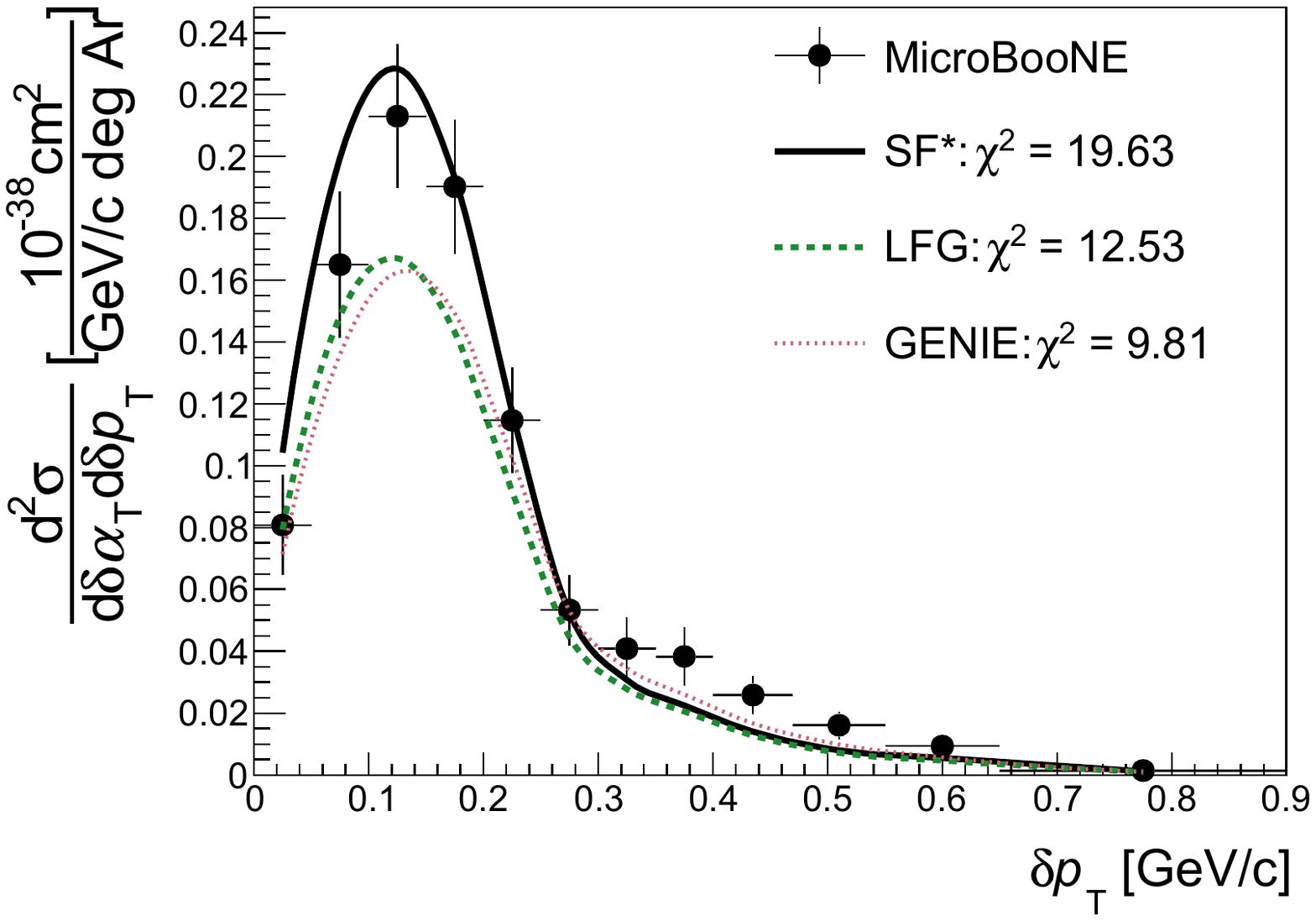}};
    \draw (1.9, 0) node {$45^\circ<\dalphat<90^\circ$};
\end{tikzpicture}
\begin{tikzpicture}
    \draw (0, 0) node[inner sep=0] {\includegraphics[width=8cm]{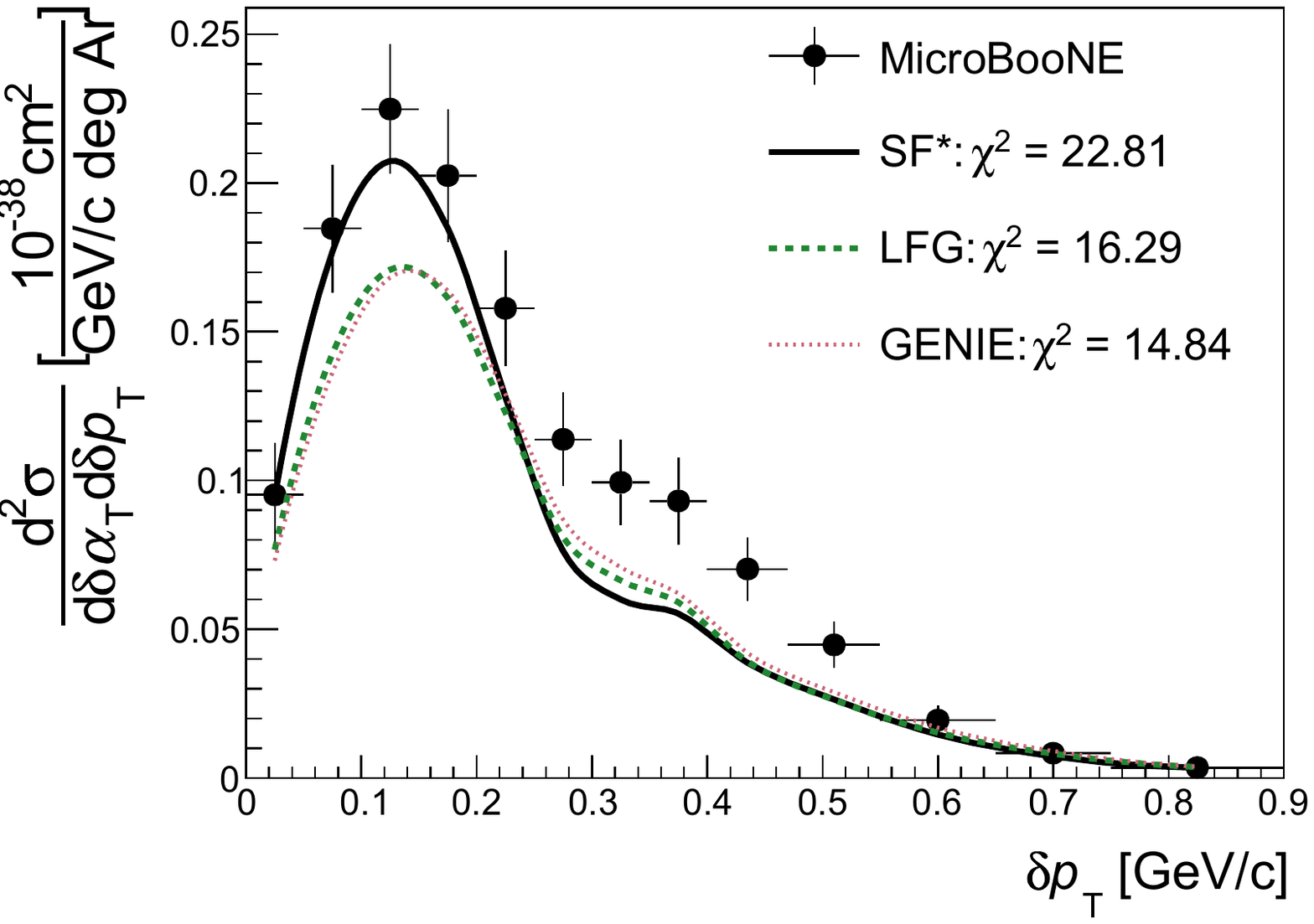}};
    \draw (1.8, 0) node {$90^\circ<\dalphat<135^\circ$};
\end{tikzpicture}
\begin{tikzpicture}
    \draw (0, 0) node[inner sep=0] {\includegraphics[width=8cm]{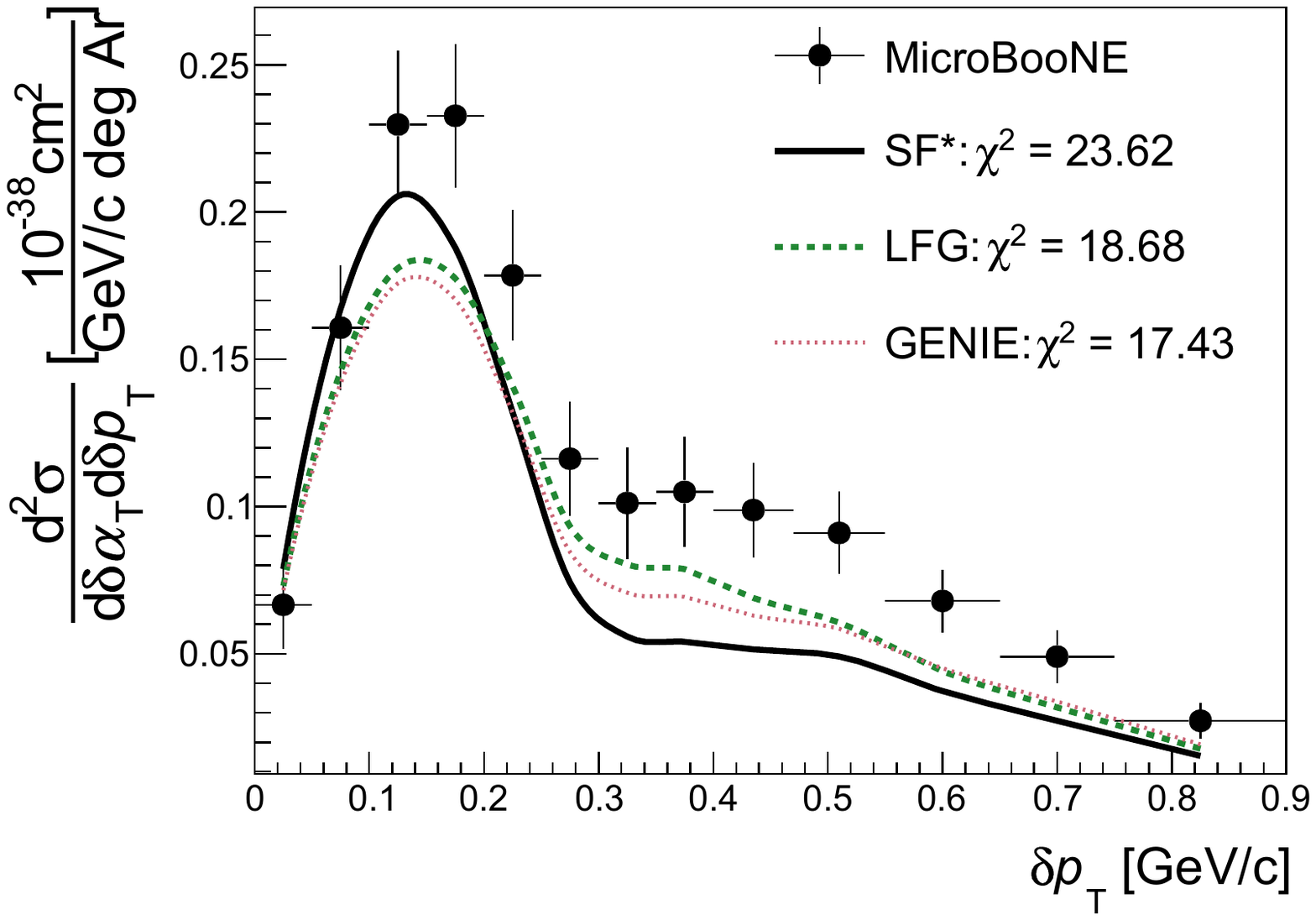}};
    \draw (1.8, 0) node {$135^\circ<\dalphat<180^\circ$};
\end{tikzpicture}

\caption{Multi-differential cross-section measurement as a function of $\dpt$ and $\dalphat$ from MicroBooNE, compared with the SF*, NEUT LFG and the GENIE \argenie predictions.}
\label{fig:dptvsdatGenComp}
\end{figure*}

The predictions from the different generators are shown in \autoref{fig:gencomp_all} for the $\dpt$ measurements by T2K, MINERvA and MicroBooNE, and in \autoref{fig:dptvsdatGenComp} for the multi-differential $\dpt$ measurement from MicroBooNE. The obtained $p$-values are summarized in \autoref{tab:chi2_pValues_genComp}. The NEUT LFG moel and GENIE prediction with the \argenie model yield better $p$-values with the SF/SF* models, giving reasonable agreement with all T2K and MicroBooNE measurements (including when the measurement is split into slices of $\dalphat)$ but all models fail to describe the MINERvA measurements. 

In the case of T2K, the GENIE and the NEUT LFG models are roughly equivalent, which is expected given their almost identical treatment of the nuclear model and the low impact of 2p2h and pion-absorption processes. The agreement is slightly worse for the NEUT SF model, primarily driven by the transition region between the tail and the bulk. Since the modelling of 2p2h and pion-absorption processes is similar between the different simulations, the difference in shape in this region may be driven by SRCs. The latter are included in the NEUT SF model and are approximated in the GENIE model, but completely absent from NEUT LFG.

In the case of the MicroBooNE measurement, the GENIE and NEUT LFG models yield the best overall agreement with the measurement, although the SF* model is not excluded. Both the GENIE and the NEUT LFG model fall below the SF* prediction in the bulk of the MicroBooNE measurement, which can be explained by the higher cross section predicted by the SF* model, as was discussed in \autoref{sec:nucmodel}. Although the SF* model seems to better agree visually with the $\dpt$ bulk, it yields the lowest $p$-value ($p$-value=0.12) due to the shape of the $\dpt$ tail and the transition between the bulk and the tail, which the measurement appears very sensitive to. In the MicroBooNE multi-differential comparison, shown in \autoref{fig:dptvsdatGenComp}, we observe the same trends for $\dalphat<90^\circ$, whereas for $\dalphat>90^\circ$ the impact of the FSI model in the tail is more visible. This region highlights the shortcomings in the modelling of FSI processes from all three generators (NEUT, GENIE and NuWro), as the agreement between the measurement and the simulations worsens with increasing $\dalphat$, albeit at different rates. An important difference between these models concern the total cross section of 2p2h processes, for which the GENIE simulation uses the SuSAv2 model as opposed to the Valencia model used in NEUT LFG and SF. SuSAv2 predicts a total cross section (before phase space constraints) which is $\sim$30\% higher than that predicted by the Valencia model at MicroBooNE energies. Although part of this increase is visible (notably in the $\dpt$ tail at $135^\circ<\dalphat<180^\circ$), it is insufficient to cover the discrepancy between the simulation and the measurement. The variation tested in \autoref{sec:2p2h} was indeed larger than the difference between the SuSAv2 and Valencia models at MicroBooNE and still proved to be insufficient. 

Finally, the MINERvA measurement presented in \autoref{fig:gencomp_all} excludes all models. As previously discussed, the difference between the NEUT LFG and GENIE models in terms of QE processes is very small. The largest differences between the generator predictions stem from the modelling of 2p2h and pion absorption processes. However, most models broadly reproduce the tail, and the model-measurement agreement is mainly driven by the simulation of the QE-dominanted bulk and the transition region between the bulk and the tail, where correlations between adjacent bins play a major role.

\begin{table}[]
\centering
\begin{tabular}{l|c|ccc}
Measurement                           & $N_{bins}$ & SF/SF*                       & LFG                          & GENIE                        \\
\hline
T2K $\dpt$                            & 8          & 0.08                         & 0.69                         & 0.47                         \\
\hline
MINERvA $\dpt$                        & 24         & \cellcolor[HTML]{FFCCC9}0.00 & \cellcolor[HTML]{FFCCC9}0.00 & \cellcolor[HTML]{FFCCC9}0.00 \\
\hline
MicroBooNE $\dpt$                     & 13         & 0.12                         & 0.42                         & 0.73                         \\
MicroBooNE $\dpt$ low $\dalphat$      & 11         & 0.26                         & 0.23                         & 0.28                         \\
MicroBooNE $\dpt$ mid-low $\dalphat$  & 12         & 0.07                         & 0.40                         & 0.41                         \\
MicroBooNE $\dpt$ mid-high $\dalphat$ & 13         & \cellcolor[HTML]{FFCCC9}0.04 & 0.23                         & 0.32                         \\
MicroBooNE $\dpt$ high $\dalphat$     & 13         & \cellcolor[HTML]{FFCCC9}0.03 & 0.13                         & 0.18                         \\
\end{tabular}
\caption{$p$-values obtained from $\chi^2$ under a Gaussian error approximation between different models and measurements. GENIE here is \argenie. $N_{bins}$ gives the number of bins for each measurement. $p$-values below 0.05, broadly indicating model rejection, are marked in red.}
\label{tab:chi2_pValues_genComp}
\end{table}

\section{Conclusions}
\label{sec:conclusions}

A joint analysis of measurements of TKI variables by T2K, MicroBooNE and MINERvA has been shown to provide a unique opportunity to highlight and disambiguate issues in modelling neutrino-nucleus interactions in the few-GeV regime. In particular, measurements of $\dpt$ have shown sensitivity to variations of Fermi motion, 2p2h and FSI (both on nucleons and to absorb pions), whilst $\dalphat$ has shown sensitivity mostly to FSI, but with some sensitivity to 2p2h. Furthermore, the multi-differential measurement of TKI variables by the MicroBooNE collaboration allows further untangling of nuclear effects, such as distinguishing the impact of altering 2p2h interactions from alterations to FSI properties. Comparisons of MicroBooNE and T2K measurements are sensitive to how nuclear effects scale with nuclear target, whilst comparisons of T2K and MINERvA are sensitive to their energy dependence. 

Quantitatively (assuming the validity of the Gaussian uncertainties in the covariance matrices provided by experiments), no model or systematic variation considered can describe all measurements and in particular the MINERvA measurements of $\dpt$ and $p_N$ reject all of them. Conversely, many of the models are allowed by MicroBooNE measurements, although those with a weaker FSI strength are disfavoured.

In terms of nuclear models, the RFG model is clearly excluded both qualitatively and quantitatively by all $\dpt$ measurements. The LFG model and the SF (and SF*) model achieve much better agreement with the experimental measurements. However, conclusions related to the performance of a nuclear model are coupled to the modelling of hadron transport through the nucleus and must be interpreted with caution. 

LFG or SF simulations of T2K and MINERvA measurements, from the NEUT generator, broadly find a good description of the proportion of events in the bulk and tail of $\dpt$. However, in the MicroBooNE case, all simulations lack significant strength in the tail. Alterations of FSI strength on top of the NuWro-based SF* model are insufficient to cover the discrepancy. Moreover, any attempt to raise the tail with FSI depletes the bulk to the detriment of measurement-model agreement. Changing NuWro's FSI model to NEUT's improves agreement much more than variations to NuWro's FSI, despite the models having similar overall nuclear transparency. This is because of differences in the way the FSI alters the nucleon kinematics. Still, even with NEUT's FSI model, the simulations still under-predict the tail, particularly at intermediate $\dalphat$.

The other primary means to enhance the tail relative to the bulk is to vary the poorly-known 2p2h contribution, and indeed stronger 2p2h is preferred by MicroBooNE, contrary to the cases for T2K and MINERvA. Given that even large variations of FSI cannot give simulations the strength needed to match MicroBooNE's observation of the $\dpt$ tail (and certainly not without breaking agreement with the bulk), there appears to be reasonable evidence for a mis-modelling of 2p2h strength differences on carbon and argon. We note again that the GiBUU event generator predicts a carbon/argon cross-section ratio often more than twice that of the 2p2h models considered in this work~\cite{Buss:2011mx,Dolan:2018sbb}. Conversely, the good agreement with the relative tail-bulk strength between T2K and MINERvA implies a reasonable modelling of 2p2h energy dependence, but it should be noted that this is degenerate with possible variations of pion absorption FSI (which affects MINERvA much more than T2K).

Whilst confronting the generator configurations used by current and future experiments with the experimental measurements, it is found that MINERvA measurements exclude all of them. The T2K and MicroBooNE measurements are broadly compatible with the LFG-based configurations, whereas the SF* model is excluded at high values of $\dalphat$ by the MicroBooNE measurements, indicating an insufficient FSI strength. 

In conclusion, a comparative analysis between T2K, MINERvA and MicroBooNE measurements reveals:
\begin{itemize}
    \item evidence for stronger 2p2h contributions for neutrino-argon interactions;
    \item considerable sensitivity to FSI and particularly how it changes outgoing nucleon kinematics;
    \item a clear preference for more sophisticated nuclear ground state models, like LFG or SF rather than RFG. 
\end{itemize}

The statistical power and granularity of existing measurements, as well as the lack of predictive power for hadron kinematics for non-QE processes in models, prevents unambiguous conclusion on how exactly each process should change. Future measurements offer an opportunity to further lift degeneracies between nuclear effects. In particular, multi-differential measurements of TKI variables from MINERvA and T2K (in particular those using the latter's upgraded near detector with tracking thresholds comparable to MicroBooNE) offer opportunities to complement those of MicroBooNE. Higher statistics measurements from SBND will allow increasingly differential measurements (for example using calorimetric energy as an additional separator of nonQE processes) whilst higher energy measurements from ICARUS will allow an evaluation of the scaling of nuclear effects up to energies more relevant for DUNE. Additional measurements of TKI in other topologies are also promising, for example considering exclusively interactions with more than one proton or with/without tagged neutrons to target specific QE or nonQE enhanced topologies, allowing further disambiguation of nucleon FSI, pion absorption and 2p2h.

\begin{acknowledgments}

All authors would like to acknowledge support from the T2K, MicroBooNE and MINERvA collaborations in the completion of this work. We offer particular thanks to Afroditi Papadopoulou, Luke Pickering, Kevin McFarland and Ulrich Mosel for feedback on preliminary versions of this work. We further thank Afroditi for technical support using the MicroBooNE measurement's data release. We additionally thank the EP-NU group at CERN both for funding WF's summer internship and for numerous discussions. An important thanks is given to Ciaran Hasnip for providing important technical insights. LM~and~SD also thank Bongo Joe buvette, Geneva.

\end{acknowledgments}

\bibliography{apssamp}

\end{document}